%% file: thesis.tex
\documentclass[oneside,11pt]{prlthesis}
\usepackage{latexsym,epsf}
\usepackage[dvips]{graphicx}
\usepackage{subeqnar}
\usepackage{psfig}

\begin{document}
\title{Studies in Topics Going Beyond 
The Standard Electroweak Model}
\author{Vempati Sudhir Kumar} 

\dept{Physical Research Laboratory}
\guide{Prof. Saurabh D. Rindani}
\submitdate{November 2000}


\copyrightfalse
\coguidefalse
\figurespagefalse
\tablespagefalse
\beforepreface

\input {ack.tex}

\afterpreface

\input def.tex
\input chap1.tex

\input chap2.tex

\input chap3.tex

\input chap4.tex

\input chap5.tex

\input chap6.tex

\input chap7.tex

\end{document}

%% file: ack.tex
\begin{center}
{\underline{\sc Acknowledgment}}
\end{center}

\noindent
Most of this thesis is based on work I had done with Prof. Anjan Joshipura.
In addition to his insights in physics, his nice and kind nature made
me cherish every moment of my interaction with him. I feel fortunate
to have been able to work with him.

\noindent
Prof. Saurabh Rindani had been extremely supportive and encouraging
through out my thesis period. I would like to thank him for his
concern and constant encouragement.  Dr. Subhendra Mohanty's influence
on me extends not only on physics but also in several other spheres
of life. I feel greatly indebted to him.

\noindent
I would also like to thank Prof. J. C. Parikh, Prof. A. C. Das, 
Prof. A. R. Prasanna, Dr. D. R. Kulkarni
 and Dr. Sai Iyer who taught us several courses over the years. 
Discussions with Prof. U. Sarkar, Dr. R. Rangarajan and Dr. H. Misra
have been highly educative and I would like to thank them for their
patience. I would also like to thank Prof. V. B. Sheorey and 
Prof. D. P. Dewangan for their constant support. My discussions with
Prof. D. P. Roy, Prof. P. N. Pandita, Prof. M. Drees and Prof. S. F. King
have taught me several new things during my collaborations with them. 
I am thankful to them. 

\noindent
It is a pleasure to thank Dr. V. Ravindran, Dr. G. Dutta, Dr. S. Sahu,
Dr. M. S. Santhanam, Dr. A. Gupta, Dr. P. Stockinger, Dr. S. Goswami, 
Dr. S. Pandit, Mr. K. V. Shajesh and Mr. R. Vaidya from whom 
I have learnt several things. 

\noindent
The staff of computer center and the library have been extremely helpful
to me. I am very thankful to them.

\noindent
I enjoyed every moment of my stay with my `batchmates', AdS, Alok, Dipu, Esfan,
Jitti, Koushik, Mittu, Muthu, Nanda, Rajesh, Rajneesh, Sankar, Som, Sunish 
and Vinay. My sincere gratitude lies with them. I would also like
to thank Arun, Jayendra, Sarika, Chandan, Tarak, Jay and RP for making the
seventh floor lively. 

\noindent
I am also greatful to my parents, sisters, babai, aunts, cousins and 
friends at home, who have been patient, supportive and caring. 

\noindent
{\sl Sudhir Vempati}

%% file: def.tex
\def\ap#1#2#3{           {\it Ann. Phys. (NY) }{\bf #1} (19#2) #3}
\def\arnps#1#2#3{        {\it Ann. Rev. Nucl. Part. Sci. }{\bf #1} (19#2) #3}
\def\cnpp#1#2#3{        {\it Comm. Nucl. Part. Phys. }{\bf #1} (19#2) #3}
\def\apj#1#2#3{          {\it Astrophys. J. }{\bf #1} (19#2) #3}
\def\asr#1#2#3{          {\it Astrophys. Space Rev. }{\bf #1} (19#2) #3}
\def\apjl#1#2#3{         {\it Astrophys. J. Lett. }{\bf #1} (19#2) #3}
\def\ass#1#2#3{          {\it Astrophys. Space Sci. }{\bf #1} (19#2) #3}
\def\jel#1#2#3{         {\it Journal Europhys. Lett. }{\bf #1} (19#2) #3}

\def\ib#1#2#3{           {\it ibid. }{\bf #1} (19#2) #3}
\def\nat#1#2#3{          {\it Nature }{\bf #1} (19#2) #3}
\def\nps#1#2#3{          {\it Nucl. Phys. B (Proc. Suppl.) } {\bf #1} (19#2) #3}

\def\np#1#2#3{           {\it Nucl. Phys. }{\bf #1} (19#2) #3}

\def\pl#1#2#3{           {\it Phys. Lett. }{\bf #1} (19#2) #3}
\def\pl#1#2#3{           {\it Phys. Lett. }{\bf #1} (19#2) #3}
\def\pr#1#2#3{           {\it Phys. Rev. }{\bf #1} (19#2) #3}
\def\prep#1#2#3{         {\it Phys. Rep. }{\bf #1} (19#2) #3}
\def\prl#1#2#3{          {\it Phys. Rev. Lett. }{\bf #1} (19#2) #3}
\def\pw#1#2#3{          {\it Particle World }{\bf #1} (19#2) #3}
\def\ptp#1#2#3{          {\it Prog. Theor. Phys. }{\bf #1} (19#2) #3}
\def\rpp#1#2#3{         {\it Rep. on Prog. in Phys. }{\bf #1} (19#2) #3}
\def\ptps#1#2#3{         {\it Prog. Theor. Phys. Suppl. }{\bf #1} (19#2) #3}
\def\rmp#1#2#3{          {\it Rev. Mod. Phys. }{\bf #1} (19#2) #3}
\def\zp#1#2#3{           {\it Zeit. fur Physik }{\bf #1} (19#2) #3}
\def\fp#1#2#3{           {\it Fortschr. Phys. }{\bf #1} (19#2) #3}
\def\Zp#1#2#3{           {\it Z. Physik }{\bf #1} (19#2) #3}
\def\Sci#1#2#3{          {\it Science }{\bf #1} (19#2) #3}
\def\n.c.#1#2#3{         {\it Nuovo Cim. }{\bf #1} (19#2) #3}
\def\r.n.c.#1#2#3{       {\it Riv. del Nuovo Cim. }{\bf #1} (19#2) #3}
\def\yf#1#2#3{           {\it Yad. Fiz. }{\bf #1} (19#2) #3}
\def\zetf#1#2#3{         {\it Z. Eksp. Teor. Fiz. }{\bf #1} (19#2) #3}
\def\zetfpr#1#2#3{    {\it Z. Eksp. Teor. Fiz. Pisma. Red. }{\bf #1} (19#2) #3}
\def\pl#1#2#3{           {\it Phys. Lett. }{\bf #1} (19#2) #3}
\def\pr#1#2#3{           {\it Phys. Rev. }{\bf #1} (19#2) #3}
\def\prep#1#2#3{         {\it Phys. Rep. }{\bf #1} (19#2) #3}
\def\prl#1#2#3{          {\it Phys. Rev. Lett. }{\bf #1} (19#2) #3}
\def\pw#1#2#3{          {\it Particle World }{\bf #1} (19#2) #3}
\def\ptp#1#2#3{          {\it Prog. Theor. Phys. }{\bf #1} (19#2) #3}
\def\rpp#1#2#3{         {\it Rep. on Prog. in Phys. }{\bf #1} (19#2) #3}
\def\ptps#1#2#3{         {\it Prog. Theor. Phys. Suppl. }{\bf #1} (19#2) #3}
\def\rmp#1#2#3{          {\it Rev. Mod. Phys. }{\bf #1} (19#2) #3}
\def\zp#1#2#3{           {\it Zeit. fur Physik }{\bf #1} (19#2) #3}
\def\Zp#1#2#3{           {\it Z. Physik }{\bf #1} (19#2) #3}
\def\Sci#1#2#3{          {\it Science }{\bf #1} (19#2) #3}
\def\n.c.#1#2#3{         {\it Nuovo Cim. }{\bf #1} (19#2) #3}
\def\r.n.c.#1#2#3{       {\it Riv. del Nuovo Cim. }{\bf #1} (19#2) #3}
\def\sjnp#1#2#3{         {\it Sov. J. Nucl. Phys. }{\bf #1} (19#2) #3}
\def\yf#1#2#3{           {\it Yad. Fiz. }{\bf #1} (19#2) #3}
\def\zetf#1#2#3{         {\it Z. Eksp. Teor. Fiz. }{\bf #1} (19#2) #3}
\def\zetfpr#1#2#3{         {\it Z. Eksp. Teor. Fiz. Pisma. Red. }{\bf #1} (19#2)
 #3}
\def\jetp#1#2#3{         {\it JETP }{\bf #1} (19#2) #3}
\def\mpl#1#2#3{          {\it Mod. Phys. Lett. }{\bf #1} (19#2) #3}
\def\ufn#1#2#3{          {\it Usp. Fiz. Naut. }{\bf #1} (19#2) #3}
\def\ppnp#1#2#3{           {\it Prog. Part. Nucl. Phys. }{\bf #1} (19#2) #3}
\def\cnpp#1#2#3{           {\it Comm. Nucl. Part. Phys. }{\bf #1} (19#2) #3}
\def\ijmp#1#2#3{           {\it Int. J. Mod. Phys. }{\bf #1} (19#2) #3}
\def\ic#1#2#3{           {\it Investigaci\'on y Ciencia }{\bf #1} (19#2) #3}
\def\tp{these proceedings}
\def\pc{private communication}
\def\ip{in preparation}

\newcommand{\TeV}{\,{\rm TeV}}
\newcommand{\GeV}{\,{\rm GeV}}
\newcommand{\MeV}{\,{\rm MeV}}
\newcommand{\keV}{\,{\rm keV}}
\newcommand{\eV}{\,{\rm eV}}
\newcommand{\Tr}{{\rm Tr}\!}
\renewcommand{\arraystretch}{1.2}
\newcommand{\be}{\begin{equation}}
\newcommand{\ee}{\end{equation}}
\newcommand{\bea}{\begin{eqnarray}}
\newcommand{\eea}{\end{eqnarray}}
\newcommand{\ba}{\begin{array}}
\newcommand{\ea}{\end{array}}
\newcommand{\bmat}{\left(\ba}
\newcommand{\emat}{\ea\right)}
\newcommand{\refs}[1]{(\ref{#1})}
\newcommand{\ler}{\stackrel{\scriptstyle <}{\scriptstyle\sim}}
 \newcommand{\ger}{\stackrel{\scriptstyle >}{\scriptstyle\sim}}
\newcommand{\lag}{\langle}
\newcommand{\rag}{\rangle}
\newcommand{\ns}{\normalsize}
\newcommand{\cm}{{\cal M}}
\newcommand{\gr}{m_{3/2}}
\newcommand{\pa}{\partial}
\def\si{\psi}
\def\Si{\Psi}

\def\321{$SU(3)\times SU(2)\times U(1)$}
\def\tl{{\tilde{l}}}
\def\tL{{\tilde{L}}}
\def\bd{{\overline{d}}}
\def\tL{{\tilde{L}}}
\def\a{\alpha}
\def\b{\beta}
\def\g{\gamma}
\def\c{\chi}
\def\d{\delta}
\def\D{\Delta}
\def\db{{\overline{\delta}}}
\def\Db{{\overline{\Delta}}}
\def\e{\epsilon}
\def\l{\lambda}
\def\n{\nu}
\def\m{\mu}
\def\nt{{\tilde{\nu}}}
\def\p{\phi}
\def\P{\Phi}
\def\x{\xi}
\def\k{\kappa}
\def\r{\rho}
\def\s{\sigma}
\def\t{\tau}
\def\th{\theta}
\def\ne{\nu_e}
\def\nm{\nu_{\mu}}
\def\snui{\tilde{\nu_i}}
\def\nmu{\n_\m}
\def\ne{\n_e}
\def\barnmu{\bar{\nu}_\mu}
\def\barne{\bar{\nu}_e}
\def\dm21{\Delta {\mbox m}^2_{21}}
\def\dm32{\Delta {\mbox m}^2_{32}}
\def\dm{\Delta {\mbox m}^2}
\def\nmut{\stackrel{(-)}{\nu_{\mu}}}
\def\net{\stackrel{(-)}{\nu_{e}}}

%% file: chap1.tex
\renewcommand{\le}{\left(}
\newcommand{\ri}{\right)}
\newcommand{\cov}{{\cal{D}}}
\def\nm{\nu_{\mu}}
\def\snui{\tilde{\nu_i}}
\def\lm{{\cal L}}
\def\row{\rightarrow}

\chapter{Introduction}

``What lies beyond the Standard Model ?" is the question which
intrigues many particle physicists today. It is surprising
that such a question should arise when most of the existing
data agree very well with the predictions of the Standard Model.
However, there are strong theoretical arguments suggesting existence
of new physics beyond the Standard Model. Independent of these arguments, 
there are some recent experimental results which may be considered as 
the first indications of some new physics. Put together, these arguments
constitute a `strong evidence' for the presence of physics beyond
the Standard Model and thus the above question. 

At present, the question has not been answered. Future 
experiments like LHC (Large Hadron Collider) are likely to
provide an answer. To satisfy some of the theoretical arguments
against it, the Standard Model is usually extended by assuming an 
additional symmetry called supersymmetry. One of the most popular of 
such models is the Minimal Supersymmetric Standard Model (MSSM). The 
popularity of this model lies in the fact that it not only satisfies 
one of the major theoretical prejudices, namely the hierarchy problem, 
but is also testable in the future colliders. On the other hand, the 
experimental `evidence' for physics beyond the Standard Model comes from
the recent results of the atmospheric neutrino experiments
which indicate that the neutrinos may have small masses and non-zero mixing 
among them. The Standard Model which
has no provision for neutrino masses thus has to be extended in 
some sector {\it viz.,} with additional symmetries, additional particles 
etc. These extensions are achieved in many cases, independent of the theoretical
prejudices, thus leading to a different set of extensions of the 
Standard Model. 

In this thesis, we assume supersymmetry as the physics beyond the
standard model. We then study the generation of neutrino masses 
within supersymmetric standard models and implications from the 
results of the solar and atmospheric neutrino experiments on these
models. But, 
before proceeding further, we try to motivate our work in this
chapter, along with a summary of salient features of the MSSM. 

\section{The Standard Model} 

As a starting point, we here briefly review the salient features
of the Standard Model. The Standard Model (SM)  \cite{smold} is a
spontaneously broken Yang-Mills quantum field theory describing the 
strong and electroweak interactions.  The theoretical assumption on 
which the Standard Model rests on is the principle of local gauge 
invariance with the gauge group given by,
\be
\label{gsm}
G_{SM} \equiv SU(3)_c \times SU(2)_L \times U(1)_Y
\ee

The particle spectrum and their transformation properties
under these gauge groups are given as,

\bea
\label{smspectrum}
Q_i \equiv \bmat{c} {u_L}_i \\ {d_L}_i \emat \sim 
\le 3,~ 2,~ {1 \over 6} \ri
&\;\;\;\;& U_i \equiv {u_R}_i \sim \le 3,~ 1,~ {2 \over 3} \ri \nonumber \\
&\;\;\;\;& D_i \equiv {d_R}_i \sim \le 3,~ 1,~ -{1 \over 3}\ri \nonumber \\
L_i \equiv \bmat{c} {\n_L}_i \\ {e_L}_i \emat \sim \le 1,~ 2,~ -{1 \over 2} \ri
&\;\;\;\;& E_i \equiv {e_R}_i \sim  \le 1,~ 1,~ -1 \ri \nonumber \\
\eea
In the above $i = 1,2,3$ stands for the generation index.
$Q_i$ represents the left handed quark doublets, $L_i$ represents left
handed lepton doublet, $U_i,~D_i,~E_i$ represent right handed up-quark, 
down-quark and charged lepton singlets respectively. 
The numbers in the parenthesis represent the transformation properties of
the particles under $G_{SM}$ in the order given in eq.(\ref{gsm}). 
In addition to the fermion spectra represented above, there is also 
a fundamental scalar called Higgs whose transformation properties 
are given as,
\be
\label{smhiggs}
H \equiv \bmat{c} H^+ \\ H^0 \emat \sim \le 1,~2,~{1 \over 2} \ri
\ee

There are also gauge boson fields which enter the spectrum through
the requirement of local gauge invariance.  The total lagrangian with 
the above particle spectrum and gauge group can be represented as, 

\be
\lm_{SM} = \lm_F + \lm_{YM} + \lm_S + \lm_{yuk}.
\ee

\noindent
The fermion part, $\lm_F$ is given as,
\be
\label{lfer}
\lm_{F} = i \bar{\Si} \g^\m \cov_\m \Si
\ee
with
\be
\Si = \le Q_i~U_i,~ D_i,~ L_i,~ E_i \ri
\ee
where $\cov_\m$ represents the covariant derivative of the field given as,
\be
\label{cova}
\cov_\m  = \partial\m - i g_s G_\m^A \l^A - i{g \over 2} W_\m^I \tau^I  
- i g' B_\m Y 
\ee 

Here $A= 1,..,8$ with $G_\m^A$ representing the $SU(3)_c$ gauge bosons,
$I = 1,2,3$ with $W_\m^I$ representing the $SU(2)_L$ gauge bosons. The
$U(1)_Y$ gauge field is represented by $B_\m$. The self interactions of 
the gauge fields are given by,
\be
\label{lym}
\lm_{YM} = - { 1\over 4} G^{\m\n A} G_{\m\n}^A -{1 \over 4} W^{\m\n I} 
W_{\m\n}^I - {1 \over 4} B^{\m\n} B_{\m\n}
\ee
with 
\bea
G_{\m\n}^A&=&\pa_\m G_\n^A - \pa_\n G_\m^A + g_s ~f_{ABC} 
G_\m^B G_\n^C \nonumber \\
F_{\m\n}^I&=&\pa W_\n^I - \pa_\n W_\m^I + g~ f_{IJK} W_\m^J W_\n^K \nonumber\\
B_{\m\n}&=& \pa_\m B_\n - \pa_\n B_\m ,
\eea
where $f_{ABC (IJK)}$ represent the structure constants of the 
$SU(3)(SU(2))$ group.  

\noindent
The scalar part of the lagrangian is given by,
\be
\label{lscalar}
\lm_{S} =   \le \cov_\m H \ri^{\dagger} \cov_\m H - V(H), 
\ee where
\be
V(H)  =  \m^2 H^\dagger H + \l  \le H^\dagger H \ri^2
\ee
and finally the Yukawa part is given by,
\be
\label{lyuk}
\lm_{yuk} = h^u_{ij} \bar{Q}_i U_j \tilde{H} + h^d_{ij} \bar{Q}_i D_j H + 
h^e_{ij} \bar{L}_i E_j H + H.c
\ee
where $\tilde{H} = i \sigma^2 H^\star$. The symmetry $G_{SM}$ is 
not realised in nature. It has to be broken to $G_{SM}~ \rightarrow~ 
SU(3)_c \times U(1)_{em}$. This is achieved in the Standard Model in
a  renormalisable way by spontaneous symmetry breaking. The Higgs field 
attains a vacuum expectation value ({\it vev}) in this mechanism called
the `Higgs mechanism'.  Fermions (except neutrinos) and gauge bosons 
attain mass through this mechanism. Only one degree of freedom remains
for the Higgs field, the Higgs boson. 

The Standard Model is renormalisable \cite{thooft} and anomaly free 
\cite{jackiw}. 
The fermion spectra of the SM has been completely  discovered \cite{databook}
with the recent discovery of Tau-neutrino ($\n_\tau$). In addition 
to verifying the predicted neutral current interactions, experiments are 
now able to probe higher order radiative corrections with precision 
measurements \cite{prelang}.  The results from these experiments agree 
with the predictions of the SM very well \cite{smstatus}. There has been 
no indication so far for deviations from the Standard Model physics 
\footnote{Please see \cite{zwirner}, for a recent fit to precision 
electroweak observables and their comparison to SM predictions.} in this case.
 On the other hand, the scalar sector of the SM is not very well known. 
The Higgs boson is yet to be discovered. Precision observables suggest a 
light Higgs scalar $\sim 60-200$ GeV , whereas the present experimental 
bounds put it within 106 GeV $~< m_H <~$ 235 GeV \cite{marciano}.  
The first number comes from the non observation of $e^+e^- \rightarrow ZH$
at LEPII, whereas the upper bound comes from a global fit to the data 
\cite{erler}. Thus we see that the Standard Model 
stands on a firm footing both theoretically and experimentally, 
except for the scalar sector which is yet to be discovered. Incidentally,
it is the existence of a fundamental scalar in the Standard Model that 
gives rise to most of the conceptual problems with the SM, when SM is 
being incorporated in a bigger theory. But before that we will see the
need for extending the Standard Model.

In spite of the success of the Standard Model in explaining most of the
observed phenomena, there are several features of the SM which still 
need to be understood. The SM contains about 20 parameters which are
fixed by the experiments. Thirteen of them are related to the masses
of the quarks and leptons ( nine fermion masses, three mixing angles
and one phase of the CKM matrix). All these masses which are generated
by the Yukawa part of the lagrangian, $\lm_{yuk}$ (eq.~\ref{lyuk}) are
hierarchical in nature in the generation space. At present, we have no
understanding of the origin of this hierarchical nature. Similarly, 
we have no understanding of why there are only three generations within
the SM. All these are collectively known as the `flavour problem' of
the SM. As a solution to this problem, one typically extends the SM
with an additional `flavour symmetry'. The quantum numbers of the 
SM fields under this symmetry would now decide the magnitude of the
masses these particles take. 

Whereas the above arguments require an extension of the SM for a 
deeper understanding of the Standard Model, recent results from 
the atmospheric neutrino experiments make this requirement almost a  
necessity. The atmospheric neutrino experiments measure the
flux of the electron and muon neutrinos produced in the atmosphere 
due to cosmic ray collisions with the air molecules. These flux 
measurements (or rather the ratio of the fluxes ) are anomalous with 
respect to the predicted flux from various calculations. This 
reduction of the measured flux is generally understood in terms of 
`neutrino oscillations', which require neutrinos to have small masses. 
Recently, super-Kamiokande experiment has reported evidence for the existence
of neutrino oscillations \cite{evid}. If taken seriously, these 
results would imply
existence of neutrino masses. In addition to these signals from the
atmospheric neutrinos, there are other neutrino experiments like the
solar neutrino experiments which also suggest neutrinos to have small
masses. As we have seen earlier, the SM does not have right handed 
neutrinos in its spectrum eq.(\ref{smspectrum}), thus  denying neutrinos 
any mass through $\lm_{yuk}$ \footnote{Neutrinos cannot attain majorana 
masses either, as lepton number is `accidentally' conserved in the SM.}. 
Thus one has to  extend the SM in some sector 
(symmetries, particles or both) to generate mass for the neutrinos. 
One of the most standard methods to generate neutrino mass is to add 
right-handed neutrinos in to the Standard Model particle spectrum. 
Gauge invariance would allow not only Dirac mass terms but also 
Majorana mass terms for the right handed neutrino fields. The interplay
between these terms gives rise to a `see-saw' mechanism, 
in which the left handed neutrinos attain small majorana masses.

Thus we have seen that the SM has to be extended in order to satisfy
requirements coming from neutrino experiments as well as for a deeper
theoretical understanding. There is also a search for a Grand Unified
Theory (GUT) \cite{gutmohap}  which would unify the strong and the 
electroweak forces with a single coupling constant. It is interesting to
note that in-addition to unification of forces,  neutrino masses can also be
naturally incorporated within the Grand Unified theories \cite{gutseesaw}
through the `see-saw' mechanism discussed above.
The Standard Model would then be an effective theory valid up to 
some scale relevant to the weak scale. The scale at which the GUT 
theories would take over from the SM is estimated from the life
time of proton and `Renormalisation Group' running of the three 
coupling constants to be $\sim 10^{16}$ GeV \cite{langprep} 
\footnote{~eV-scale neutrino masses in some GUT theories would also 
require a scale of same order.}. 
Inclusion of the gravitational forces would push this scale further up to
the Planck scale, $M_{planck} \sim 10^{19}$ GeV. This would imply that 
the SM has to be valid right from the weak scale up to the GUT-scale 
typically, fourteen orders of magnitude. This would create a conceptual
problem within these theories called the hierarchy problem. 

The hierarchy problem arises due to the existence of a fundamental
scalar within the Standard Model \cite{gildener}. A typical Grand 
Unified Theory uses a single gauge group and a single
coupling constant to describe the physics at the GUT scale, $M_{GUT} \sim 
10^{16}$ GeV. This symmetry is then spontaneously broken at those
scales to the symmetry of the Standard Model,  $G_{SM}$ given in 
eq.(\ref{gsm}). As a consequence of this breaking, the gauge bosons 
mediating the grand unified interactions acquire heavy masses of the 
order of $M_{GUT}$ and thus would not be relevant for the physics at 
the weak scale. The gauge bosons of the Standard Model remain massless
as $G_{SM}$ remains unbroken. The fermion masses also would not be 
affected by this breaking as they are protected by chiral symmetries. 
But, the scalar mass is unprotected by any symmetry. There is no
reason to assume the scalar would not gain mass of the order of
$M_{GUT}$ due to the spontaneous breaking of $G_{GUT} \rightarrow G_{SM}$ 
\cite{gildener}. Such a large mass for the scalar is disastrous 
phenomenologically as the observed masses of the gauge bosons 
typically require a scalar mass of $O(100)$ GeV. To solve this `hierarchy' 
problem, one can either bring down the scale of the unification or
introduce an extra symmetry to protect the Higgs mass.  The former would
require introduction of extra space time dimensions \cite{add} whereas the
symmetries which would protect the Higgs mass are called supersymmetries. 
There is also one more approach in the literature which assumes that 
the Higgs boson is a composite particle of fermions thus ruling out the
existence of a fundamental scalar in the theory \cite{technicolor}. 

In the present thesis, we follow the approach of supersymmetry. 
Supersymmetry in addition to protecting the Higgs mass from attaining 
large values has other attractive features. A minimal supersymmetric 
version of the standard model can be built which is predictable and 
testable at present and future colliders. The unification of gauge
coupling constants at high scales is exact in this case and so, 
Grand Unified Theories also prefer the existence of supersymmetry.   
More fundamental theories like string theories may also
require supersymmetry to be present as low energy effective theories
\cite{peskin}. Thus supersymmetric standard models  make an attractive
framework as the  physics beyond the Standard Model. Below, we review some
salient features of the supersymmetric version of the Standard Model. 

\section{The Minimal Supersymmetric Standard Model}

Supersymmetry is a symmetry that transforms a fermion in to a boson and
vice versa. To understand how it protects the Higgs mass, let us consider
the hierarchy problem once again. The Higgs mass enters as a bare mass
parameter in the Standard Model lagrangian, eq.(\ref{lscalar}). 
Contributions from the self energy diagrams of the Higgs are 
quadratically divergent pushing the Higgs mass up to cut-off scale 
\cite{susskind}.  In the absence of any other new physics at intermediate 
energies, the cut-off scale is typically $M_{GUT}$ or $M_{planck}$.  
Cancellation of these divergences with the bare mass parameter 
would require fine-tuning of order one part in $10^{-36}$ rendering 
the theory `unnatural' \cite{natural}. On the other hand, if one has an 
additional contribution from a fermionic loop, with the fermion carrying 
the same mass as the scalar, the contribution from this diagram would now 
cancel the quadratically divergent part, with the total contribution now
being only logarithmically divergent \cite{mdine}. This would require
a symmetry which would relate a fermion and a boson with same mass. 
Such symmetries are known as supersymmetries. 

Supersymmetries were first introduced in the context of string 
theories by Ramond \cite{ramondtext}. In quantum field theories,
this symmetry is realised through fermionic generators \cite{haag} thus
escaping the no-go theorems of Coleman and Mandula \cite{colm}. The simplest 
Lagrangian realising this symmetry in four dimensions was built by  Wess
and  Zumino \cite{wzmodel} which contains a spin ${1 \over 2}$ fermion 
and a scalar. To build interaction lagrangians one normally resorts to 
the superfield formalism of Salam and Strathdee \cite{salamst}, as 
superfields simplify addition and multiplication of the representations
\cite{wessbagger}. It should be noted however that the component fields 
may always be recovered from superfields by a power series expansion
in grassman variable.

A chiral superfield contains a weyl fermion,  a scalar and and an auxiliary
scalar field generally denoted by F. A vector superfield contains a spin 
1 boson, a spin 1/2 fermion and an auxiliary scalar field called D. 
A minimal supersymmetric extension of the Standard Model \cite{csaba} 
is built by replacing every standard model matter field
by a chiral superfield and every vector field by a vector superfield.
Thus the existing particle spectrum of the Standard Model is doubled. 
The particle spectrum of the MSSM \footnote{The same
notation is followed in the entire thesis.} and their 
transformation properties under $G_{SM}$ is given by,

\bea
\label{mssmspectrum}
Q_i \equiv \bmat{cc} u_{L_i} & {\tilde u}_{L_i} \\ 
d_{L_i} & \tilde{d}_{L_i} \emat \sim 
\le 3,~2,~{1 \over 6} \ri
&\;\;& U_i^c \equiv \bmat{cc} u_{R_i}^{~~c}  & \tilde{u}^{c}_i \emat
 \sim \le 3,~ 1,~ {2 \over 3} \ri \nonumber \\
&\;\;& D_i \equiv \bmat{cc} d_{R_i}^{~~c} & \tilde{d}^{c}_i  \emat
\sim \le 3,~ 1,~ -{1 \over 3} \ri \nonumber \\
L_i \equiv \bmat{cc} \n_{L_i} & \tilde{\n}_{L_i}  \\ e_{L_i}& 
\tilde{e}_{L_i} \emat 
\sim \le 1,~ 2,~ -{1 \over 2} \ri
&\;\;& E_i \equiv \bmat{cc} e_{R_i}^{~~c} & \tilde{e}^{c}_i \emat \sim 
 \le 1,~ 1,~ -1 \ri \nonumber \\
\eea
The scalar partners of the quarks and the leptons are typically named
as `s'quarks and `s'leptons. For example, the scalar partner of the top
quark is known as the `stop'. In the above, these are represented by
a `tilde' on their standard counterparts. A distinct feature 
of the supersymmetric standard models is that they require two Higgs 
fields with opposite hypercharges to cancel the anomalies and to 
give mass to both up-type and 
the down-type quarks \footnote{The hermitian conjugate of the superfield 
is not allowed into the superpotential due to its  holomorphic nature.} 
These transform as

\bea
H_1 &\equiv& \bmat{cc} H^{-}_1 & \tilde{H}_1^{-} \\ H^{0}_1 & 
\tilde{H}_1^0 \emat 
\sim \le 1,~2,~-{1 \over 2} \ri  \nonumber \\
H_2 &\equiv & \bmat{cc} H^+_2 & \tilde{H}_2^+  \\ H^0_2 & \tilde{H}_2^0 \emat
\sim \le 1,~2,~{1 \over 2} \ri 
\eea
The fermionic parts of the Higgs superfields are known as Higgsinos. 
The gauge bosons and their fermionic partners, gauginos are represented 
as,

\bea
\label{vsf}
V_s^A & : & \bmat{cc} G^{\m A} & \tilde{G}^A \emat \nonumber \\
V_w^I & : & \bmat{cc} W^{\m I}& \tilde{W}^I \emat \nonumber \\
V_y & :  &\bmat{cc} B^{\m} & \tilde{B} \emat
\eea

\noindent
In the above, the fermionic (susy) counter parts of the Higgs and 
the gauge bosons have been represented by a `tilde' on their bosonic 
notations. The total Lagrangian of the MSSM is of the form :
\be
\label{mssmlag}
\lm_{MSSM} = \int d \th^2 ~W(\Phi) + \int d \th^2~ d {\bar \th}^2 ~
\Phi_i^\dagger~ e^{g V}~ \Phi_i + \int d \th^2~ {\cal W}^\a {\cal W}_\a. 
\ee 
In the above, $W(\Phi)$ is a function of the chiral superfields  called
the superpotential, $\Phi$ representing a generic superfield. The 
renormalisable superpotential is of dimension three or less. For the MSSM, 
the superpotential invariant under $G_{SM}$ is given as, 
\be
\label{super}
W  =  W_1 + W_2, \\
\ee
\noindent
where
\bea
\label{superyuk}
W_1 &=& h^u_{ij} Q_i U_j^c H_2 + h^d_{ij} Q_i D_j^c H_1 + 
h_{ij}^e L_i E_j^c H_1 + \m H_1 H_2 \\
\label{superrpv}
W_2&=& \e_i L_i H_2 + \l_{ijk} L_i L_j E_k^c + \l'_{ijk} L_i Q_j D_k^c 
+ \l''_{ijk} U_i^c D_j^c D_k^c .
\eea

The second term in the RHS of eq.(\ref{mssmlag}) represents the 
matter-gauge boson coupling, with $ V = \le V^A_s, V^I_w, V_y \ri$, 
with the appropriate coupling constants. The last term in the RHS of 
the eq.(\ref{mssmlag}) represents the self interaction of the gauge fields.
 The Lagrangian in component form can be found by expanding the superfields 
as noted above. The scalar potential can be derived by eliminating the 
auxiliary fields $D$ and $F$ which appear in the definitions of the 
vector superfield and chiral superfield respectively.

\section{R-parity}

In the above, we have seen the minimal supersymmetric version of the
Standard Model. Comparing eq.(\ref{super}) with the Standard Model 
Lagrangian, we see that in additional to doubling of the particle
spectrum, new interactions which violate baryon number and individual
lepton number are now allowed in the lagrangian. This is because
the Higgs superfield, $H_1$ and the lepton superfields $L_i$ have
the same quantum numbers under $G_{SM}$. Though they carry the
same quantum numbers in the SM also, these interactions are not
possible as the fermions and the Higgs transform as different
representations under Lorentz Group. Within the MSSM the distinction
is lost since the superfields corresponding to the  leptons and the Higgs
transform identically under supersymmetry and gauge symmetry
 and these interactions appear as they are gauge invariant.

The first three terms of the second part of the superpotential $W_2$
(eq.(\ref{superrpv})), are lepton number violating whereas the last term 
is baryon number violating. 
The simultaneous presence of both these interactions can lead to proton
decay for example, through a squark exchange. Stringent limits can be placed
on the products of these couplings from the life time of 
proton \cite{vissani}. To avoid proton decay, one
can either remove both these couplings or assume one set (either baryon
number violating or lepton number violating couplings) to be zero. The
former is normally arrived at by imposing a discrete symmetry on the
lagrangian called R-parity. R-parity  has been originally
introduced as a discrete R-symmetry \footnote{R-symmetries are symmetries
under which the $\th$ parameter transform non-trivially.} by Ferrar and
Fayet \cite{fayetferrar} and then later realised to be of the following 
form by Ferrar and Weinberg \cite{weinferra}:
\be
R_p = (-1)^{3 (B-L) + F},
\ee
where  B and L represent the Baryon and Lepton number respectively and
F represents the Fermion parity given as -1 for fermions and + 1 for
bosons. Under R-parity the transformation 
properties of various superfields can be summarised as:

\bea
\label{rptrans}
\{ V_s^A , V_w^I , V_y \} &\row& \{ V_s^A , V_w^I , V_y \} \nonumber \\
\th &\row& - \th^{\star} \nonumber \\
\{Q_i, U^c_i, D^c_i, L_i, E_i^c \}&\row&
 -\{Q_i,U^c_i,D_i^c,L_i,E_i^c \}\nonumber \\
\{ H_1, H_2\}&\row& \{H_1,H_2\}
\eea

Imposing R-parity has an advantage that it provides a natural
candidate for dark matter. This can be seen by observing that
R-parity distinguishes a particle from its superpartner. Thus
the lightest supersymmetric particle (LSP) cannot decay and remains
stable \cite{ellis}. R-parity has also been motivated as a remnant 
symmetry in a large class of supersymmetric theories,  especially in 
theories with an additional $U(1)$ symmetry. It has also been pointed out 
that though R-parity constraints baryon and lepton number violating 
couplings of dimension four, dim 6 operators which violate baryon and 
lepton numbers are still allowed by R-parity \cite{rpvwein}. 

\section{Supersymmetry breaking}

The MSSM lagrangian built so far is invariant under supersymmetry. 
Supersymmetry breaking has to be incorporated in the MSSM
to make it realistic. In a general lagrangian,  supersymmetry can be broken 
spontaneously if the auxiliary fields D or F appearing in the definitions 
of the chiral and vector superfields respectively attain a vacuum expectation
value ({\it vev}). Incorporation of this kind of breaking in MSSM 
using the MSSM superfields leads to phenomenologically inconsistent 
scenarios, like for example existence of a very light scalar \footnote{This
can be seen from the mass sum rules which appear as a consequence of 
spontaneous supersymmetry breaking \cite{palumbo}.}.  An alternative 
approach to incorporate supersymmetry breaking is using the `hidden-sector'
 scenarios. 

The hidden-sector is a sector of superfields which do not carry any 
quantum numbers under ordinary gauge interactions, i.e, $G_{SM}$. The
only interactions they share with the visible (MSSM) sector superfields 
is through gravity. Supersymmetry can then be broken spontaneously in 
the hidden-sector and this information is then passed on to the MSSM
sector through gravitational interactions. Since gravitational interactions
play an important role only at very high energies, 
$M_{p} \sim O(10^{19})$ GeV, the breaking information is passed on
to the visible sector only at those scales. The end effect of this mechanism
is that explicitly supersymmetry breaking terms are now added in to the
lagrangian. These `soft' terms do not reintroduce quadratic divergences back 
into the theory. The strength of the soft terms is characterised by, 
$M_S^2 / M_{planck}$, where $M_S$ is the typical scale of supersymmetry
breaking. These masses can be comparable to weak scale for $M_S \sim 10^{10}$
 GeV \cite{hpn}.  The `soft' supersymmetric breaking part of the lagrangian
can be classified to contain the following terms \cite{giraldello}:

\begin{itemize}
\item a) Mass terms for the gauginos, $M_1,M_2,M_3$.

\item b) Mass terms for the scalar particles, $m^2_{\phi_i}$ with
	$\phi_i$ representing the scalar partner of chiral 
	superfields of the MSSM. 

\item c) Trilinear scalar interactions, $A_{ijk} \phi_i \phi_j \phi_k$
	corresponding to  the cubic terms in the superpotential. 

\item d) Bilinear scalar interactions, $B_{ij} \phi_i \phi_j$ corresponding
	to the bilinear terms in the superpotential. 

\end{itemize}

\noindent
The total MSSM lagrangian is thus equal to 
\be
\lm_{total} = \lm_{MSSM} + \lm_{soft}
\ee
with $\lm_{MSSM}$ given in eq.(\ref{mssmlag}).

\subsection{Universality and CMSSM}
The above mechanism of supersymmetry breaking is called minimal
supergravity (mSUGRA) inspired supersymmetry breaking. As we have
seen above this type of breaking introduces several new soft 
parameters in to the theory. Typically the number of these parameters
is large $\sim 105$. Moreover, large flavour changing neutral currents
(FCNC) are also introduced by this kind of breaking \cite{susyflavmas}. 
To reduce the number of parameters as well as remove the large FCNC
contributing terms, an ansatz is usually followed at the high scale
where the soft terms are introduced in to the theory. This ansatz
is called universality and it assumes the following: 

\begin{itemize}
\item All the gaugino mass terms are equal at the high scale.\\
	$$M_1 = M_2 = M_3 = M$$
\item All the scalar mass terms at the high scale are equal.\\
	$$m_{\phi_i}^2 = m_0^2$$
\item All the trilinear scalar interactions are equal at the high
	scale. \\
	$$A_{ijk} = A$$

\item All bilinear scalar interactions are equal at the high scale. \\
	$$B_{ij} = B$$ 	
\end{itemize}

It is possible to construct supergravity models which can give rise
to such kind of strong universality \cite{hmura}. This approximation 
now drastically reduces the number of parameters of the
theory to about five, $m_0, M$~(or equivalently $M_2$), ratio of the
{\it vevs} of the two Higgs, tan$\beta$, $A$, $B$. Thus, these models
are also known as `Constrained' MSSM \cite{workgp} in literature. 
The low energy mass spectrum of the soft terms is now determined by
the Renormalisation Group scaling of those parameters. The 
supersymmetric mass spectrum of these models has been extensively 
studied in literature \cite{sugradrees}. The Lightest Supersymmetric
Particle (LSP) is mostly a neutralino in this case. 

\subsection{Gauge Mediated Supersymmetry breaking}

In the above we have seen that supersymmetry is broken at a high scale
and is communicated through gravity to the normal particle sector. It
induces large FCNC's and a large number of parameters which can be
corrected by assuming a `strong' universality at the $M_{GUT}$. 
An alternative mechanism has been proposed which tries to avoid
these problems in a natural way. The key idea is to use gauge interactions
instead of gravity to mediate the supersymmetry breaking from the 
hidden (also called secluded sector sometimes) to the visible MSSM sector
\cite{dineoriginal}. In this case supersymmetry breaking can be communicated 
at much lower energies $\sim 100$ TeV. 

A typical model would contain a susy breaking sector called `messenger 
sector' which contains a set of superfields transforming under a gauge
group which `contains' $G_{SM}$. Supersymmetry is broken spontaneously 
in this sector and this breaking information is passed on to the 
ordinary sector through gauge bosons and their fermionic partners in loops. 
The end-effect of this mechanism also is to add the soft terms in to the
lagrangian. But now these soft terms are flavour diagonal as they are
generated by gauge interactions. The soft terms at the messenger scale 
also have simple expressions in terms of the susy breaking parameters. 
In addition, in some models of gauge mediated supersymmetry breaking, 
only  one parameter can essentially determine the entire soft spectrum. 

In a similar manner as in the above, the low energy susy spectrum is 
determined by the RG scaling of the soft parameters. But now the high 
scale is around 100 TeV instead of $M_{GUT}$ as in the previous case. 
The mass spectrum of these models has been studied in many papers 
\cite{rattazi}. The lightest supersymmetric particle in this case is
mostly the gravitino in contrast to the mSUGRA case. 

\subsection{$SU(2) \times U(1)$ breaking}
As we have seen earlier, the supersymmetric version of the Standard Model 
is a two Higgs doublet model. A consistent incorporation of the 
$SU(2)_L \times U(1)_Y$ breaking puts constraints  relating various
parameters of the model. To see this, consider the neutral Higgs
part of the scalar potential. It is given as 

\bea
\label{vscalar}
V_{scalar}&=& (m_{H_1}^2 + \m^2) |H_1^0|^2 + (m_{H_2}^2 + \m^2 ) |H_2^0|^2 -
(B_\m \m H_1^0 H_2^0 + H.c) \nonumber \\
&+&  {1 \over 8} ( g^2 + g^{\prime 2}) ({H_2^0}^2 - {H_1^0}^2)^2 + \ldots,
\eea

\noindent
where $H_1^0, H_2^0$ stand for the neutral Higgs scalars and we have 
parameterised the bilinear soft term $B \equiv B_\m \m$.
The existence of a minima for the above potential would require 
at least one of the Higgs mass squared to be negative.
In both gravity mediated as well as gauge mediated supersymmetry
breaking models, such a condition at low energies can be naturally
incorporated. The soft parameters which appear in the above potential
are generally positive at the high scale. But radiative corrections
significantly modify the low scale values of these parameters,
making one of the Higgs mass to be negative at the weak scale. This 
mechanism is called radiative electroweak symmetry breaking. 
In addition to ensuring the existence of a minima, one would also 
require that the minima should be able to reproduce the standard model 
relations i.e, correct gauge boson masses.  This would give rise to 
constraints on the parameters known as the minimisation conditions. 
These are given as
\bea
\label{minime}
{1 \over 2} M_Z^2 &=& {m_{H_1}^2 - \tan^2 \b~ m_{H_2}^2 \over
\tan^2 \b  - 1} - \m^2  \\
\mbox{Sin} 2 \b &=& { 2 B_\m ~\m~ \over m_{H_2}^2 + m_{H_1}^2 + 2 \m^2 },
\eea

where $\tan \b = v_2 /v_1 $ is the ratio of the vacuum expectation
values of $H_2^0$ and $H_1^0$ respectively  \footnote{The above
minimisation conditions are given for the `tree level' potential only.
The minimisation conditions for the one-loop effective potential are
given in \cite{takamini}.}. In addition to the above conditions care
should also be taken such that charge and color breaking minima are
absent. 

\section{Motivation and Outline of thesis}

So far in this chapter we have seen the reasons for believing in 
the existence of physics beyond the Standard Model. A strong signal
comes from the recent neutrino oscillation experiments. On the other
hand, extension scenarios with unification ideas are generally plagued
by the hierarchy problem. A solution to this problem is assuming
supersymmetry just above the weak scale. Supersymmetric standard models
are built which can incorporate $SU(2) \times U(1)$ breaking naturally.  
The question then remains is whether one can incorporate neutrino masses 
and mixing within these theories in a natural way.

Supersymmetric Standard Model, unlike the Standard electroweak model has
a natural source of lepton number violation. Since, there is also 
baryon number violation in these theories, one imposes R-parity to remove
both these set of couplings. But R-parity is ad-hoc. One can always
assume symmetries other than R-parity like for example baryon parity 
\cite{bparity} which can remove the baryon number violating couplings,
leaving us with lepton number violating couplings only. In the presence
of these lepton number violating couplings, neutrinos attain majorana
masses \footnote{Even with R-conservation, higher dimensional R-parity 
violating operators would give rise to small neutrino masses 
\cite{rpvwein}.}. The generation of the neutrino masses in these 
theories and whether they are of the correct order to satisfy the 
solar and atmospheric neutrino masses is the main subject of this 
thesis. In addition to neutrino masses R-parity violating theories
have a different set of experimental signatures in complete contrast
to the standard MSSM signatures \cite{herbi}. We also study some such
experimental signatures in the context of some recent experimental 
results of HERA detector.

The outline of the thesis is as follows. In chapter 2, we discuss
the various neutrino experiments and some standard neutrino mass
models. In chapter 3, we study the standard Renormalisation group
analysis of the MSSM and then try to understand the effect of RG
analysis on the neutrino mass spectrum. In chapter 4, we discuss
a specific model of R-violation namely bilinear
R-violation within the framework of gauge mediated models 
supersymmetry breaking. In chapter 5,
we discuss models with trilinear lepton number violation and the feasibility
 of   simultaneous solutions to solar and atmospheric neutrino problems. 
In chapter 6, we derive bounds on the trilinear $\l'_{1jk}$ couplings 
from neutrino mass and study its implications for the HERA anomalies.
 We end with conclusions and future outlook in chapter 7.

%% file: chap2.tex
\chapter{Neutrino Anomalies and Mass Models}
Neutrinos are one of the most abundant Standard Model particles in
the universe. In addition to the various natural radioactive sources,  
neutrinos  are produced in the hydrogen burning process in
the stars as well as when a star dies in supernova explosions. 
They are also produced when energetic cosmic rays collide with
the air molecules. There is also a cosmic neutrino background. 
Though neutrinos are abundant in the universe, it is an irony 
that they are the least understood. In the Standard Model, neutrinos 
have no mass, spin ${1 \over 2}$ and carry no electric charge.  They take 
part only in weak interactions,  which makes them extremely difficult 
to detect as their cross-sections are much smaller compared to 
electro-magnetic cross-sections \cite{kayser}. 

Of the above properties of the neutrinos, the spin and charge of the
neutrinos are experimentally well known. The mass of the neutrino
is relatively unknown. Experiments which put kinematic limits on
the neutrino mass directly are difficult to conduct and put weak limits
\cite{neutpdg}. However, the abundant sources of neutrinos, the
stars and the atmosphere help us in understanding the properties of
neutrinos further. As mentioned earlier, neutrino cross-sections are
extremely small. Thus to detect them one would need in addition to
large fluxes (which are naturally provided by these sources), large
detectors too. These detectors measure the number of neutrinos 
produced for example, in the cosmic ray showers in the atmosphere. The 
results from these experiments over the years consistently pointed
towards the phenomena of `neutrino oscillations', which in turn indicates 
existence of neutrino masses. 

The above neutrino oscillation experiments can be broadly divided as
a) solar neutrino experiments,  b) atmospheric neutrino experiments 
and c) laboratory experiments, ~depending on the source of neutrinos. 
Below, we consider some details of these experiments and how they
determine the pattern of the neutrino mass matrix. 

\section {Neutrinos from the Sun and the Solar Neutrino Anomaly}

As we have seen above, neutrino are produced in the stars which burn
hydrogen fuel. In the Sun, this process produces as a byproduct 
electron neutrinos, $\n_e$. Since the $\n_e$ interact
weakly with the solar atmosphere, they can escape the Sun without much
changes in their flux or energy and thus making it possible to measure
their flux and energy on the earth. For over thirty years starting from
1967 \cite{raydavis}, this flux of $\n_e$ has been measured on the earth. 
It is found that the measured flux is only about one-third 
of the standard expectation.  This discrepancy constitutes 
the {\it solar neutrino problem}. Several other experiments like 
Kamiokande, SAGE, GALLEX and super-Kamiokande have since then 
confirmed the existence of this problem, the last one with very 
high statistics. 

\vskip 0.8cm
\begin{tabular}{cccc}
Source & Flux & Average Neutrino & Maximum Neutrino \\
& ($cm^{-2} s^{-1})$& Energy (MeV)&Energy (MeV) \\
\hline
\hline\\[6pt]
{\it pp}& $(5.94 \pm 0.06) \times 10^{10}$&0.2668&$0.423\pm 0.03$\\[6pt]
{\it pep}&$(1.39 \pm 0.01) \times 10^8$&1.445&1.445\\[6pt]
$^7Be$&$(4.80 \pm 0.43)\times10^9$&0.3855&0.3855\\[2pt]
&&0.8631&0.8631\\[6pt]
$^8B$&$(5.15^{+0.98}_{-0.72})\times 10^6$&$6.735 \pm 0.036$&$ \sim 15$\\[6pt]
{\it hep}&$2.10 \times 10^3$&9.628&18.778\\[6pt]
$^{13}N$&$(6.05^{+ 1.15}_{-0.77})\times 10^8$&0.7063&$1.1982 \pm 0.0003$\\[6pt]
$^{15}O$&$(5.32^{+1.17}_{-0.80})\times 10^8$&0.9964&$1.7317 \pm 0.0005$\\[6pt]
$^{17}F$&$(6.33^{+0.76}_{-0.70})\times 10^6$&0.9977&$1.7364 \pm 0.0003$\\[6pt]
\hline
\end{tabular}
\vskip 0.3cm
{\bf Table 2.1}~~{\it In the above, we present the total flux of the
neutrinos coming from various interactions along with their average
and maximum energies \cite{grimus}}.

In the Sun, the main reactions which produce the energy can be
grouped as the {\bf pp} cycle and the {\bf CNO} cycle. The final outcome
of both these cycles can be given as,
\be
4~ p + 2~ e^-  \rightarrow 2 ~\n_e + ^4 \mbox{He}_2 + Q
\ee

Here $Q = 26.72$ MeV represents the energy released in this process. 
A major part of the energy is in the form of photons. A
small part ($\sim 2\%$) is taken by the neutrinos. Neutrinos are 
produced in various intermediate reactions within the {\bf pp} and 
the {\bf CNO} cycles. In the {\bf pp} cycle, these intermediate 
reactions are named ( after their reactants ) as
{\em pp, pep, $^7 Be$, $^8 B$, hep}. 
In the CNO cycle the corresponding reactions involve
{\em $^{13} N$, $^{15} O$,$^{17} F $}. The cross-sections of these
reactions are determined by the standard model weak interaction physics.
But the total flux of the neutrinos emanating from these reactions
depends on the chemical abundances in the solar interior which in turn
is dependent on the solar physics. 

These total fluxes are determined by a Standard Solar Model (SSM) developed
by J. Bahcall and his collaborators \cite{jnbbook}.
The model assumes that a) the Sun is in Hydrostatic equilibrium, b) Energy
transport happens in Sun by photons or by convective motions, c) energy
generation is done by nuclear reactions and d) changes in the chemical
abundances are solely due to nuclear reactions. With these assumptions,
the model then predicts the flux of the neutrinos to be 
observed at various experiments. It should be however noted that the
flux of the neutrinos also changes with the neutrino energy. Infact,
neutrinos produced from different reactions would have a different
energy spectrum. In the {\bf Table 2.1} we present, the total
fluxes of the neutrinos from various reactions as predicted
by the SSM and their corresponding average and maximum energies.

\vskip 0.8cm
\begin{tabular}{cccc}
Experiment&Result&Theory&${\hbox{Result} \over \hbox{Theory}}$\\[10pt]
\hline
\hline \\[6pt]
{\sc Homestake}&$2.56 \pm 0.16 \pm 0.16$ &
$7.7^{+ 1.2}_{-1.0}$&$ 0.33^{+0.06}_{-0.05}$\\[6pt]
{\sc Gallex}&$77.5 \pm 6.2^{+4.3}_{-4.7}$&
$129^{+8}_{-6}$&$ 0.60\pm 0.07$\\[6pt]
{\sc Sage}&$75.4^{+7.0 ~~+3.5}_{-6.8~~-3.0}$&
$147^{+8}_{-6}$&$0.51\pm0.07$\\[6pt]
{\sc Kamiokande}&$2.80 \pm 0.19 \pm 0.33$&
$5.15^{+1.0}_{-0.7}$&$0.54 \pm 0.07$ \\[6pt]
{\sc super-Kamiokande}&$2.40\pm0.03^{+0.08}_{-0.07}$&
$5.15^{+1.0}_{-0.7}$&$0.465 \pm 0.005 $\\[8pt]
\hline
\end{tabular} \\[4pt]
\noindent
{\bf Table 2.2:} {\it The results of solar neutrino experiments
confronted with the corresponding theoretical predictions.
The results of Homestake, GALLEX and SAGE experiments are
expressed in terms of event rates in SNU units
 (1 SNU $\equiv 10^{-36}$ $\mbox{events}$ $\mbox{atom}^{-1} s^{-1}$),
 whereas the results of the Kamiokande and Super-Kamiokande are expressed
in terms of the $^8 B$ neutrino flux in the units of
$10^6 \mbox{cm}^{-2} s^{-1}$. The first experimental error
is statistical and the second is systematic  \cite{grimus, nu2ksol}. }

\vskip 0.8cm
\noindent

The first experiment to measure the flux of the neutrinos coming
from the Sun was the chlorine {\sc Homestake} experiment which
measured about one-third of the standard expectation \cite{raydavis}.
This experiment has a high threshold ($\sim$ MeV) and hence was 
able to measure flux of neutrinos coming from $^8 B$ and $^7 Be$ only. 
It reported a deficit of about one-third of the SSM prediction. 
In 1988, the {\sc Kamiokande} experiment using a real time water cherenkov 
detector could point out the directionality of the incoming neutrinos and 
confirmed that neutrinos are indeed coming from the Sun. This experiment
too had a high threshold ($\sim 6.5$ MeV) and hence was able to see
only the $^8 B$ neutrino flux. This experiment reported a deficit
about one-half of the SSM prediction. In 1992, {\sc Gallex} and {\sc Sage}
with thresholds of order $\sim 200$ KeV were able to measure neutrinos
coming from the {\it pp} reactions also, confirming that the source of
Sun's energy is indeed through thermonuclear fusion reactions. They too
measured a deficit of about one-half. Recently, {\sc Super-Kamiokande}
experiment announced results with very high statistics a deficit of
one-half in the Boron neutrino flux to which the experiment is sensitive.
In {\bf Table 2.2} we have listed the latest results from 
various experiments along with the corresponding SSM predictions.
It should be however noted that the SSM is also being 
constantly updated over the years with latest inputs from various
experimental results and computational techniques. The theoretical 
predictions presented in {\bf Table 2.2}
are based on the latest version of the model called BP98 \cite{bp98}. 
An independent confirmation of the Standard Solar Model was recently
achieved by the helioseismological data which confirms the predictions
of the SSM for sound velocities in the solar interior to a very high
accuracy \cite{langacker}. 

In the above, we have seen that the measured solar neutrino flux
is about half the expectations based on the Standard Solar Model (SSM). 
One of the possible reasons for this deficit could be the expectations
themselves which are based on a model. However, it can be shown that 
independent of the SSM, one still faces problem to explain the data 
from various experimental results. Also model independent constraints
like luminosity constraint etc, which when used in conjunction with
the experimental data would lead to scenarios of the Sun which are
unphysical or severely constrained by other observations like helioseismology. 
For example,  using the luminosity constraint one can show that data
from {\sc Gallex} or {\sc Sage} would fit well if all the neutrinos
are coming from {\it pp} reactions only. Such a scenario which suppresses
other sources of the neutrinos is severely constrained by helioseismological
data. Similarly, a simultaneous fit for {\sc Super-Kamiokande} data and
{\sc Homestake} data would leave no room for the flux of neutrinos coming
from $^7 Be$. This result is again severely constrained by helioseismological
data \cite{grimus,phyrept}. Thus it looks like independent of  Standard
Solar  Model, one still has a problem in explaining the data coming from
solar neutrino experiments. In the present thesis, we believe that the 
SSM is correct and there exists a solar neutrino problem with respect
to the expectations based on the SSM. 

\section{Neutrinos from the Skies and the Atmospheric Neutrino Anomaly}

The other natural source of neutrinos comes from the cosmic rays.
Cosmic rays which have led to many fundamental discoveries in particle
physics early this century, have now contributed to our knowledge of
neutrinos significantly. Neutrinos are produced  when intergalactic
cosmic rays interact with the air molecules in the atmosphere. The
production mechanism can be summarised as follows:

\noindent
(1).~ Primary cosmic rays interact with the air molecules producing
kaons ($K^{\pm}$) and pions ($\pi^{\pm}$).\\
(2). These pions then decay to form a part of the neutrino ($\nmut$) flux
and muons ($\mu^{\pm}$). \\
(3). Lastly muons decay to give rest of the neutrino ($\nmut$) flux and 
the ($\net$) flux. \\

The actual flux which reaches the surface of the earth depends on a 
number of other factors like the properties and composition of 
the primary cosmic rays, modulations due to Solar wind and the 
geomagnetic field cut-off. These fluxes can be calculated within
some uncertainties. Whereas the modulations due to Solar wind 
and the earth's geomagnetic field can be incorporated within these
calculations, the major uncertainties comes from the large errors
in primary cosmic ray flux measurements \cite{hondagra}. 
However, one can still make
strong predictions of the neutrino flux which can be summarised 
as follows :

\vskip 0.2 cm
\noindent
(a) Both $\ne,\nmu$ and $\barne, \barnmu$ are produced in these
showers.

\vskip 0.2cm
\noindent
(b). One can also predict with very less uncertainties the ratio of 
$\nmu$ to $\ne$  in these fluxes. This is typically of order:
\be
\label{atmratio}
{ \nmu + \barnmu \over \ne + \barne} \approx 2
\ee
However, this would vary from place to place due to the geomagnetic
cut-off and with energy of the neutrinos.

\vskip 0.2cm
\noindent
(c). Flux of neutrinos with  energy greater than 5 GeV is not
affected either by geomagnetic cut-off or by solar modulation. 
\vskip 0.2cm 
\noindent
(d). One can also  predict fairly well the angular dependence of the 
neutrino flux at a given experimental site. Taking the 
direction of the neutrino to be the direction of the incoming cosmic 
ray \footnote{ This is a good approximation for neutrinos with energy
 $ \ger 100$ KeV.}, the angular dependence of the neutrino flux is 
characterised by two angles, the zenith angle ($\th$) and the azimuth
angle ($\phi$). The zenith angle dependence is caused by the increase
in the length of the air column (more probability of decay and thus
enhanced flux) at angles $\th \sim {\pi \over 2}$. But since this 
dependence is symmetric over $\pi \row \pi - \th$, this dependence
is characterised only by $|cos \th|$ and would not induce an up-down
asymmetry. The geomagnetic field can induce both zenith and azimuth
angle dependence on the neutrino flux. But, as we have seen earlier
for neutrino energies $\ger 5$ GeV, this effect is negligible. Thus
these effects would not be able to induce an up-down asymmetry in
the neutrino flux. 

\vskip 0.2cm
\noindent
Though neutrinos from the atmosphere have been detected long time ago 
at experiments at Kolar Gold Fields, India and in South Africa, 
detailed measurements of these fluxes have started only about
a decade ago. At the energies of the atmospheric neutrinos, the typical
detection process is through DIS (Deep Inelastic Scattering):
\be
\stackrel{(-)}{\n_l} + N \row l^{\pm} + X,
\ee 
where $l$ is the corresponding charged lepton, $N$ is the nucleon and 
$X$ is the remnant of the scattering. The lepton is then detected
either through water cherenkov detectors as in IMB, Kamiokande, 
super-Kamiokande or through iron calorimeters like Baksan and MACRO. 
As we have seen earlier that even though there are large uncertainties
in the primary cosmic ray fluxes ($\sim 20\%$), the predictions for
the ratio, eq.(\ref{atmratio}) are free from such uncertainties. Thus
the experimental results are usually presented in the form of 
`ratio of ratios' given as,
\be
R = {\tilde{R}_{exp} \over \tilde{R}_{MC}},
\ee
where 
\be
\tilde{R}_{exp(MC)}=
 \le { \nmu + \barnmu \over \ne + \barne} \ri _{exp(MC)} 
\ee

One would expect this ratio to be one. In the following Table, 
we present the results from various experiments.

\begin{tabular}{ccc}
Experiment & Ratio & Energy \\
\hline
\hline \\[6pt]
Kamiokande& $0.60^{+ 0.07}_{-0.06} \pm 0.05$ & sub-GeV \\[4pt]
Kamiokande & $0.57^{+0.08}_{-0.07}\pm 0.07$& multi-GeV\\[4pt]
super-Kamiokande&$0.652\pm0.019\pm0.051$&sub-GeV\\[4pt]
super-Kamiokande&$0.668\pm0.034\pm0.079$&multi-GeV\\[4pt]
IMB&$0.54\pm0.05\pm0.11$&\\[4pt]
Soudan-2&$0.61\pm0.15\pm0.05$&\\[4pt]
\hline
\end{tabular}
\vskip 0.4cm 
\noindent
{\bf Table 2.3} In the above we present results from the various
atmospheric neutrino experiments \cite{grimus,nu2katm}.

\noindent
From the above we see that the measured ratios are much different
from unity. This constitutes the {\it atmospheric neutrino anomaly}. 

\section{Neutrino Oscillations}

In the above we have seen that both solar and atmospheric neutrino 
experiments measure a deficit in the neutrino flux. This deficit
can be understood in terms of neutrino oscillations, requiring
neutrinos to have small masses. The main idea is that similar to 
$K^0 - \bar{K}^0$ oscillations, the current eigenstates of neutrinos
are not its mass eigenstates. This is expressed as,
\be
\le \ba{c} \n_e\\ \n_\m \\ \n_\tau \ea \ri = U 
\le \ba{c} \n_1\\ \n_2 \\ \n_3 \ea \ri ,
\ee

\noindent
where $\n_i$ are the mass eigenstates and $U$ is the mixing matrix. 
The neutrinos are produced and detected in their current basis, but
they propagate in the physical mass basis. Thus a  neutrino produced 
can now change its flavour as it propagates a distance L where it is
detected. Making simplifying assumptions about relativistic nature of 
the neutrino, a simple survival probability formula can be derived 
using the schr\"{o}dinger equation \cite{palash}. For example, for
the electron neutrino survival probability, this is given as \footnote{
For intricacies involving the neutrino oscillation formula, please see
 \cite{peter}.},
\be
\label{oscform}
P_{\n_e \n_e} = 1 -  4 \sum_{\stackrel{i,j}{i>j}} ~ U_{ei}^2 U_{ej}^2 Sin^2 \le 
{ \Delta m^2_{ij} L \over 4 E} \ri, 
\ee
where the LHS represents the survival probability. $\Delta m^2_{ij}$ 
represents the mass squared difference between $i,j$ neutrino mass 
eigenvalues and L represents the distance traveled. If there
are only two generations involved, this formula reduces to,

\be
\label{2flav}
P_{\n_e \n_e} = 1 -  Sin^2 2 \th Sin^2 \le { \Delta m^2 L \over 4 E} \ri,
\ee

\noindent
where now the mixing matrix is represented as a rotation by an angle $\th$.  
The solar and atmospheric neutrino anomalies can be understood in terms
of neutrino oscillations, if the mass squared differences of the neutrinos
and the mixing angles are of the right order. The experimental data now 
constraints regions in the parameter space of the neutrino oscillation
formula. The deficit in the data of a particular anomaly can be caused 
either by oscillations in two generations or in three generations. 
Another possibility is that the neutrinos oscillate in to a sterile neutrino.  

\vskip 0.3cm
\noindent
{\it CHOOZ and simultaneous solutions}
\vskip 0.3cm
In the present thesis, we consider that there are only three active
neutrinos and the solar and atmospheric neutrino anomalies have 
solutions through oscillations among these neutrinos. This would
generally need a three flavour analysis of the entire solar and
atmospheric neutrino data. However, assuming a hierarchical pattern 
for the neutrino masses \footnote{For a discussion
on the patterns of neutrino mass matrix allowed by the experiments,
including degenerate spectra please see \cite{asjpra}.}, an important 
constraint on the neutrino mixing matrix comes from the CHOOZ experiment
which simplifies such an analysis \cite{mohan}. 

The standard two flavour oscillation solutions for the solar and 
atmospheric neutrino data show an hierarchy in $\Delta m^2$. Assuming
neutrino masses also have an hierarchy \footnote{Either 
$m_{\n_1} \sim m_{\n_2} \ll m_{\n_3}$ or $m_{\n_1} \ll m_{\n_2} 
\ll m_{\n_3}$.} the general three flavour 
oscillation formulae for the solar and the atmospheric neutrino 
problems decouple and reduce to two flavour oscillation formulae 
as the mass squared differences are not linearly independent 
\footnote{$\Delta m_{21}^2 + \Delta m_{32}^2 + \Delta m_{13}^2 = 0$.}. 
These are given as,
\bea
P_{\n_e \n_e} &=& 1 - 4 U_{e1}^2 U_{e2}^2 Sin^2 \le 
{ \Delta m^2_{21} L \over 4 E} \ri - 2 U_{e3}^2 (1 - U_{e3}^2) \\
P_{\n_\m \n_\m} &=& 1 - 4 U_{\m 2}^2 U_{\m 3}^2  Sin^2 \le 
{ \Delta m^2_{32} L \over 4 E} \ri.
\eea

The mixing matrix element $U_{e3}$ is constrained by the CHOOZ experiment
to be small. The CHOOZ experiment which falls in the category of 
laboratory neutrino experiments \cite{choose},  rules out 
oscillations for the electron neutrino in the mass squared difference 
range, $\Delta m^2 \ger 10^{-3}$ with a large mixing $\sim 1$. For 
hierarchical pattern of neutrino masses, this in turn constraints
$U_{e3} \leq 0.22$. The above formulae are identical to the 2-flavour
oscillation formula, eq.(\ref{2flav}),
in the limit of small $U_{e3}$. In this case, we can concentrate only on
two flavour solutions to the neutrino anomalies. 

The standard solutions for the  the solar neutrino anomaly comprise of
three `regions'  in the oscillation parameter space of $\dm$ and $Sin^2 2\th$. 
One of the regions is called the `Vacuum Oscillations' or 
`just so'. The $\dm$ required is very small $\sim 10^{-11} \mbox{eV}^2$ 
and the mixing angle $\sim {\pi \over 4}$. The other two solutions
require a much larger $\dm$, typically of $O(10^{-6} \mbox{eV}^2)$.  
These two solutions consider matter effects on neutrino propagation 
whilst the neutrino traverses the distance between the core of the Sun 
and its surface. They allow for matter enhanced resonant conversion (MSW
mechanism) \cite{mswmech} of the neutrinos from electron species to another 
in specific density regions of the Sun. One of the solutions allows 
for a large mixing angle, $ \th \sim {\pi \over 4}$ - Large Angle MSW. 
The other region allows for a small mixing angle $\th \sim 10^{-3}$  
and is called Small Angle MSW \cite{krasbs}. In the case of atmospheric 
neutrino anomaly, the results are much more constrained. The analysis of
super-Kamiokande results \cite{sk98} allow solutions for regions
in $\Delta m^2 \sim 10^{-3}$ and $Sin^2  2\th \sim 1$. These results
are summarised in the table below:

\vskip 0.4cm
\begin{tabular}{cccc}
\hline
Anomaly&Solution&$\Delta m^2$& $Sin^2 2\th$\\[3pt]
&&$eV^2$&\\[3pt]
\hline
\hline
Solar& SAMSW&$(4-12)~10^{-6}$&$(3-11)~10^{-3}$\\[3pt]
& LAMSW&$(0.8-3)~10^{-5}$& 0.7-1. \\[3pt]
& VO&$(6-11)~10^{-11}$&0.47-1. \\[3pt]
Atmosphere&&$(4-6)~10^{-3}$&1\\[1pt]
\hline
\end{tabular}
\vskip 0.4 cm
\noindent
{\bf Table 2.4} In the above we present constraints on 2 flavour oscillation
solutions for solar and atmospheric neutrino anomalies. 

Whereas the above results have been valid during the period where
much of the thesis work has been done, recent results from the 
super-Kamiokande experiment do not favour some regions of the parameter
space presented above. These results have been systematically analysed
recently by \cite{garciavalle} and we summarise them in the following
table \footnote{These results are presented in terms of $tan^2 \th $ as
a convenience for the `dark side' of the parameter space \cite{dark}.}: 

\vskip 0.4cm
\begin{tabular}{cccc}
\hline
Anomaly&Solution&$\Delta m^2$& $ tan^2 \th$\\[3pt]
&&$eV^2$&\\[3pt]
\hline
\hline
Solar& SAMSW&$(1-10) \times 10^{-6}$&$(1-20) \times 10^{-4}$\\[3pt]
& LAMSW&$2  \times 10^{-5} - 10^{-3}$& 0.2~-~ 3. \\[3pt]
& LOW-QVO& $6 \times~10^{-7}- 5 \times 10^{-10}$&0.1 - 7. \\[3pt]
Atmosphere&&$(1.6-5.4) \times~10^{-3}$&0.43 - 2.7\\[3pt]
CHOOZ &~ &~ &~ $\ler$ 0.043\\[1pt]
\hline
\end{tabular}
\vskip 0.4 cm
\noindent
{\bf Table 2.5} Constraints on oscillation parameters in the light
of recent super-Kamiokande data are presented above \cite{garciavalle}.

As we have see from the above Table, the vacuum oscillation solution
is no longer favoured by the latest solar neutrino data. This is because of an 
independent observable called day-night spectrum which aids in 
distinguishing different possible oscillation scenarios for the solar
neutrinos. A `flat' day-night spectrum as such observed 
would be difficult to reconcile with the vacuum oscillation solutions.
However, there are other solutions of the type `Quasi-Vacuum Oscillations
(QVO)' which contain small matter effects are allowed by the recent
data. Most of the work in this thesis was done much before the recent
results from super-Kamiokande were announced. We will comment on the 
implications due to the new data at relevant places. 

\subsection{Evidence for neutrino oscillations}

Till now we have been assuming neutrino oscillations as a solution
to the solar and atmospheric neutrino anomalies. This assumption 
can be verified in the experiments both with laboratory sources 
as well as natural sources. One of the first indirect experimental 
observation of neutrino oscillations has been reported by the 
super-Kamiokande experiment recently \cite{skprl}. As we have seen earlier, 
the incoming atmospheric neutrino flux is independent of the zenith angle,
for energies  $\ger 5$ GeV. This is because the geomagnetic field
would not affect the primary cosmic ray fluxes at these energies. 
Zenith angle dependence  can also be detected if the involved 
neutrinos are undergoing oscillations. From eq.(\ref{oscform}), we
see that oscillation probability varies with the distance traveled
by the neutrino $L$ and the energy of the neutrino $E$. Neutrinos
coming from the atmosphere to the detector in the super-Kamiokande
experiment travel distances within a range of 15 Kms - 13,000 Kms.
The first number is for the neutrinos which are entering the detector
vertically (zenith angle $\sim 0$ )  whereas the latter number is for 
neutrinos which travel vertically upwards (zenith angle $\sim \pi$). 
If there are no neutrino oscillations, the number of neutrinos entering
the detector vertically downwards should be equal to the number of
neutrinos traveling upwards. This is quantified by the up-down asymmetry
given as

\be
A_\m = { U - D \over U + D} ,
\ee

\noindent
where $U$ represents the number of upward moving neutrinos and $D$ the
number of downward neutrinos. Super-Kamiokande experiment has measured
this number for the case of muon neutrinos and found it significantly
different from zero. The actual experimental number is \cite{skprl},
\be
A_\m = -0.296 \pm 0.048 \pm 0.01
\ee

\noindent
The non-zero up-down asymmetry provides first indirect evidence 
for the existence of neutrino oscillations.

At present, we do not have a strong experimental signature for the
existence of neutrino oscillations for solar neutrino experiments. 
The major reason being that different experiments measure different
regions in energy of the solar neutrino spectrum. While {\sc Super-Kamiokande}
has had extremely high statistics, it has been sensitive only to 
high energy neutrinos from $^8 B$ and {\it hep} neutrinos only. Most
of the neutrinos ($\sim 97 \%$) from the Sun are at very low energies 
({\it pp} neutrinos ) which have not been measured with high statistics. 
While efforts are on for the measurement of such neutrinos, results 
are awaited from the neutrino experiments like {\sc SNO} and {\sc Borexino}
which would help us to point at a single oscillation solution to the
solar neutrino problem and  clear the ambiguities \cite{smirsno}. 
The {\sc SNO} experiment
has a facility to accurately measure the neutral current interactions of
the neutrinos also. Thus, if the electron neutrinos from the Sun are
 converted to some other type of neutrinos (active), then the flux
of the other type of neutrinos can be detected by neutral current interactions. 
This way, we would get a measurement of the total $^8 B$ neutrino flux and
of course a clear signal of neutrino oscillations. The {\sc Borexino} 
experiment would be able to measure $^7 Be$ flux accurately. This would help
to conduct a model independent analysis of the solar neutrino flux. 
Thus, in the next few years we hope to have a single solution to the 
solar neutrino problem with accurate values of $\dm$ and $Sin^2 2 \th$. 

\vskip 0.4cm 
\noindent
{\it LSND and KARMEN}
\vskip 0.3cm 
\noindent
Till now we have concentrated only on solar and atmospheric neutrino 
anomalies.  In addition to these anomalies, there also has been an 
laboratory experiment called the Liquid Scintillator Neutrino Detector 
(LSND) which has reported evidence for neutrino oscillations. One 
characteristic feature which differentiates the LSND from the 
experiments discussed so far is that it is an `appearance' experiment. 
The LSND has reported evidence for $\bar{\n}_\m \rightarrow \bar{\n}_e$ 
transitions by observing an {\it excess} of $\n_e$ events \cite{lsnd} 
and thus a direct evidence for the evidence of neutrino oscillations. 
In terms of oscillation parameters, these results
require $\Delta m^2 \sim 1 eV^2$, $Sin^2 2 \th \sim 10^{-3}$. Whereas
the above results have been from experiments with $\m$ decay at rest,
later experiments with $\m$ decay in flight have been consistent with
their earlier results.  Incorporation of the LSND results would require
an additional neutrino species \footnote{See however, Barenboim and Scheck
\cite{sch} for scheme involving only three neutrinos.}, as it has a 
characteristically different mass-squared 
difference which cannot be incorporated within three neutrino species 
which allow only two mass-squared differences. 

The KARMEN experiment was originally built to verify the LSND 
results. The KARMEN experiment found negative results for the some 
of the allowed parameter space of the LSND results. However, it 
was realised that the KARMEN experiment would not be able to verify
the entire parameter space allowed by the LSND experiment \cite{grimus}. 
In this thesis where we consider only  three active neutrino species,
we do not consider the results from LSND. 

\section{Mechanisms of Neutrino Mass Generation }

We have seen that neutrino anomalies can be understood in terms
of neutrino oscillations if the neutrinos have masses \footnote{Alternative
scenarios are also considered in literature \cite{pack}. We do not
consider them here.}. In the hierarchical scenario, typically there 
are two distinct mass
scales needed for simultaneous solutions of solar and atmospheric
neutrino anomalies. The atmospheric mass scale $ \sqrt \Delta_A$ is 
$\sim 0.1$ eV, whereas the solar scale, $\sqrt \Delta_S$ can be either
$\sim 10^{-3}$ eV or $ \sim 10^{-5}$ eV. The mixing matrix has atleast
one large mixing in the 2-3 sector. Thus we see that neutrino mass
spectrum is characteristically different from the quark mass spectrum.

As we have seen in Chapter 1, the Standard Model has to be extended in some
sector to incorporate neutrino masses. In addition to explaining the smallness
of the neutrino masses, these models should also incorporate large mixing
as required by the solutions to the neutrino anomalies. Some of the
extensions may also have experimental signatures which may be stringently 
constrained. Thus a consistent model of neutrino masses should be able 
satisfy all the above requirements. Many such models have been proposed
in literature \cite{modelsmir}. Here instead of considering specific
models, we consider two of the most popular mechanisms which are often
used in literature to build models. 
 
\subsection{See-Saw Mechanism}
The most natural way of generating neutrino masses in the 
Standard Model is to add right handed neutrinos in to the model 
and demand neutrinos attain Dirac masses in the same
spirit as other fermions of the SM. But, in this case, it becomes
difficult to justify the  smallness of the neutrino mass. A natural
way of generating small neutrino masses in this manner was proposed
by Gellman, Ramond, Slansky \cite{grs} and 
Mohapatra, Senjanovic \cite{rnmp} and is called the 
see-saw mechanism. The key idea in to use to the majorana masses
for the right handed neutrinos, which are naturally allowed by the
gauge symmetry to suppress the masses for the neutrinos.  

Representing the three left handed fields by a column vector $\n_L$ and 
the three right handed fields by $\n_R$, the Dirac mass terms are 
given by,
\be
-\lm^D = \bar{\n}_L {\cal M}_D \n_R + H.c
\ee
where ${\cal M}^D$ represents the Dirac mass matrix. The majorana 
masses for the right handed neutrinos are given by, 
\be
-\lm^R = {1 \over 2} \bar{\n}_R^c {\cal M}_R \n_R + H.c
\ee
The total mass matrix is given as, 
\be
-\lm^{total} = {1 \over 2} \bar{\n}_p {\cal M} \n_p
\ee
where the column vector $\n_p$ is given as,
\be
\n_p = \le \ba{c} \n_L \\ \n_R^c \ea \ri
\ee
The matrix ${\cal M}$ is given as, 
\be
{\cal M}= \le \ba{cc} 0 & {\cal M}_D^T \\ {\cal M}_D & {\cal M}_R \ea \ri 
\ee

Diagonalising the above matrix, one sees that the left handed neutrinos
attain majorana masses of order, 
\be
{\cal M}^\n = - {\cal M}_D^T~ {\cal M}_R^{-1} {\cal M}_D
\ee 
This is called the seesaw mechanism. Choosing for example the 
Dirac mass of the neutrinos to be typically of the order of
charged lepton masses or down quark masses, we see that 
for a heavy right handed neutrino mass scale, (left handed) 
neutrinos  masses are suppressed. This way the smallness of
the neutrino masses can be explained naturally in this mechanism. 
The see-saw mechanism can be naturally incorporated in Grand
Unified Theories like SO(10) and also in left-right symmetric
models \cite{rnmp}. Though the actual scenarios are model dependent,
broadly for a range of $M_R \sim 10^5 - 10^{15}$ GeV, one attains 
light neutrinos in these models.  Large mixing \cite{asjsmir} and
degenerate spectra \cite{asjmoh} can also be realised in these models.
Recently an extensive analysis has been reported which studies
proton decay and neutrino masses in SO(10) GUT theories \cite{babuwilz}.

\subsection{Radiative Mechanisms}

In the above we have introduced additional fermions  with a heavy mass 
scale to generate small neutrino masses. One can instead modify the
scalar sector of the model, which is anyway not well understood. 
Neutrino masses are now generated radiatively and thus are naturally
small. This model is called the Zee model after A. Zee who first
proposed it \cite{zee}. 

Within the Standard Model neutrinos can attain majorana masses by
modifying the scalar sector. This can be seen by considering the
operator, $\e_{ab} L_i^a~ C~ L_j^b$, where $C$ is the charge
conjugation matrix and  $a,b$ are the $SU(2)$ indices and $i,j$ are
the generation indices. This can couple to a field transforming 
either as a singlet or a triplet under $SU(2)$. Models with triplet 
Higgs are considered unattractive as they contribute to $\rho$ 
parameter \cite{utpal}. Instead here we consider models with
a singlet field $h^+$. The coupling of the lepton fields to this
field is given as, $f^{ab} \e_{ab} L_i^a~ C~ L_j^b h^+$.

Though the above coupling violates lepton number one can always
conserve it by defining $h^+$ to have a lepton number of -2. 
However if one introduces an additional scalar $\phi_2$ (in-addition
to the already existing $\phi_1$), a new coupling of the form
$M_{\alpha \beta} \phi_\alpha \phi_\beta h^+$  is possible which
violates lepton number exactly by two units as required for 
neutrino mass generation. In this model which is named as 
$\{\phi_1 \phi_2 h\}$ model, neutrino attain masses at the one-loop
level and thus are naturally suppressed. One interesting fact
about this model is that the couplings $f_{ab}$ are antisymmetric
due to the $SU(2)$ metric. This would lead to an interesting texture
of the neutrino mass matrix whose diagonal elements are zero. 
Instead of adding an additional doublet one can as well add a doubly
charged singlet in to the model, $k^{++}$. In this case, neutrinos
attain masses at the two-loop level. This is popularly known as
Babu model in literature \cite{babu2l}. Including both the additional
doublet as well as the doubly charged singlet would lead to neutrino
masses both at the 1-loop level as well as at the 2-loop level. Such
a scenario may be required to understand neutrino anomalies in these
models with discrete symmetries like $L_e - L_\m -L_\tau$ 
\cite{asjsau}. 

In this thesis, we consider an alternative method to generate
neutrino masses. In these models neutrinos attain masses employing
both the `see-saw type' mechanism as well as radiative mechanisms. 
We will discuss them in detail in further chapters.

%% file: chap3.tex
\chapter{RG scaling and R violation}
Renormalisation Group (RG) Methods have been introduced by Gellman and Low
\cite{gmlow} to tackle the limitations of
perturbation theory in calculating processes at high momenta. Since
then they have been applied widely in particle physics and in
statistical mechanics \cite{kogut}. The main idea is that a change in 
the renormalised 1PI function $\Gamma_r$ \footnote{ We denote the 
renormalised function by $\Gamma_r(E,x,g,\m)$ where $E$ is the scale 
of energy of the process, 
$x$ collectively denotes the various momenta, ratios of momenta etc,
 $g$ are the coupling constants involved and $\m$ is the arbitrary scale
 introduced during regularisation like dimensional regularisation.}
 due to a change in scale of momentum can be compensated
by an appropriate change in the value of the coupling constant \cite{rgwein}. 
This would require the coupling constants
to be scale dependent. In a general regularisation process
one introduces an arbitrary mass scale $\m$. Using this scale, the 
coupling constants are now defined on a `sliding renormalisation scale'. 
But, the unrenormalised function $\Gamma$ is still independent of 
this arbitrary mass parameter $\m$ \cite{ryder,ramond}.  
Requiring the renormalised $\Gamma_r$ to be invariant under a change of 
scale (in momenta) and the unrenormalised 
$\Gamma$ to be invariant under a change of $\m$ would lead to a 
differential equation called the `Renormalisation Group Equation'\cite{ryder}. 
This equation now depicts the fact that a change in scale can be compensated
with a change in the coupling constant.  The end product of this analysis
are differential equations for the coupling constants with respect
to the momentum scale which are typically of the form:
\be
t {dg \over dt} = \beta(g)
\ee
where t represents the scale change in momenta. The functions $\beta$
are known as the beta functions of the theory. Thus, knowing the coupling
constant at one particular scale, the value of the coupling constant
at another scale can be known by integrating the above equation. 

Renormalisation group methods are used in field theory to study the 
asymptotic limits of the theory. For example, asymptotically free field
theories are those theories for which the $\beta$-function
vanishes at a high-scale ($E \rightarrow \infty$).
Various other features of the renormalisation group equations
(RGE) like fixed points, singularities, sign of the $\beta$-functions would help
to understand the nature of the theory at high energies. In particle physics,
renormalisation group studies have been used to mainly probe the nature
of physics beyond the Standard Model. Used in conjunction with the Grand Unified
models, renormalisation group equations can probe coupling constant
unification at high energies \cite{georgi}. The renormalisation group
studies of the Standard Model have been recently described in \cite{ramondsm}. 
In the present thesis, we concentrate on Renormalisation Group Studies in
the context of supersymmetric Standard Model (MSSM) \cite{revkaz} 
with and without R-parity violation. 

\section{Renormalisation Group and MSSM}
As we have seen renormalisation group studies allow us to probe the
scale of grand unification. Assuming only Standard Model fields to 
be present up to $M_{GUT}$, detailed analysis have shown that 
the gauge couplings do not `meet' at the same point in the momentum
scale, where the new theory can take over \cite{ramondsm}. 
The MSSM doubles the particle spectrum of the SM. This significantly
modifies the evolution of the gauge couplings leading to unification
of the gauge couplings at a scale, $M_{GUT} \approx 10^{16} $ GeV.
This can be easily verified from the renormalisation group equations 
of the gauge couplings which are presented in Appendix for the
MSSM case. The solutions of these equations are given as,
\be
\tilde{\alpha}_i(0) = {\tilde{\alpha}_i(tz) \over 
\left(1 - b_i \tilde{\alpha}_i(tz) \right) }
\ee 

\noindent
where the parameters appearing in the above are defined in the Appendix. 
Choosing approximate values for the gauge couplings at $M_Z$ as,
\be
\alpha_1(M_Z) \approx {1 \over 99};\;\; \alpha_2(M_Z) \approx {1 \over 30};\;\;
\alpha_3(M_Z) \approx {1 \over 8.5}
\ee

\noindent
A simple algebra leads us to see that \footnote{The factor ${5 \over 3}$
is required for normalisation of the $U(1)$ group. See \cite{mohapictp}.},
\be
{5 \over 3} \alpha_1(0) = \alpha_2(0) = \alpha_3(0) \approx {1 \over 24}
\ee

\noindent
for $M_{GUT} = 3 \times 10^{16} $ GeV. The actual analysis, including
two loop RGE and threshold effects predicts $\alpha_3(tz) = 0.129$ which
is slightly higher than the observed value. The couplings meet at the
value $M_X = 2 \times 10^{16}$ GeV \cite{mssmunif}. The `exact' unification
of the gauge couplings within the MSSM may or may not be an accident. But
it provides enough reasons to consider supersymmetric standard models 
seriously  as it links supersymmetry and grand unification in an 
inseparable manner \cite{mohapictp}. 

Along with the gauge coupling unification, some grand unified theories 
also predict bottom quark Yukawa $Y_b$ and tau lepton Yukawa, $Y_\tau$ 
unification at $M_{GUT}$. Renormalisation Group 
studies of the MSSM Yukawa couplings show $Y_b-Y_\tau$ unification 
\cite{schrempp} for large ranges
in tan$\beta$. In the Appendix, we have given the RGE for $Y_t, Y_b$
and $Y_\tau$. We have not presented here the RGE for the first two 
generation Yukawa couplings. Since the masses of these particles are small, 
the effect of their Yukawa couplings in the RG studies 
is generally small compared to the third generation Yukawa couplings. 
We have neglected them in most of the  analyses presented in this thesis 
except in the chapter 5 (chapter 6) where the second (first) generation 
Yukawa couplings play an important role. 

Renormalisation Group studies also play an important role in determining
the weak scale spectra of the MSSM soft sector. As we have seen in chapter 1,
supersymmetry is generally broken in a hidden sector. The breaking information
is then transferred to the visible sector through gravitational interactions
(mSUGRA). The result of this mechanism is that supersymmetry breaking 
soft terms are added in to the Lagrangian at a high scale $\sim M_{GUT}$. 
At the high scale `strong' universality is assumed with the respective
soft parameters being equal. The weak scale spectrum is modified  
due to radiative corrections involving Yukawa couplings and gauge parameters. 
The corresponding renormalisation group equations of the parameters 
reflect this phenomena. Using the RGE, one can determine the entire weak
scale spectrum in terms of the five basic parameters of the model,
tan $\beta, m_0,~ M ,~ A ~$and$~ B$.  Some salient features of these 
renormalisation group equations are :

\vskip 0.15cm
\noindent
(a). The gaugino masses evolve in the same manner as the gauge couplings. 
Thus the relation :
\be
{M_1 \over g_1} = {M_2 \over g_2 } = {M_3 \over g_3} 
\ee
holds true for most of the scales up to small two-loop effects \cite{spm}.
\vskip 0.15cm
\noindent
(b). The superpotential parameters are constrained by non-renormalisable 
theorems and thus the RGE of these parameters are proportional to 
themselves \cite{spm}.
\vskip 0.15cm
\noindent
(c). The soft parameters are unconstrained unlike the superpotential
 parameters.  Thus, even if the soft parameters are small or zero at 
the high scale, they can acquire large values at the weak scale due 
to RG scaling. 

\vskip 0.3 cm
\noindent
{\it Radiative electroweak symmetry breaking}:
\vskip 0.3 cm

As we have seen in Chapter 1, the MSSM weak scale scalar potential 
requires at-least one of the Higgs mass squared to be negative to generate 
vacuum expectation values to the Higgs. Starting with a positive mass 
squared for the Higgs at the high scale, large radiative corrections 
from the top quark Yukawa can turn the Higgs mass squared negative at the 
weak scale. This can be seen from the solution to the RGE for the $m_{H_2}^2 $ 
which can be approximately written as \footnote{$t=0$ characterises the
high scale in our notation.}, 
\be
m_{H_2}^2 (tz) \approx m_{H_2}^2(0) -{ 3 h_t^2 \over 4 \pi^2} 
M_{SUSY}^2 \ln \left( {M_{GUT}^2 \over M_Z^2 } \right)
\ee

\noindent
where the weak scale is characterised by the mass of the $Z$ boson and
$M_{SUSY}$ characterises the typical SUSY breaking mass scale $\sim$ 1 TeV
which appears in the equations ( for example, squark masses ). This scale 
together with the large logarithmic factor $\approx 66$ appears with a
negative sign in the above solution. Thus even with a positive mass squared
at the high energies, the Higgs mass squared can turn negative at low
energies. This mechanism is called radiative electroweak symmetry breaking
\cite{ibanez}.  Within the mSUGRA inspired MSSM, with universal boundary
conditions, radiative $SU(2) \times U(1)$ breaking helps in reducing the
number of parameters of the theory. Consider the minimisation conditions
which we have already seen in chapter 1: 

\bea
\label{redparam}
\m^2 &=& {m_{H_1}^2 - \tan^2 \b~ m_{H_2}^2 \over
\tan^2 \b  - 1} - {1 \over 2} M_Z^2 \\
B_\m &=& {\mbox{Sin} 2 \b \left( m_{H_2}^2 + m_{H_1}^2 + 2 \m^2 \right)
 \over 2~\m~ }
\eea

The parameters $m_{H_2}^2, m_{H_1}^2$ are determined at the weak scale
by tan$\beta$,~$m_0,~M$~(or equivalently $M_2$)~ $A$. Using
the above equations one can determine $\m$ and $B_\m$ at the weak scale.
Only $Sign(\m)$ remains as a free parameter. On the other hand, one can
trade $B_\m$ to determine tan $\beta$ or $\m$ to determine $m_0$ etc.
In the analyses presented in the thesis we have used both the sets of 
parameters. This mechanism is effective for large ranges in tan $\beta$ 
and other MSSM parameters. Radiative electroweak symmetry breaking can
also be incorporated in models with gauge mediated supersymmetry breaking.
In this case, the smallness of the logarithmic factor is compensated
by the largeness of the squark masses \cite{radewgmsb}.

\subsection{Analytical and semi-analytical solutions}

In a typical model of supersymmetry breaking, the soft terms
are added at a high scale. The soft masses and the couplings
at the high scale are the input parameters of the model. RG
evolution would change these masses and couplings at the weak
scale significantly. However, one would like to express the 
weak scale soft masses and couplings also in terms of the 
basic input parameters of the model. This would require to 
have solutions for the RGE for the soft parameters analytically. 
From the RGE given in the appendix we see that except for the
third generation sfermion masses, rest of the differential
equations can be solved analytically. The third generation 
sfermion masses are in general coupled and have to be solved
numerically. However, Ibanez and Lopez \cite{iblpnp} have shown
that the solutions of these equations can be expressed in terms
of `semi-analytical' formulae, thus making the dependence on 
the basic parameters transparent. For example, in mSUGRA inspired
MSSM with universality at a high scale, the solutions for the
sfermion masses would be typically of the form:
\be
\label{solform}
m_{\tilde{f}}^2 (t_) = D_{m_0} m_0^2 + D_M~ M^2 + D_{mMA}~ m~ M ~A + 
D_{mA}~ m^2 ~A^2 
\ee 
where $m_0,~ M,~ A,$ stand for the standard mSUGRA inspired MSSM
parameters. The coefficients $D$'s  are given in terms integrals
which can be solved numerically. In this section, we present the
solutions of the RGE systematically.

\noindent
As we have seen earlier, the RG equations for the gauge coupling
constants can be solved analytically. Since the gauginos also 
follow similar type of equations, their evolution can also be
represented with analytical solutions of the RGE. 
These solutions are given as 
\bea
\tilde{\alpha}_i(t)&=&\tilde{\alpha}_i(0) z_i(t)\nonumber \\
M_i(t)&=&M_i(0) z_i(t),
\eea
where $z_i(t)$ is defined as:
$$z_i(t) = {1 \over ( 1 + b_i~\tilde{\alpha}_i(0)~ t)}$$  
with $i=1,2,3$ over the generations and $b_i$ are defined in the 
Appendix. Similarly, in the limit where the first two generation Yukawa
couplings are neglected, the equations for the corresponding soft masses,  
eqs.(\ref{rgemass12}), depend only the gauge couplings and the gauginos. 
The general solutions of these equations can be represented as:

\be
\label{solmass12}
m_{\tilde{f}_j}^2 (t) = m_{\tilde{f}_j}^2 (0) +  \sum_i \left( C_i^j~ 
k_i(t)~ M_i(0)^2 \right),
\ee 

where $j= \left( Q_{1,2},U^c_{1,2},D^c_{1,2}, L_{1,2},E^c_{1,2} \right)$
and $C_i^j$ are the coefficients appearing in the RHS of the  
respective differential equations. The functions $k_i$ appearing in the 
above solutions are given as,

\be
k_i(t) = {1 \over 2~b_i } \left( 1 - z_i(t)^2 \right)
\ee

As an example, the solution for the RGE of the first two generation
 left-handed squark masses is given as,  
\be
m^2_{Q_{1,2}}(t) = m^2_{Q_{1,2}}(0) + {16 \over 3} M_3(0)^2 k_3(t) +
3  M_2(0)^2 k_2(t) + {1 \over 15} M_1(0)^2 k_1(t)
\ee  
Similarly, the solutions for the rest of the first two generation soft
masses can be read off from their RGE. Thus we see that the RGE for the 
first two generation soft masses have analytical solutions as the 
corresponding Yukawa couplings are small and thus can be neglected. 
But as mentioned above, 
such an approximation is not valid for the third generation Yukawa
couplings as they are large and can induce large corrections to 
the soft masses through RG evolution. This makes the set of 
equations coupled and in general has to be solved numerically. 
However, one can still find semi-analytical formulae for the set 
of equations when tan $\beta$ is small \cite{iblpnp}. In this limit, $Y_b$ and 
$Y_\tau$ are small compared to the Top quark Yukawa and thus can be neglected.
Neglecting the $Y_b$ and $Y_\tau$ couplings in the equation for the
third generation Yukawa couplings, eq.(\ref{rgeyukawa}), we 
can arrive at the following solution \cite{arfken} for $Y_t$ :
\be
\label{yt}
Y_t(t) = {Y_t(0) E_1(t) \over Y_{den}(t)},
\ee
where the function $E_1(t)$ is given as,
\be
E_1(t) = z_3(t)^{-16 /3 b_3}~ z_2(t)^{-3/b_2}~ 
z_1(t)^{-13/15 b_1}
\ee
with the functions $z_i(t)$ defined above. The function $Y_{den}(t)$ is
given as,
\be
Y_{den}(t) =  1 + 6 ~Y_t(0)~ F_1(t) 
\ee
where
\be
F_1(t)  =  \int_0^t E_1(t') dt' 
\ee

\noindent
$Y_t(0)$ is the Top quark Yukawa at the high scale which is determined
in-terms of the Top quark mass at $m_t$ \cite{koide}. Defining
$tmt = 2~\ln \left({M_X \over m_t} \right)$, $Y_t(0)$ is given as,
\bea
Y_t(0) &=& {Y_t(tmt) \over E_1(tmt) - 6~ Y_t(tmt)~ F_1(tmt)}, \nonumber \\
Y_t(tmt)&=&{ m_t(m_t)^2 \over 16~ \pi^2~ v^2~ sin \beta^2}. \nonumber \\
\eea

\noindent
In this approximation, the solutions of $Y_b$ and $Y_\tau$ are given 
as,
\bea
Y_b(t)&=&{Y_b(0) E_2(t) \over Y_{den}^{1/6}} \nonumber \\
Y_\tau(t)&=&Y_\tau(0) E_3(t)
\eea
With $Y_b(0)$ and $Y_\tau(0)$ determined in a similar manner as 
$Y_t(0)$ in terms of $Y_b(tmb)$ and $Y_\tau(tmtau)$ with $tmb$ and 
$tmtau$ having analogous definitions of $tmt$. The functions $E_2(t)$
and $E_3(t)$ are given as,
\bea
E_2(t)&=&  z_1(t)^{2/5 b_1} E_1(t), \nonumber \\
E_3(t)&=& z_2(t)^{-3/b_2} z_1(t)^{-9/5b_1}.
\eea

\noindent
The integral $F_1$ is generally solved numerically. Thus, we call these
solutions as semi-analytical solutions. In a similar manner, 
in deriving the solutions for
the rest of the parameters we completely neglect the effects of bottom
and tau-lepton Yukawa couplings. This scheme is sometimes known as `top
Yukawa dominance' approximation. In this case, we neglect the bottom 
and tau-lepton
Yukawas appearing in the equations for the soft A-parameters, 
eqs.(\ref{rgeAparameters}), $\m$ and $B_\m$ terms, eqs.(\ref{rgemu}), 
the third generation squark and slepton masses, eqs.( \ref{rgemass3}) and 
the Higgs mass terms, eqs.(\ref{rgehiggs}).  Since $A_b$ and $A_\tau$ 
are always accompanied by their respective Yukawa couplings in these
equations, we will not worry about their solutions here. 
The solution for $A_t$ is given as,
\be
A_t(t) = {1 \over  Y_{den}(t)} \le A_t(0) -  \int_0^t Y_{den}(t') 
\kappa(t') dt' \ri
\ee 
where the function $\kappa$ is given as,
\be
\kappa(t) = {16 \over 3}~ \tilde{\alpha}_3(0)~ M_3(0) z_3(t)^2 + 
 3~ \tilde{\alpha}_2(0)~ M_2(0) z_2(t)^2 + 
{13 \over 15}~ \tilde{\alpha}_1(0)~ M_1(0) z_1(t)^2. 
\ee 
The integral on the RHS of the above formula is solved
numerically. In the case of third generation sfermion masses, the
observation that only $m_{Q_3}^2,~ m_{U_3}^2$ and $m_{H_2}^2$ depend
on the top quark Yukawa,  reduces the task significantly. The RGE 
for $m_{D_3}^2,~ m_{L_3}^2$ and $m_{E_3}^2$ now have the
same form as those for the first two generation sfermion masses.
Thus their solutions too have the same form as given in
eq.(\ref{solmass12}). The coefficients on the RHS of these equations
would now form the constants $C_i^j$.  $m_{H_1}^2$ evolves in the same
manner as $m_{L_3}^2$ in this approximation and thus has the same solution. 

To find the solutions for the remaining third generation sfermion 
masses, ie,  $m_{H_2}^2(t), m_{Q_3}^2(t)$ and $m_{U_3}^2(t)$ we observe
that though independently they cannot be solved analytically, there are
combinations of these masses which can be solved analytically. These
combinations do not contain the top quark Yukawa \footnote{Please see
Ibanez-Lopez \cite{iblpnp} for a different set of combinations which
are valid when all the three third generation Yukawa couplings are considered.}.
\bea
\label{relat}
m_4^2(t)&=& m_{U_3}^2(t) - 2 m_{Q_3}^2(t) \nonumber \\
m_5^2(t)&=& m_{H_2}^2(t) - 3 m_{Q_3}^2(t) \nonumber \\
2~m_6^2(t)&=&2~m_{H_2}^2(t) - 3 m_{U_3}^2(t) 
\eea
Using the RGE for these three mass parameters, eqs.(\ref{rgemass3}), 
on the RHS of the above, these functions are determined as,
\bea
\label{combi}
m_4^2(t)&=&m_{U_3}^2(0) - 2~m_{Q_3}^2(0) - {16 \over 3} M_3^2(0) k_3(t)
- 3 M_2^2(0) k_2(t) - {14 \over 15} M_1^2(0) k_1(t) \nonumber \\
m_5^2(t)&=&m_{U_3}^2(0) - 2 m_{Q_3}^2(0) - 16 M_3^2(0) k_3(t) - 
9 M_2^2(0) k_2(t) - {1 \over 5} M_1^2(0) k_1(t) \nonumber \\
2~m_6^2(t)&=&2~ m_{H_2}^2(0) - 3 m_{U_3}^2(0) - 16 M_3^2(0) k_3(t) 
- {16 \over 5} M_1^2(0) k_1(t)
\eea
Using the above relations the differential equations for $m_{H_2}^2(t),~ 
m_{Q_3}^2(t),~m_{U_3}^2(t)$ parameters
can be rewritten in the approximation of top Yukawa dominance  as,
\bea 
\label{newrgemass3}
{d m_{Q_3}^2(t) \over dt} + 6 Y_t(t) m_{Q_3}^2(t)&=& L_1(t) \nonumber \\
{d m_{U_3}^2(t) \over dt} + 6 Y_t(t) m_{U_3}^2(t)&=& L_2(t) \nonumber \\
{d m_{H_2}^2(t) \over dt} + 6 Y_t(t) m_{H_2}^2(t)&=& L_3(t)
\eea
with the functions $L_i$ are given by, 
\bea
\label{li}
L_1(t)&=& {16 \over 3} \tilde{\alpha}_3(t)~ M_3(t)^2 + 3 \tilde{\alpha}_2(t)
~M_2(t)^2 + {1 \over 15} \tilde{\alpha}_1(t)~M_1(t)^2 \nonumber \\
&-& Y_t(t) \le -{2 \over 3} m_6^2(t) + {5 \over 3} m_5^2(t) + A_t(t)^2 \ri \\
L_2(t)&=& {16 \over 3} \tilde{\alpha}_3(t)~ M_3(t)^2 + 
 {16 \over 15} \tilde{\alpha}_1(t)~M_1(t)^2 \nonumber \\
&-& 2~Y_t(t) \le -{1 \over 2} m_4^2(t) + m_6^2(t) + A_t(t)^2 \ri \\
L_3(t)&=&  3 \tilde{\alpha}_2(t) ~M_2(t)^2 + {3 \over 5} \tilde{\alpha}_1(t)
~M_1(t)^2 \nonumber \\
&-&3~ Y_t(t) \le  m_4^2(t) - m_5^2(t) + A_t(t)^2 \ri 
\eea

Solving any one of the equations in eqs.(\ref{newrgemass3}), 
one can get the solutions of the other parameters from the relations 
given in eqs.(\ref{relat},\ref{combi}). For example, the solution 
for the equation for $m_{H_2}^2$ is given as,
\be
\label{mh2sol}
m_{H_2}^2(t)= {1 \over Y_{den}(t)} \le m_{H_2}^2(0) + \int_0^t L_3(t') 
Y_{den}(t') dt' \ri
\ee

The integral appearing on the RHS of the above equation can be solved
numerically. Expanding the function $L_3(t)$ as given in eqs.(\ref{li}),
we see that the solution, $m_{H_2}^2(t)$ has the desired form 
of eq.(\ref{solform}). Using this solution and
the solutions in eqs.(\ref{combi}), we can derive the solutions for 
$m_{Q_3}^2(t)$ and $m_{U_3}^2(t)$ as per the combinations given in 
eqs.(\ref{relat}). Analogously, we can start with solutions of either
$m_{Q_3}^2$ or $m_{U_3}^2$ and arrive at the other two solutions 
using the combinations listed above. 

The parameters $\m$ and $B_\m$ are generally determined at the 
weak scale. Using the values at the weak scale, one finds the
values of these parameters at the $M_{GUT}$. The solutions for
these parameters at any scale $t$ using weak scale boundary
conditions are given as,
\be
\m(t) = \m(tz) \le { Y_{den}(t) \over Y_{den}(tz) } \ri ^{ - 1/4} 
\le { z_1 (t) \over z_1(tz)} \ri ^{3 / 10 b1} 
\le { z_2 (t) \over z_2(tz)} \ri ^{3 / 2 b2} 
\ee 
The functions $z_i$ and $Y_{den}$ have been defined earlier. In the 
same notation the equation for $B_\m$ takes the form:
\bea
B_\m(t)&=& B_\m(tz) -3 \int_{tz}^t A_t(t') Y_t(t') dt' - 3 
{ M_2(0) \over b_2} \le z_2(tz) - z_2(t) \ri \nonumber \\
& - & {3 \over 5} { M_1(0) \over b_1} \le z_1(tz) - z_1(t) \ri
\eea

As mentioned earlier, in deriving the above solutions, we have 
followed the approach of Ibanez and Lopez \cite{iblpnp}. An
 alternative approach has been followed by 
Barbieri et al \cite{rgbarbieri}. Recently, these formulae were
also given by Carena et al \cite{carena}.
All the above solutions hold good only in the top-quark Yukawa
dominance approximation. This corresponds to regions in the
tan $\beta$ parameter space around $2 - 20$ approximately. For
larger values of tan $\beta$ Yukawa couplings of bottom and the
tau leptons would also become comparable and subsequently would
have to be taken in to account. Recently, Kazakov and collaborators
have presented analytical solutions to these RGE even in the
limit of large tan $\beta$ \cite{kazmoul}.

\section{R violation and RG evolution}

As we have seen in chapter 1, R-violating schemes in supersymmetric 
theories have a rich phenomenology of their own. In these models,
additional couplings which violate lepton and baryon number are
present in the superpotential. The RGE presented earlier for
the case of MSSM would be significantly modified to take in to 
consideration these additional couplings. The presence of these
couplings leads in general to processes which are constrained
severely by experiments. These experimental limits can be converted
into bounds on lepton number and baryon number violating couplings
at the low scale. Using the RGE the bounds at the low scales
can be converted to bounds at the high scale \cite{dedes}. Moreover,
since these couplings also contain flavour violation, the presence
of these couplings at the high scale can generate additional 
flavour violating contributions through RG scaling \cite{carloswhite,fb}.
Recently, infrared fixed point studies have also been conducted in the 
presence of these couplings \cite{pandita}. 

In this thesis, we concentrate our studies to neutrino mass
structure in the presence of these couplings. As we will see later, the
typical magnitude of the R-violating parameters required in this
case is very small, for example, $\sim 10^{-4}$ for the dimensionless 
$\l'$ couplings. The presence of such small R-violating parameter would
not modify the RG evolution of the MSSM parameters significantly 
\cite{dicus}. Thus in this thesis, we consider that the RG evolution of 
standard MSSM parameters would not be effected by the presence of 
R-parity violating couplings.  

However the RG evolution of these couplings can significantly affect 
the neutrino mass spectrum in these models. To study the 
effect of RG evolution on neutrino mass spectrum, we divide the 
R-parity violating couplings as,

\noindent
(a). Bilinear lepton number violation (dimensionful $\epsilon_i$ terms) \\
(b). Trilinear lepton number violation (dimensionless 
$\l'_{ijk},\l_{ijk}$ terms). 

\noindent 
We study the structure of neutrino mass spectrum in both of these 
cases separately. Assuming only one set of the parameters to 
be present at a time {\it i.e} either only bilinear or only trilinear
couplings, we derive the RGE for the R-parity violating 
parameters present both in the superpotential and the soft potential.
 To derive these equations,
we have used general formulae given by Falck \cite{falckrge}. In
this work, the general formulae were derived by using effective
potential method. Martin and Vaughn \cite{martinvaughn} presented
general formulae up to 2-loop order. They used both diagrammatic and
effective potential method to arrive at their results. We have used
their results in chap.4 where we have derived two loop RGE for
R-violating parameters. The equations for the R-violating superpotential
and soft potential parameters are presented in the Appendix for each
case separately. The equations presented here agree with those 
presented by Carlos and White \cite{carloswhite}. We have also checked
our equations with the equations presented in other papers \cite{rpvrge}.
 
\section{RG scaling of dimensionless L-violation and soft sector}

As we have seen earlier, one of the important characteristics of the 
RG evolution of the soft sector is that,  even though these parameters 
are small at the high scale, they can acquire large values at
the weak scale. This characteristic plays an important role
in supersymmetry breaking models like Minimal Messenger Model
of Gauge Mediated Supersymmetry Breaking, where the soft parameters
$B_\m$ and $A$ vanish at the high scale. However, these parameters
acquire non-zero values at the weak scale due to RG scaling thus
facilitating electroweak symmetry breaking \cite{radewgmsb}. This can
be clearly seen from the RG equation for the $B_\m$ parameter, 
eq.(\ref{rgemu}). In the presence of non-zero gaugino masses and Yukawa 
couplings, a non-zero $B_\m$ can be generated at the weak scale. 
Similarly even if the bilinear lepton number violating soft parameters 
are absent at the high scale, they can be generated at the weak scale 
through RG scaling. This would require non-zero lepton number violating
trilinear couplings at the high scale \cite{carloswhite,en,vr1}.

As mentioned earlier, we would divide the R-violation studies
interms of dimensionful $\e_i$ terms and dimensionless $\l',\l$ 
couplings. Consider an R-violating model with only dimensionless
trilinear $\l'$ or $\l$ couplings. In this case, the soft potential
at the high scale would not contain any bilinear lepton number
violating couplings. However, these couplings can be generated at
the weak scale due to RG scaling. To see this consider the RG evolution
of the bilinear lepton number violating soft terms. These are represented
as $B_{\e_i}$ and $m_{\n H_1}^2$. The corresponding equations for these
terms in the presence of only $\l'$ or $\l$ couplings are given in 
eqs.(\ref{lprimerge},\ref{lambdarge}). From these equations we can see
that even if these couplings vanish at the high scale, non-zero
values can be acquired at the weak scale in the presence of non-zero
$\l'$ or $\l$ couplings. 

The above generation of bilinear soft lepton number violating terms
at the weak scale has important consequences for neutrino phenomenology.
In general, the presence of dimensionless $\l',~\l$ couplings in the
superpotential is believed to give rise to neutrino masses only at the
1-loop level. But, RG scaling can generate additional contribution to 
neutrino masses which is much larger. This happens as follows. The bilinear
soft lepton number violating terms generated at the weak scale lead to
generation of {\it vevs} for the sneutrinos. This in-turn leads to a non-zero
mixing between neutrinos and neutralinos leading to a `tree level' mass 
for the neutrino. Thus, RG scaling can induce a `tree level' mass to the 
neutrino which could be significantly larger than the 1-loop level mass. The
effect of this RG scaling in a realistic supersymmetric breaking model 
and its consequences for the neutrino mass spectrum have been worked out
in chapters 5 and 6. 

\subsection{Semi-analytical solutions}

In the case of  MSSM without R-violation, we have seen that the weak 
scale soft spectrum can be expressed as  semi-analytical formulae in terms
of the four basic parameters of the mSUGRA model, $m_0, ~M_2, A $ and 
tan $\beta$.  The coefficients of these parameters involve simple 
integral which can
be solved numerically in the limit of low tan $\beta$ where only the top
quark Yukawa dominates. A similar analysis can be done for the R-parity 
violating couplings discussed so far. As an example, we present here
analytical and semi-analytical solutions for the case where only trilinear
$\l'_{ijk}$ couplings are present in the superpotential. Moreover, we
choose there is only one coupling of the form $\l'_{ijk}$ present in 
the superpotential. The analysis can be extended for the presence of 
additional couplings but would be cumbersome.  

In this limit of neglecting $Y_b, Y_\tau$, the solutions for the $\l'$
 parameters take the following simple form,
\be
\l'_{ijk}(t)=\l'_{ijk}(0) z_3(t)^{-8/ 3 b_3} z_2(t)^{-3 / b_2} 
z_1(t)^{-7/30 b_1} 
\ee

As mentioned above the soft bilinear lepton number violating parameters,
$B_{\e_i}$ and $m^2_{\n_i H_1}$ are generated due to RG scaling in 
the presence of trilinear $\l'$ couplings. The solutions for the
RGE for these parameters have the generic form of that of the soft masses.
 The solution for the RGE for $B_{\e_i}$ in this case is given by,  
\be
B_{\e_i}(t) = p(t) \le B_{\e_i}(0) + {3 \over 16 \pi^2} \int_0^t 
q_{ij}(t') {1 \over p(t')} dt' \ri
\ee
The functions $p(t)$ and $q_{ij}(t)$ are given as,
\bea
p(t)&=& z_2(t)^{3 /2 b_2} z_1(t)^{-3 / 10 b_1} \nonumber \\
q_{ij}(t)&=& \m(t) \l'_{ijj}(t) h^d_{jj}(t) \le {1 \over 2} B_\m(t) +
A^{\l'}_{ijj}(t) \ri
\eea
For the $m^2_{\n_i H_1}$, the  RGE leads to the following solution:
\be
m^2_{\n_i H_1}(t) = m^2_{\n_i H_1}(0) - {3 \over 32 \pi^2}
\int_0^t r_{ij}(t') dt'
\ee
The function $r_{ij}(t)$ is given by,
\be
r_{ij}(t) = m_{H_1}^2(t) + m_{L_i}^2(t) + 2 m_{Q_j}^2(t) + 2 A^{\l'}_{ijj}(t)
A^D_{jj}(t) + 2 m_{D_j}^2(t)
\ee

The solution for the trilinear lepton number violating coupling, 
$A^{\l'}_{ijj}$ which appears in the above solutions is given as, 

\bea
A^{\l'}_{ijj}(t)&=&A^{\l'}_{ijj}(0) - {3 \over 2} \int_0^t A_t(t') Y_t(t') 
\delta_{j3} dt' - {7 \over 30} M_1(0) \tilde{\alpha}_1(0) \xi_1(t) \nonumber \\
&-& 2 M_2(0) \tilde{\alpha}_2(0) \xi_2(t) - {16 \over 3}
  M_3(0) \tilde{\alpha}_3(0) \xi_3(t)
\eea

The functions $\xi(t)$ are given as,
\be
\xi(t)_i = {1 \over 2 b_i} \le 1 - z_i(t) \ri
\ee

Using the solutions for the various soft parameters in the above, 
one can determine the weak scale values of these parameters in terms 
of the supersymmetry breaking soft parameters at the high scale which 
are input parameters of the model. Using these
expressions the entire neutrino spectrum can now be expressed in terms
of the few supersymmetry breaking parameters. In all the analyses in 
this thesis, we have solved these integrals numerically using 
$\mbox{MATHEMATICA}^{\copyright}$ software.

\subsection{Numerical Methods}
The above semi-analytical formulae hold true only in the limit 
of small tan $\beta$. In calculations involving models which
allow large tan $\beta$, the RGE have to be solved numerically.  
This set of first order differential equations is coupled and 
non-linear and thus have to be solved simultaneously. We have
used the standard Rung-Kutta algorithm \footnote{We have employed
fourth order RK method. Sometimes we have also employed standard
RK routines provided by IMSL.} \cite{numrec,collatz} which
is generally recommended for this purpose.

\begin{figure}
\label{flow}
\centerline{\psfig{figure=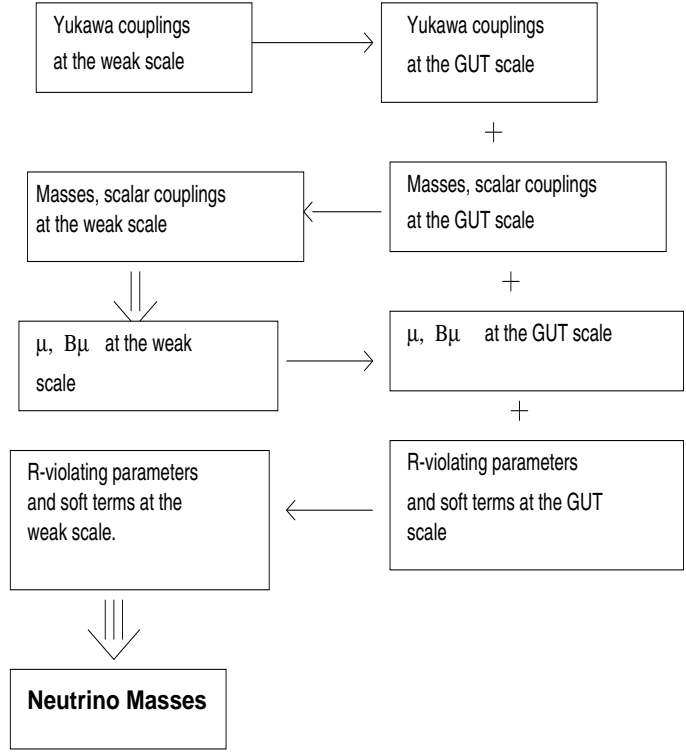,height=10cm,width=9cm}}
\caption{Flow chart for numerical analysis of RG equations.}
\end{figure}

The methodology employed for the numerical analysis is presented
in a pictorial manner in Fig.(\ref{flow}). Masses of the top
quark, bottom quark and the tau-lepton determine the input parameters
for the third generation Yukawa couplings at the weak scale (characterised
by the Z boson mass ) for a given tan $\beta$. These will be then used
to determine the Yukawas at the high scale. At the high scale, the masses
and the scalar couplings are given as input parameters. Running them down
along with the Yukawas determine the weak scale values of those parameters.
Using the weak scale values of $m_{H_1}^2$ and $m_{H_2}^2$ one can 
determine the $\m$ and $B_\m$ parameters at the weak scale. In turn we use
these parameters to determine their high scale values. The R-parity violating
soft terms contain $\m$  and $B_\m$ in their RGE. The values of these 
parameters are known at $M_{GUT}$. Running down the entire set of RGE
to the weak scale one finds the weak scale values of the R-parity violating
terms. These are then used to find out the neutrino masses. 

\subsection{Appendix }
\noindent
{\bf Standard MSSM Equations:}
The MSSM RGE are presented in many papers \cite{listrge}. In writing
down the below set of RGE we follow \cite{fb1}. 

\vskip 0.2cm
\noindent
Here we use the notation $ t = 2 ln({M_X \over Q})$ where 
$M_X$ corresponds to the high scale and $Q$ corresponds to
the low scale. In mSUGRA inspired MSSM $M_X$ is taken to
be $M_{GUT} \equiv 3 \times 10^{16}$ GeV in our calculations of 
chapters 5 and 6 whereas $M_X$ varies with $\Lambda$ 
in Gauge Mediated Models discussed in chapter 4. $tz$ always
corresponds to $2 ln ({M_X \over M_Z})$, where $M_Z$ is the
$Z$ boson mass.

\vskip 0.3 cm
\noindent
{\it Definitions}\\
\noindent
We also use the following definitions with $g_i$ representing
the gauge coupling and $h_f$ representing the Yukawa coupling
of the fermion $f$. 
\bea
\label{rgedef} 
\tilde{\a}_i &=& {g_i^2 \over 16 \pi^2} \\
Y_f &=& {h_f^2  \over 16 \pi^2}
\eea

\vskip 0.3 cm
\noindent
{\it Gauge couplings}
\bea
\label{rgegauge}
{d \tilde{\a}_3(t) \over dt} &=& 3 \tilde{\a}_3(t) \\
{d \tilde{\a}_2(t) \over dt} &=& - \tilde{\a}_2(t)\\
{d \tilde{\a}_1(t) \over dt} &=& -{33 \over 5} \tilde{\a}_1(t)
\eea

\noindent
{\it Gauginos}
\bea
\label{rgegauginos}
{d M_3(t) \over dt} &=& 3 \tilde{\a}_3(t) M_3(t) \\
{d M_2(t) \over dt} &=& - \tilde{\a}_2(t) M_2(t) \\
{d M_1(t) \over dt} &=& -{33 \over 5} \tilde{\a}_1(t) M_1(t)
\eea

\noindent
{\it Yukawa couplings}
\bea
\label{rgeyukawa}
{d Y_t (t) \over dt} &=&  Y_t(t) \le {16 \over 3} \tilde{\a}_3(t) + 
3 \tilde{\a}_2(t) + {13 \over 15} \tilde{\a}_1(t) - 6 Y_t(t) - Y_b(t) \ri  \\
{d Y_b (t) \over dt} &=&  Y_b(t)\le{16 \over 3}\tilde{\a}_3(t)+ 
3 \tilde{\a}_2(t) + {7 \over 15} \tilde{\a}_1(t) - Y_t(t) - 6 Y_b(t)-
 Y_\tau (t)\ri  \\
{d Y_\tau (t) \over dt} &=&  Y_\tau(t) \le  3 \tilde{\a}_2(t) + {9 \over 5} 
\tilde{\a}_1(t) - 3 Y_b(t) - 4 Y_\tau(t) \ri 
\eea

\noindent
{\it A-parameters}
\bea
\label{rgeAparameters}
{d A_t (t) \over dt} &=& -{16 \over 3} \tilde{\a}_3(t)~ M_3(t) - 
3 \tilde{\a}_2(t)~ M_2(t) - {13 \over 15} \tilde{\a}_1(t)~ M_1(t) \nonumber\\
&-&6 Y_t(t)~ A_t(t) - Y_b(t)~ A_b(t) \\
{d A_b (t) \over dt} &=& -{16 \over 3}\tilde{\a}_3(t)~ M_3(t) - 
3 \tilde{\a}_2(t)~ M_2(t) - {7 \over 15} \tilde{\a}_1(t)~M_1(t) \nonumber \\
&-& Y_t(t)~ A_t(t) - 6 Y_b(t)~ A_b(t) - Y_\tau (t)~ A_\tau(t) \\
{d A_\tau (t) \over dt} &=& -3 \tilde{\a}_2(t)~ M_2(t) - {9 \over 5} 
\tilde{\a}_1(t)~ M_1(t) - 3 Y_b(t)~A_b(t)  - 4 Y_\tau(t)~ A_\tau(t)
\eea

\noindent
{\it $\m$ and $B_\m$ parameters}
\bea
\label{rgemu}
{d \m(t) \over dt} &=& \m(t) \le -{3 \over 2} Y_t(t) - {3 \over 2} Y_b(t) 
- {1 \over 2} Y_\tau(t) + {3 \over 2} \tilde{\a}_2(t) + {3 \over 10} 
\tilde{\a}_1(t) \ri \\
{d B_\m (t) \over dt}&=& -3 \tilde{\a}_2(t)~M_2(t) - {3 \over 5} 
\tilde{\a}_1~ M_1(t) - A_\tau~Y_\tau -3 A_b~ Y_b - 3 A_t~ Y_t
\eea

\noindent
{\it First two generation squark and slepton masses}
\bea
\label{rgemass12}
{d m_{Q_{1,2}}^2(t) \over dt}&=& {16 \over 3} \tilde{\a}_3(t)~ M_3^2(t) + 
3 \tilde{\a}_2(t)~M_2^2(t) + {1 \over 15} \tilde{\a}_1(t)~ M_1^2(t) \\ 
{d m_{U_{1,2}}^2(t) \over dt}&=& {16 \over 3} \tilde{\a}_3(t)~ M_3^2(t) +
 {16 \over 15} \tilde{\a}_1(t)~ M_1^2(t) \\
{d m_{D_{1,2}}^2(t) \over dt}&=& {16 \over 3} \tilde{\a}_3(t)~ M_3^2(t) +
 {4 \over 15} \tilde{\a}_1(t)~ M_1^2(t) \\
{d m_{L_{1,2}}^2 (t)\over dt}&=& 3 \tilde{\a}_2(t)~ M_2^2(t) +
 {3 \over 5} \tilde{\a}_1(t)~ M_1^2(t) \\
{d m_{E_{1,2}}^2(t) \over dt}&=& {12 \over 5} \tilde{\a}_1~ M_1^2(t)
\eea

\noindent
{\it Third generation sfermion masses}
\bea
\label{rgemass3}
{d m_{Q_3}^2(t) \over dt}&=& {16 \over 3} \tilde{\a}_3(t)~ M_3^2(t) 
+ 3 \tilde{\a}_2(t)~M_2^2(t) + {1 \over 15} \tilde{\a}_1(t)~
 M_1^2(t) \nonumber \\ 
&-& Y_t(t) \le m_{Q_3}^2(t) + m_{U_3}^2(t) + m_{H_2}^2(t) + A_t(t)^2 \ri 
\nonumber \\
&-& Y_b(t) \le m_{Q_3}^2(t) + m_{D_3}^2(t) + m_{H_1}^2(t) + A_b(t)^2 \ri \\
{d m_{U_3}^2(t) \over dt}&=& {16 \over 3} \tilde{\a}_3(t)~ M_3^2(t) +
 {16 \over 15} \tilde{\a}_1(t)~ M_1^2(t) \nonumber \\
&-& 2 Y_t(t) \le m_{Q_3}^2(t) + m_{U_3}^2(t) + m_{H_2}^2(t) + A_t(t)^2 \ri \\
{d m_{D_3}^2(t) \over dt}&=& {16 \over 3} \tilde{\a}_3(t)~ M_3^2(t) + 
{4 \over 15} \tilde{\a}_1(t)~ M_1^2(t) \nonumber  \\
&-& 2 Y_b(t) \le m_{Q_3}^2(t) + m_{D_3}^2(t) + m_{H_1}^2(t) + A_b(t)^2 \ri \\
{d m_{L_3}^2 (t)\over dt}&=& 3 \tilde{\a}_2(t)~ M_2^2(t) +
 {3 \over 5} \tilde{\a}_1(t)~ M_1^2(t) \nonumber \\
&-& Y_\tau \le m_{L_3}^2(t) + m_{E_3}^2(t) + m_{H_1}^2(t) + A_\tau^2(t) \ri\\
{d m_{E_3}^2(t) \over dt}&=& {12 \over 5} \tilde{\a}_1(t)~ M_1^2(t)
- Y_\tau(t) \le m_{L_3}^2(t) + m_{E_3}^2(t) + m_{H_1}^2(t) + A_\tau^2(t) \ri
\eea

\noindent
{\it Higgs mass parameters}
\bea
\label{rgehiggs}
{d m_{H_1}^2(t) \over dt}&=& 3 \tilde{\a}_2(t)~ M_2^2(t) +
 {3 \over 5} \tilde{\a}_1(t)~ M_1^2(t) \nonumber \\
&-& 3 Y_b(t) \le m_{Q_3}^2(t) + m_{D_3}^2(t) + m_{H_1}^2(t) +
 A_b(t)^2 \ri \nonumber \\
&-& Y_\tau(t) \le m_{L_3}^2(t) + m_{E_3}^2(t) + m_{H_1}^2(t)+ A_\tau^2(t) \ri \\
{d m_{H_2}^2 (t)\over dt}&=& 3 \tilde{\a}_2(t)~ M_2^2(t) +
 {3 \over 5} \tilde{\a}_1(t)~ M_1^2(t) \nonumber \\
&-& 3 Y_t(t) \le m_{Q_3}^2(t) + m_{U_3}^2(t) + m_{H_2}^2(t) + A_t(t)^2 \ri 
\eea

\noindent
{\it Additional RGE for MSSM with lepton number violation}\\
Below we present RGE for the lepton number violating superpotential
and soft potential parameters. These are presented separately for the
case of dimensionful and dimensionless parameters. In writing them,
we have neglected terms higher order in the L-violating parameter. 
This is permissible as we are interested only in small L-violation
as required by neutrino masses.  
\vskip 0.3 cm

\vskip 0.3cm
\noindent
{\it Bilinear R-violation}\\
\vskip 0.2cm 
\noindent
\bea
{d \e_i(t) \over dt}& =& \e_i(t) \le {3 \over 2} \tilde{\alpha}_2(t) +
{3 \over 10} \tilde{\alpha}_1(t) - {1 \over 2} Y^E_{ii}(t) -{3 \over 2} 
Y_t(t) \ri \\
{d B_{\e_i}(t) \over dt}&=& B_{\e_i}(t) \le -{ 1\over 2} Y^E_ii(t) - 
{3 \over 2} Y_t(t) + {3 \over 2} \tilde{\alpha}_2(t) + {3 \over 10} 
\tilde{\alpha}_1(t) \ri - \e_i \le 3 \tilde{\alpha}_2(t) 
M_2(t) \right. \nonumber \\
&+& \left.  {3 \over 5} \tilde{\alpha}_1(t)M_1(t) + 3 A_t(t) Y_t(t) 
+ Y^E_{ii}(t) A^E_{ii}(t) \ri
\eea

\vskip 0.3cm
\noindent
{\it Trilinear R-violation ($\l'$)}\\
\vskip 0.2cm 
\noindent

\bea 
\label{lprimerge}
{d \l'_{ijk}(t) \over dt}&=& \l'_{ijk}(t) \left( - Y^E_{ii}(t) - Y^D_{jj}(t) 
- Y^U_{jj} (t) - 2 Y^D_{kk}(t) - 3 Y^D_{jj}(t) \delta_{jk} + 
{8 \over 3} \tilde{\a}_3(t) \right. \nonumber \\
&+& \left. 3 \tilde{\a}_2(t) + {7 \over 30} \tilde{\a}_1(t)  \right) 
 - {6 \over 32 \pi^2}~h^D_{uu}(t) \l'_{iuu}(t) h^D_{jk}(t) \delta_{jk} \\
{d B_{\e_i}(t) \over dt}&=& B_{\e_i}(t) \le -{3 \over 2} Y_b(t) - {1 \over 2} 
Y_\tau (t) + {3 \over 2} \tilde{\a}_2(t) + {3 \over 10} \tilde{\a}_1(t) \ri 
\nonumber \\
&-& {3 \over 16 \pi^2} \mu(t) \l'_{ijj}(t) 
h^d_{jj}(t) \le {1 \over 2} B_\m(t) + A^{\l'}_{ijj}(t) \ri \\
{d m_{\n_i H_1}^2 (t)\over dt}&=& m_{\n_i H_1}^2(t) \le - {3 \over 2} Y_b(t)
- {1 \over 2} Y_\tau(t) \ri - { 3 \over 32 \pi^2 } \l'_{ipp}(t) h^d_{pp}(t)
\left( m_{H_1}^2(t) \right. \nonumber \\
&+& \left. m_{L_i}^2(t) + 2 m_{Q_p}^2(t) + 2 A_{ipp}^{\l'}(t)
 A^D_{pp}(t) + 2 m^2_{D_p}(t) \right)  \\
{d A^{\l'}_{ijj}(t) \over dt}&=& - {9 \over 2} A^D_{jj}(t) Y^D{jj}(t) - 
{3 \over 2} A^{\l'}_{ijj}(t) Y^D_{jj}(t) - A^U_{jj}(t) Y^U_{jj}(t)
\nonumber \\
&-& A^E_{ii}(t) Y^E_{ii}(t) - {7 \over 30} M_1(t) \tilde{\a}_1(t)
- 2 M_2(t) \tilde{\a}_2(t) - {16 \over 3} M_3(t)\tilde{\a}_3(t) 
\eea

\vskip 0.3cm
\noindent
{\it Trilinear R-violation ($\l$)}\\
\vskip 0.2cm 
\noindent

\bea
\label{lambdarge}
{d \l_{ijk}(t) \over dt}&=& \l_{ijk}(t) \le {-1 \over 2} Y^E_{ii}(t)
-{1 \over 2} Y^E_{jj}(t) - {1 \over 2} Y^E_{kk}(t) + {3 \over 2} 
\tilde{\alpha}_2(t) + \tilde{\alpha}_1(t) \ri \nonumber \\
&-& {1 \over 2} \le  \l_{ijj}(t)  Y^E_{jj}(t) \delta_{j3} \delta_{jk} +
\l_{jii}(t) Y^E_{ii}(t) \delta_{i3} \delta_{ik} \ri \\
{d B_{\e_i}(t) \over dt}&=& B_{\e_i}(t) \le -{3 \over 2} Y_t (t) - {1 \over 2}
Y^E_i(t) + {3 \over 2} \tilde{\a}_2(t) + {1 \over 2} \tilde{\a}_1(t) \ri 
\nonumber\\
&-& {1 \over 32 \pi^2}  \m(t) \l_{idd}(t) h^E_{dd}(t) \le A^\l_{idd}(t) +
{1 \over 2} B_\m(t) \ri \\
{ d m^2_{\n_i H_1}(t) \over dt}&=&m^2_{\n_i H_1}(t) \le - {3 \over 2} Y^E_i(t)
- {1 \over 2} Y_\tau(t) - {3 \over 2} Y_b(t) \ri - {1 \over 32 \pi^2} 
\l_{ijj}(t) h^E_{jj}(t) \left( m_{H_1}^2(t) \right. \nonumber \\
&+& \left. 2 ~m_{L_i}^2(t)~m_{L_j}^2(t) + 2 A^{\l}_{ijj}(t) A^E_{jj}(t)
+ 2 m_{E_j}^2 \right) \\
{d A^{\l}_{ijj}(t) \over dt} &=& -3 M_1(t) \tilde{\alpha}_1(t) - 3 M_2(t) 
\tilde{\alpha}_2(t) - {7 \over 2} A^E_{jj}(t) Y^E_{jj}(t) - {3 \over 2}
A^{\l}_{ijj}(t) Y^E_{jj}(t)
\eea

%% file: chap4.tex
\chapter{Bilinear R violation and Neutrino Anomalies }

\section{Introduction:}
Among the Standard Model fermions, neutrinos are the only ones 
which do not carry any electric charge. This leads to the possibility
that these particles may be of majorana nature. Majorana masses to
the neutrinos would violate lepton number by two units. To generate
majorana masses to the neutrinos within the Standard Model, one has
to modify significantly its scalar sector. On the other hand, 
Supersymmetric Standard Model naturally allows for lepton number
violation. However, these couplings are generally removed by imposing
R-parity. To allow the lepton number violating couplings back in to
the superpotential, one can discard R-parity and impose other symmetries
like baryon parity. The  R-breaking couplings now present in the
superpotential would give rise to neutrino masses \cite{hall}. Another
approach is to assume R-parity is spontaneously broken \cite{masvalle}.
This approach however leads to a low scale massless mode called 
majoron \cite{majoron} which can be cured \cite{romaovalle}. In 
this thesis, we follow the earlier approach where explicit lepton
number violating couplings are present in the superpotential.

In the present chapter, we look at a specific model of R-parity 
violation and study the neutrino mass spectrum in these models.
As has been noted earlier,  lepton number violation is characterised
either by dimensionful $\e_i$ terms or the dimensionless $\l, \l'$
terms in the superpotential. In the present chapter, we consider
that only dimensionful $\e_i$ terms are the source of lepton number
violation in the superpotential. Such a scenario is sometimes known
as bilinear R-parity violation in the literature.

\be \label {bi}
W_{\not{L}} = \epsilon_i L'_i H_2
\ee
 
There are several theoretical motivations to consider such a scenario.
Theories with spontaneous breaking of lepton number can be closely
identified with these models if one assumes $\e = h \n_R$ where
$h$ is the Yukawa coupling of the gauge singlet and the singlet
field is represented by $\n_R$ \cite{romaovalle}. The $\e$  couplings
can be identified as the {\it vevs} of the singlet fields setting the scale
of the lepton number violation. Alternatively one could imagine   
a  generalised Peccei-Quinn symmetry whose spontaneous breaking leads to $\mu$
and $\e_i$ terms at the weak scale through dim 5 operators \cite{tamvakis}. 
Moreover, it is possible to choose the PQ charges of the different fields in
such a way that the generation of the effective trilinear operators
is suppressed  but bilinear terms are allowed \cite{asjbabu}. 
A study of neutrino masses in bilinear R violating models
is thus theoretically well motivated.  Several works have been done
to this extent in the recent past \cite{hall,asjbabu,rhemp,massbrpv}. 

As we have seen in chapter 2, with the advent of super-Kamiokande 
results on the atmospheric neutrinos with very high statistics, 
a definitive and clear indication
for neutrino masses and mixing has been achieved. Though there are
still ambiguities associated with the solar neutrino 
experiments, the structure of neutrino mass matrix has achieved more clarity
with one large mixing as a necessary ingredient if one would like to have
simultaneous solutions for solar and atmospheric neutrino problems with
three active neutrinos. It has been known for sometime that neutrino mass 
matrix and mixing are completely calculable in terms of standard 
supersymmetry breaking parameters and R violating parameters in 
supersymmetry models with R violation. The question then arises
is, whether it is possible to have a neutrino mass matrix which would be
able to provide simultaneous solutions to both solar and atmospheric
neutrino problems within bilinear R violating models which have a minimality
of only three R violating couplings. Solutions to neutrino anomalies 
have been extensively studied by Hempfling \cite{rhemp} and Joshipura and
Babu \cite{asjbabu} within these models though they had not concentrated
on simultaneous solutions. These studies have been done within the framework
of standard supergravity inspired MSSM with universal boundary conditions
at the high scale $\sim M_{GUT}$. In the present work we extend the analysis
and look for simultaneous solutions to solar and atmospheric neutrino problems.
We also confine our study to models where supersymmetry breaking is 
communicated through gauge interactions, in particular to the popular 
Minimal Messenger Model (MMM) \cite{gm1}. 
We conduct our study \cite{ggm} within the framework
of Joshipura and Babu \cite{asjbabu} which is more analytical. 

\section{Structure of Neutrino Masses:}

The bilinear lepton number violating couplings have a unique feature
associated with them. By a simple redefinition of the fields one 
can remove these dimensionful couplings from the superpotential. 
This is possible because the superfields $L_i, H_1$ carry the 
same quantum numbers under $SU(2)_L \times U(1)_Y$ and such a redefinition 
would leave the action invariant \cite{weinvol1c7}. For example, 
considering only one dimensionful coupling in the superpotential 
$\e_3$, a suitable redefinition of the type:
\be
L_3 \rightarrow { \e_3 L_3 + \m H_1 \over \sqrt{ \e_3^2 + \m^2 }} \;\;\;\;
H_1 \rightarrow { -\m H_1 + \e_3 L_3 \over \sqrt{ \e_3^2 + \m^2 }}
\ee
would lead to absence of dimensionful lepton number violation in the
superpotential. It should be however noted that such a redefinition 
is not possible in the Standard Model as the leptonic fields and the 
Higgs fields transform as two different representations under the
Lorentz group. This distinction is removed in the MSSM.

The above feature of bilinear R violation had led to the general belief
that these couplings would not be of interest as, by a general redefinition
one can transform the dimensionful lepton number violation to the 
dimensionless lepton number violation in the superpotential. 
A crucial feature which works against the above arguments is the 
presence of supersymmetry breaking
soft terms. These terms do not preserve the above redefinition in 
general \cite{hall}. This complication can completely change the
phenomenology of the supersymmetry spectrum and neutrino 
masses in particular. To understand these issues in a better 
manner we follow the
framework given by Joshipura and Babu. In this framework most of the 
features of the bilinear R parity violating models are transparent 
unlike the other studies which were more numerical. 

\vskip 0.3cm
\noindent
{\it The framework:}\\

\noindent
Here we consider a redefinition of the fields at the weak scale.
A similar redefinition of the fields can be done at the high scale. 
In models with universal soft terms at the high scale, such a 
redefinition would rotate away the soft bilinear terms too. 
The superpotential with bilinear R-violation is written as :

\be \label{sup}
W= \l_{i}L_{i}'E_{i}^c H_1'+\m'H_1'H_2+\e_{i}L_{i}'H_2+\l_{i}^D
Q_{i}D^c_{i}H_1'+\l_{ij}^U
Q_{i}U^c_{j}H_2 \;, 
\ee

\noindent
where  $ i,j=1,2,3$ represents the generation index.
A specific choice of the basis is also made such that 
these fields denote the mass eigenstates of the charged lepton (down
quarks) in the absence of the $\e_i$ terms. We now
redefine the leptonic and the Higgs fields in such a way that 
the superpotential (\ref{sup})
does not contain bilinear $\e_i$-dependent terms.
The redefinition is an orthogonal transformation which we define as follows
in the unprimed basis:

\bea \label {h1}
H_1&=&c_3 H_1'+s_3(s_2(c_1L_2'+s_1L_1')+c_2L_3'), \nonumber \\
L_1&=&-s_1L_2'+c_1L_1', \nonumber \\   
L_2&=&c_2(c_1L_2'+s_1L_1')-s_2L_3',  \nonumber \\  
L_3&=&-s_3 H_1'+c_3(s_2(c_1L_2'+s_1L_1')+c_2L_3'); 
\eea
where

\be \label{angles} \ba{cc}
\e_1=\m s_1s_2s_3,\;\;\;\;&\e_2=\m c_1s_2s_3,\\
\e_3=\m s_3c_2,\;\;\;&\m'=\m c_3, \ea \ee 
and  $\m\equiv 
(\m'^2+\e_1^2+\e_2^2+\e_3^2)^{1/2}$.
The consequences of the above redefinition are two fold :
a) The lepton number violation now reappears as trilinear lepton number
violation in the superpotential. b)The originally diagonal charged lepton mass
matrix $M_l$ now acquires \cite{asjbabu} non-diagonal parts given
by \footnote{Note that we have neglected here a sub-dominant contribution
to ${\cal M}_l$ arising due to sneutrino vevs.}:
\be \label{ml}
{M_l\over <H_1>}= \left (
\begin{array}{ccc} \l_1c_1c_3&-\l_2 s_1c_3&0\\
                  \l_1s_1c_2c_3&\l_2 c_1c_2c_3&-\l_3s_2c_3\\
                  \l_1s_1s_2&\l_2c_1s_2&\l_3c_2 \end{array} \right ) 
\ee

We denote the mass basis for the charged leptons in the presence of 
non-zero $\e_i$ as $\a=e,\mu,\tau$ and are defined as:
\be \label{weak} 
L_i=(O_L^T)_{i\a} L_{\a},\;\;\;\; e^c_i=(O_R^T)_{i\a} e_{\a}^c,
\ee
where
\be\label{dia} O_LM_lO_R^T=diagonal. \ee

Note that the parameters $\l_i$ which denote charged lepton masses when
$\e_i=0$ still need to be hierarchical if $M_l$ in (\ref{ml}) is to
reproduce the charged lepton masses \footnote{This is because ${\cal M}_l$
is still diagonal even in the presence of non-zero $\e_i$.}.
One can determine \cite{asjbabu} $O_L$ by assuming
$\l_1\ll \l_2\ll \l_3$ and neglecting $\l_1$:
\be \label{ol} 
O_L^{T}\approx N_1\left( 
 \ba{ccc}                c_1&-s_1N_2&0\\
                 s_1 c_2&c_1 c_2N_2^{-1}&-s_2c_3N_1^{-1}N_2^{-1}\\
                s_1 s_2c_3&c_1s_2c_3N_2^{-1}&c_2N_1^{-1}N_2^{-1}\\ \ea
\right) 
\left( 
 \ba{ccc}                1&0&0\\
                 0&\cos\theta_{23}&-\sin\theta_{23}\\
                0&\sin\theta_{23}&\cos\theta_{23}\\ \ea
\right). \ee
where 
\bea
N_1&=&(1-s_1^2s_2^2s_3^2)^{(-1/2)}, \nonumber \\
N_2&=&(1-s_2^2s_3^2)^{(1/2)}, \nonumber \\
\theta_{23}&\approx&N_1 c_1 c_2 s_2 \left( {s_3
m_{\mu}\over c_3m_{\tau}}\right )^2 .
\eea 
$\theta_{23}$ is small even for $s_{1,2,3}\sim O(1)$.
 We shall therefore neglect it.
The trilinear terms generated in the superpotential due to rotation
 (\ref{h1}) assume the following form \cite{asjbabu} in the physical
mass eigenstate basis of charged leptons:
\be \label{w3}
W= -{\tan\theta_3\over <H_1>}[(O_L^T)_{3\a}L_{\a}]\left(
 m^l_{\b}L_{\b}e^c_{\b}+m_i^DQ_id_i^c \right).
\ee

It should be noted here that the trilinear lepton number violating
interactions generated by a rotation of bilinear couplings form only
a part of the most general set of trilinear couplings, especially
these couplings are flavour diagonal \cite{asjbabu}.  
The above trilinear terms lead to neutrino masses at 
1-loop \cite{loop}.  The other contribution to neutrino mass is 
generated by the soft supersymmetry breaking terms in a manner discussed below.

\subsection{Sneutrino vevs, neutrino masses and RG scaling }

The soft supersymmetry breaking part of the scalar potential can be
written as follows in the primed basis:
\begin{equation}\label{soft}\begin{array}{ccc}
V_{soft}&=&
m_{H_1^{\prime
2}}|H^{0\prime}_1|^2+m_{H_2^2}|H_2^0|^2+m_{\tilde{\nu}_i^{\prime^2}}  
|\tilde{\nu}_i^{\prime}|^2 \nonumber \\ 
&-&  \left[ H_2^0\left( \mu'B_{\m} H_1^{\prime 0}+\e_i
B_i\tilde{\nu'}_i\right )\;\; + H. c.\right]. \end{array}  
\end{equation}

The above soft potential contains bilinear lepton number violating
terms which are linear in the sneutrino field. The presence of such
terms can lead to the generation of vacuum expectation values for
the sneutrinos. The generation of the sneutrino {\it vev} in these
models should be however contrasted with models where lepton number 
is broken spontaneously \cite{masvalle}. 
The generation of sneutrino {\it vev} in these models is a consequence
of the explicit lepton number violating terms present in the superpotential.
In the limit these couplings tend to zero there are no sneutrino {\it vevs}
in these models. Hence these theories are not plagued by majoron problems.

Once the sneutrinos attain {\it vevs}, neutrinos mix with neutralinos through
the kinetic terms in the kahler potential 
 leading to a tree level neutrino
mass. The neutrino mass so generated is proportional to the soft lepton
number violating coupling, $B_{\e_i}$ which can be $\sim O(M_{susy})$. Thus
very large neutrino masses can be generated within these models which can
be problematic phenomenologically. It has been found that
a natural way of suppressing the neutrino mass would be to incorporate 
the theory in models where supersymmetry is
broken at a high scale and the low energy supersymmetry spectrum is determined by
renormalisation group evolution. As we have seen earlier,  in 
these cases, universality between
various soft parameters at the high scale is achievable like for example
in supergravity inspired models.  One can extend this
universality to the lepton number violating soft terms which can lead
to a suppressed sneutrino {\it vev} at the weak scale.  

The above can be more transparently seen by observing that
it is always possible to choose a minimum with zero sneutrino {\it vev} if 
$B_\m=B_i$ and $m_{H_1^{\prime
2}}=m_{\tilde{\n}_i'}^2$. These equalities can be imposed at 
the high scale through
universal conditions. But these are not generally 
satisfied at the weak scale.  This is due to
the presence of Yukawa couplings which distinguish between Higgs and 
leptons. Thus the RG equations which determine the low scale values
of these parameters depend on the Yukawa couplings. If one neglects the Yukawa
couplings of the first two generations then the soft mass parameters
related to the first two leptonic generations evolve in the same way.
Thus, one would have the following non-zero differences among low energy
parameters:
\be \label{deltas1} 
\Delta m_{L,H}\equiv (m_{\tilde{\nu}_3^{\prime^2}}
-m_{\tilde{\nu}_2',H_1'}^2)~,\;\;\;\; 
\Delta B_{L,H}\equiv (B_3-B_{2 ,\mu}). \ee
These differences would now determine the sneutrino {\it vev}. The latter 
are obtained by minimizing the scalar potential expressed in the 
redefined basis of eqs.(\ref {h1}). In this basis, one finds:
\bea \label{vfull}
V&=&(m_{H_1}^2+\m^2)|H_1^0|^2+(m_{H_2}^2+\m^2)|H_2^0|^2+m_{\tilde{\nu}_3}^2
|\tilde{\nu_3}|^2
+ m_{\tilde{\nu}_2'}^2 |\tilde{\nu_2}|^2 + m_{\tilde{\nu_1}'}^2
|\tilde{\nu_1}|^2 \nonumber\\ 
&-& \left[ \mu H_2^0\left( B H_1^0+ c_3s_3 \tilde{\nu}_3( \Delta
B_H-s_2^2 
\Delta B_L)
- c_2 s_2 s_3 \tilde{\nu}_2  \Delta B_L\right)\;\; + H. c.\right ]
\nonumber \\
&+&\left[ -c_2 s_2 \Delta m_L (s_3H_1^0+ c_3 \tilde{\nu}_3)
\tilde{\nu}_2^*
+ s_3c_3 \tilde{\nu}_3 H_1^{0*} ( \Delta m_H- s_2^2 \Delta m_L)
\;\;+\;H.c.
\right] \nonumber\\ 
&+& {1\over 8} (g^2+g'^2) ( |H_1^0|^2+|\tilde{\nu}_1|^2
+|\tilde{\nu}_2|^2+|\tilde{\nu}_3|^2-|H_2^0|^2)^2 ,
\eea
where 
\bea \label{vfull1}
m_{H_1}^2&=&m_{H_1'}^2+s_3^2\Delta m_H-s_2^2 s_3^2 \Delta m_L,
 \nonumber \\ 
m_{\tilde{\nu}_3}^2&=&m_{\tilde{\nu}_3'}^2-s_3^2\Delta m_H-s_2^2 c_3^2
 \Delta m_L, \nonumber \\
m_{\tilde{\nu}_2}^2&=&m_{\tilde{\nu}_2'}^2+s_2^2  \Delta m_L, \nonumber \\
B&=&B_{\mu}+s_3^2\Delta B_H-s_2^2 s_3^2 \Delta B_L .
\eea

The additional terms do not significantly affect the {\it vev} for the 
standard Higgs since the sneutrino {\it vevs} and  
$\Delta m_{H,L}$, $\Delta B_{H,L}$ 
are suppressed compared to $m_{SUSY}^2$. This remains true even for
$s_3 \sim O(1)$. The effect of these terms is to induce {\it vevs} for the 
sneutrinos. These can be determined by the stationary value conditions
of the above potential as:

\bea \label{omegas}
<\tilde{\nu_2}>&\approx& {c_2s_2s_3\over (m_{\tilde{\nu}_2}^2+D)}
 (v_1\Delta m_{L}-\mu v_2 \Delta B_L)~, \nonumber \\
<\tilde{\nu_3}>&\approx& {c_3s_3\over (m_{\tilde{\nu}_3}^2+D)}
 (v_1(-\Delta m_H+s_2^2\Delta m_L)-\mu v_2 (-\Delta B_H+s_2^2\Delta B_L))~.
\nonumber \\
\eea
We have neglected terms higher order in  $\Delta m_{L,H}, \Delta B_{H,L}$
while writing the above equations. Note that one of the sneutrino field 
($\equiv \tilde{\nu_1}$) does not acquire a {\it vev} in this basis. This 
{\it vev} would arise if Yukawa couplings 
of the first two generations neglected here
are turned on. It should be noted however that the above sneutrino {\it vevs}
have been derived from the RG improved `tree level' scalar potential. 
Minimisation of the full one-loop effective potential can significantly
modify these values \cite{zwir}. For the sneutrino {\it vevs}, one loop
corrections have been presented recently by Chun {\it et. al} \cite{pokchun} 
and Hirsch {\it et al} \cite{rometh}. Chun {\it et. al} use 
effective potential method
whereas, Hirsch {\it et. al} use diagrammatic method. In the limit of small
R violation, as is in our case, Chun {\it et. al} have found that the 
one-loop corrections are significant only in the regions of parameter space
where the tree level mass is suppressed and the loop mass to the neutrinos
dominates. We do not consider these corrections here. However, as we will
see below, the effect of these corrections would be negligible on our results
as in the present model, we do not encounter such a scenario in the 
parameter space. 

The sneutrino {\it vevs} are zero at the boundary scale $M_X$ corresponding
to the universal masses. Their weak scale values are determined by solving the
relevant RG equations. These RG equations can be determined by the general
set of RGE presented in chapter 3. They are given as, 
\bea \label{rgq}
{d\;\Delta m_H\over d\;t}&=&3 Y_b(t) ( m_{H_1'}^2+ m_{\tilde{b}}^2
+m_{\tilde{b^c}}^2+ A_b^2)~,   \nonumber\\ 
{d\;\Delta m_L\over d\;t}&=&- Y_{\tau}(t) ( m_{H_1'}^2+ m_{\tilde{\tau}}^2
+m_{\tilde{\tau^c}}^2+ A_{\tau}^2)~, \nonumber\\ 
{d\;\Delta B_H\over d\;t}&=&3 Y_b(t) A_b(t)~,  \nonumber\\
{d\;\Delta B_L\over d\;t}&=&- Y_{\tau}(t)  A_{\tau}~, \eea
where we follow the same notation as earlier with  
$Y_{f}\equiv {\lambda_{f}^2\over (4\pi)^2}$, $m_{\tilde{f}}$
is the mass of the sfermion concerned, $A_f$ are the trilinear soft
susy breaking terms and $t=2\;ln(M_{X}/Q)$. 

The tree level mass matrix generated due to
these {\it vev} \cite{asmarek} \footnote{In the Appendix  we present the details
of this derivation} can be written in the physical basis $\n_\a$ as:  
\be \label{mtree}
M_0= m_0 O_L \left( 
\ba{ccc}
0&0&0\\
0&s_{\phi}^2& s_{\phi} c_{\phi}\\
0&s_{\phi}c_{\phi}& c_{\phi}^2\\ \ea \right) O_L^T, \ee
where 
\be \label{phi}
 \tan \phi= {<\tilde{\nu_2}>\over <\tilde{\nu_3}>} \;\;.\ee
 $O_L$ is defined by eqs.\refs{dia}
and, 
\be \label{m0}
 m_0={\mu (c g^2+g'^2)(<\tilde{\nu}_2>^2+<\tilde{\nu}_3>^2)
 \over 2(-c\m M +2 M_W^2 c_\b  s_\b
(c+ tan^2\theta_W))}, \ee   

with $M_2$ now representing the weak scale  value of the gaugino
mass and $c=0.49$ \footnote{$M_1 = c M_2 $ is used in deriving the
above \cite{habkan}.}.
\subsection{1-loop mass}

The trilinear interactions in eq.~\refs{w3} lead to  diagrams 
involving squarks and
sleptons in the loop and generate the neutrino masses at the 1-loop
level\cite{loop}. These contributions depend upon the masses as well as
mixing between the left and the right handed squarks as well as the sleptons. 
These are however fixed in terms of the basic parameters of supersymmetry 
breaking. In the present
case, the trilinear couplings are not independent and are controlled
by the fermion masses. As a consequence, the dominant contribution
arises when the b-squark or $\tau$ slepton are exchanged in the loop. 
We shall retain only this contribution.

Let us define:
\bea \label{mix}
 \tilde{b}&=& \tilde{b}_1 \cos \phi_b+\tilde{b}_2 \sin \phi_b~, \nonumber \\
\tilde{b^c}^\dagger&=& \tilde{b}_2 \cos \phi_b-\tilde{b}_1 \sin \phi_b~.
\eea
Where, $\tilde{b}_{1,2}$ are the mass eigenstates with masses
$M_{b_1,b_2}$ respectively. The mixing angles $\phi_{\tau}$ and masses
$M_{\tau_1,\tau_2}$ are defined analogously in case of the tau slepton.
The exchange of b-squark produces the following mass
matrix for the neutrinos:
\be
(M_{1b})_{\a\b}= m_{1b} (O_L)_{\a 3}(O_L)_{\b 3}.
\ee
Due to the antisymmetry of the leptonic couplings in eq.~\refs{w3},
the exchange of the $\t$ slepton leads to the following contribution:
\be \label{mtau}
M_{1\tau}=m_{1\t}\left (\ba{ccc}
O_{L_{13}}^2&O_{L_{13}}O_{L_{23}}&0\\
O_{L_{13}}O_{L_{23}}&O_{L_{23}}^2&0\\
0&0&0\\ \ea \right). \ee
The mixing induced by these contributions is completely fixed
by the matrix $O_L$ while the overall scale of both these contributions 
is set by,
\be \label{mbt}
m_{1b,1\tau}=N_c {m_{b,\t}^3\over 16 \pi^2 v_1^2} \tan\theta_3^2 
\sin\phi_{b,\t}
\cos\phi_{b,\t} \;\;ln \left({M_{b_2,\t_2}^2\over M_{b_1,\t_1}^2}\right),
\ee
where  $N_c=3,1$ for the $\tilde{b}$ and $\tilde{\tau}$
contribution respectively. 

In the above we have considered one loop corrections to be present only
through the combinations of $\l~\l$, $\l'~\l'$ couplings. But, to derive
the one loop neutrino masses, one has to consider the complete 1-loop
corrections to the $7 \times 7$ neutrino-neutralino mass matrix 
\footnote {Please see the appendix.} . This approach has been 
followed by \cite{rhemp} and recently by  M. Hirsch {\it et.al} \cite{rometh}
where the complete 1-loop mass matrix is written in the tree level mass
basis and re-diagonalised. However, the approximation we have made in
this work can be justified following the work of Chun and collaborators 
\cite{pokchun}. This analysis is based on effective mixing matrix approach.
Using this method, one can analytically understand the dominant  contributions
to the neutrino sector. The 1-loop corrections to the neutralino sector
are of order $\sim 6 \%$ and thus can be neglected. The corrections to the
neutrino-neutralino mixing (Dirac-type) part are sub-dominant to the 
corrections induced in the neutrino mass matrix. 1-loop corrections to 
the neutrino mass matrix are through combinations of the couplings $\l~\l$,
$\l'~\l'$, $h_\tau~h_\tau$, $h_\tau~\l$ \cite{pokchun}. However the most
dominant contribution is through the combination of couplings $\l~\l$,
$\l'~\l'$ \footnote{Recently an additional diagram has been reported in
the literature \cite{habdavid}. We do not consider it here.} . 

The total mass matrix including  the 1-loop
corrections  is given by,
\be \label{total}
{\cal M}_\n=M_0+M_{1b}+M_{1\tau}. \ee

We stress that the above ${\cal M}_\n$ is in the physical basis
with diagonal charged lepton masses. This matrix assumes particularly
simple form when rotated by the matrix $O_L$:
\be \label{masses}
O_L^{T}{\cal M}_\n O_L~ \approx \left ( 
 \ba{ccc}                0&0&0\\
                 0&m_0s_{\phi}^2+m_{1\tau}N_2^{-4}c_2^2s_2^2c_3^2
&m_0s_{\phi}c_{\phi}+m_{1\tau}N_2^{-4}c_2s_2^3c_3^3\\
0&m_0s_{\phi}c_{\phi}+m_{1\tau}N_2^{-4}c_2s_2^3c_3^3&
m_0c_{\phi}^2+m_{1b}+m_{1\tau}N_2^{-4}s_2^4c_3^4\\ \ea
\right) \ee
This explicitly shows that one of the neutrinos is massless in our
approximation of neglecting Yukawa couplings of the first two generations.
The full mixing matrix analogous to the KM matrix is given by,
\be \label {fmix}
 U =O_L O_\n^T.
\ee
where  $O_\n$ is the matrix diagonalising the RHS of eq.(\ref{masses}).
As we will show the mixing angle appearing in $O_\nu$ is small due to
hierarchy in neutrino masses while, the $O_L$ can contain large mixing. 
Hence, the neutrino masses are determined by the matrix (\ref{masses}) and
mixing among neutrinos is essentially fixed by eq.(\ref{ol}). 

The above formalism shows that the neutrino masses are greatly
suppressed compared to the typical supersymmetry breaking scale if
$\Delta m_{H,L},\Delta B_{H,L}$ vanish at some scale $X$.
The weak scale values of sneutrino {\it vev} and hence neutrino masses
follow from evolution of these parameters. It is clear 
from eq.(\ref {rgq}) that the $b$ and $\tau$ Yukawa couplings control the
evolution of sneutrino {\it vev}. Similarly, the 1-loop masses following from
eq. (\ref {mbt}) are also controlled by the same couplings. As a result, 
all the effects of lepton number violating parameters $\e_i$ can be rotated away
from
the full Lagrangian when the down quark and the charged lepton Yukawa
couplings vanish. Neutrino masses also vanish in this limit. 

This formalism has been applied to the case of MSSM with universal boundary
conditions in \cite{asjbabu,rhemp}. Before studying this model in the case of
gauge mediated models which is the work of this thesis, we here review
the main results of the works of Hempfling and Joshipura-Babu. 

\vskip 0.3cm
\noindent
{\it Review of results from MSSM}:
\vskip 0.3cm

\noindent
As mentioned earlier, universality at the high scale for the soft
parameters leads naturally to small neutrino masses at the weak
scale. Constrained MSSM i.e, minimal Supergravity with universal
boundary conditions at $M_{GUT}$ provides an appropriate framework
from this point of view. The total parameters of the above model in the
CMSSM framework are the standard CMSSM parameters
$m$, $M_2$, $\tan \b$, $A$ and the sign($\mu$) along  with the
R-parity violating parameters $\e_i$ or the three angles $s_1,s_2,s_3$.

It is found that the loop mass $m_{1\tau}$ contributes negligibly to
the neutrino mass spectrum, through out the CMSSM parameter space
\cite{asjbabu,rhemp}. In the limit  $m_{1 \tau} \ll m_{1 b},
 m_0$ approximate expressions for the neutrino masses can be
 given as \cite{asjbabu}
\bea
\label{masscmssm}
m_{\n_3}&\sim& m_0 + m_{1b} \nonumber \\
m_{\n_2}&\sim& { m_0 m_{1b} s_{\phi}^2 \over m_0 + m_{1b} },
\eea
where the approximation $s_{\phi}^2 \ll 1$ is used in obtaining the second
line of the above. One of the neutrinos remains massless in this
approximation. The main features of the neutrino mass spectrum
in this framework can be summarized as follows :

\begin{itemize}
\item For most of the CMSSM parameter space, the tree level
contribution dominates over the loop contribution giving rise
to the hierarchical pattern in the mass spectrum \cite{asjbabu,rhemp}.

\item But, there also exist regions in the CMSSM parameter space, where the
two contributions to the sneutrino {\it vevs}, eq.(\ref{omegas}) cancel
each other for one particular sign of $\mu$. 
 In these regions, the loop contribution, $m_{1b}$ can become
 comparable to the tree level mass,
$m_0$ and cancellations among $m_0$ and $m_{1b}$ can take place. Here,
the two neutrinos form a pseudo-Dirac pair with a common mass
$m_0 s_\phi$ relevant for solutions of the neutrino anomalies \cite{asjbabu}.

\item The mixing matrix is given by the product $O_L O_\n^T$. Even in
the case where $O_\n$ allows only small mixing, large mixing can
be generated from $O_L$, as it depends only on the ratios of the R-parity
violating parameters \cite{asjbabu}.

\item Even though the neutrino masses are suppressed in these scenarios,
they are typically of the $O(\MeV)$. Thus, the R-parity violating parameters
 have to be chosen to be much suppressed compared to the typical
order of the supersymmetry breaking scale if neutrino masses are to be
of the right order to solve the neutrino anomalies \cite{asjbabu,rhemp}.

\item Numerical results from Hempfling \cite{rhemp} show that solutions
for solar (either with MSW conversion or vacuum oscillations) and
the atmospheric neutrino anomalies can be accommodated naturally within
these models.

\item Recently an extensive analysis has been reported by M. Hirsch {\it al}
within the framework mSUGRA inspired MSSM. They have reported to have
found no solutions for the bi-maximal mixing scenarios within these 
models with universal boundary conditions. However, with non-universal
boundary conditions one can achieve the required \cite{rometh}.

\end{itemize}

\section{Gauge mediated models and neutrino masses}

The suppression of neutrino masses due to universality of the supersymmetry
breaking parameters at some scale not only happens in the MSSM with
universal boundary conditions (CMSSM), but also in models where 
supersymmetry breaking is mediated through ordinary gauge interactions,
which have been introduced in chapter 1. To understand how it happens, 
we consider here in some detail the minimal version of gauge mediated 
supersymmetry breaking \cite{bdine}.
In this case the messenger sector contains only one pair of 
superfields $\Phi,\bar{\Phi}$ transforming as
5+$\bar{5}$ representation of the $SU(5)$ group. They couple
to a field $S$ which is a gauge singlet. Both the scalar and the
auxiliary components of $S$ attain {\it vevs}, thus introducing 
a supersymmetric mass scale $X\equiv\lambda <S>$ as well as a 
SUSY breaking (mass)$^2$
differences of order $F_S$. Models with minimal messenger sector are thus
characterized by two parameters $\Lambda\equiv{F_S\over X}$
and $x\equiv{\Lambda \over X}$. 

All the soft parameters related to MSSM fields are fixed at $X$ in
terms of $\Lambda,x$ and the gauge couplings. The gauginos attain
their masses at the 1-loop level whereas the sparticles attain their
masses at the two-loop level. These masses  have the following 
simple form \cite{r1}:
\bea \label {squ}
m_i^2(X)& =& 2 \Lambda^2 \left\{ C_3 \tilde{\alpha}_3^2(X) + C_2
\tilde{\alpha}^2_2(X) + \frac{3}{5} Y^2 \tilde{\alpha}_1^2(X) 
\right\}f(x), \nonumber \\
M_j(X)&=&\tilde{\alpha}_j(X)~\Lambda~g(x),
\eea 
where $m_i^2$ represents the scalar masses with $i$ running 
over all the scalars, whereas, $M_j$ represents the 
gaugino masses with $j$ representing the three gauge couplings. 
The functions $f(x)$ and $g(x)$ have been derived in \cite {spnew}. 
Here, 
\be
\tilde{\alpha}_j(X) = {\alpha_j(X) \over (4 \pi)};
\ee
$C_3$ = 4/3,0 for triplets and singlets of $SU(3)_C$, $C_2$ = 3/4,0 for
doublets and singlets of $SU(2)_L$ and Y = Q - $T_3$ is the hypercharge.
Since the Higgs field $H_1$ and the lepton fields $L_i$ carry the
same quantum numbers, $m_{H_1}^2(X) = m_{\n_i}^2(X)$. 
The minimal messenger model (MMM) is further characterised
by the assumption of the vanishing bilinear (B) and trilinear (A)
soft parameters at scale $X$.

The phenomenology of the Minimal Messenger Model has been extensively
studied in various papers \cite{gm1,spnew,bkw,gm2,r1,borz}. 
In this thesis, we follow the
work of Borzumati \cite{borz}. The salient features of this model 
would not change much even in the presence of the non-zero $\e_i$.
These can be summarised as follows:
Only one parameter essentially determines the entire soft spectrum
as the dependence of the boundary conditions in eq.(\ref{squ}) on
$x$ is very mild. In the present analysis we choose $x = {1 \over 2}$
following \cite{borz}. The MMM is attractive in view of the very restricted 
structure it offers. But as we will show it turns out to be too restrictive
if one wants to solve the solar and atmospheric neutrino problems
simultaneously. We shall thus consider an alternative version on
phenomenological grounds in which the  boundary conditions (\ref{squ}) 
are still imposed but the value of $B_\m$ at $\Lambda$ is not taken to be zero. 
This we call as non-minimal model of Gauge Mediated SUSY breaking. 

\vskip 0.2cm
\noindent
{\it Framework for gauge mediated models}:
\vskip 0.2cm
\noindent
The neutrino mass framework developed in the earlier section holds
good for the Gauge Mediated Models too except that the boundary 
conditions are now defined at the low scale, $X$. The formulae for
the neutrino mass matrix remain unchanged. 
The value of the $B_\m$ parameter at the weak scale
gets fixed through its running. This in turn determines both $\mu$
as well as $\tan\beta$ through the minimisation equations presented
in chapter 1:
\bea \label {mini}
\mu^2&=&{m_{H_1}^2 - m_{H_2}^2 \tan^2 \beta \over
 \tan ^2 \beta - 1} - {1 \over 2} M_Z^2 ,\nonumber \\
\mbox{Sin} 2 \beta&=&{2 B_\m \mu \over m_{H_2}^2 + m_{H_1}^2 + 2 \mu^2} .
\eea
The presence of $\e_i$ induces corrections to these equations, but
they are very small as discussed below eq.(\ref{vfull1}).
The eq.(\ref{mini}) therefore holds to a very good approximation.

In spite of the restricted structure, it is possible to self
consistently solve the above equations \cite{bkw,r1,borz} and implement
breaking of the $SU(2)\times U(1)$ symmetry at low energy. Vanishing of
the soft $B_\m$ parameter at $X$ makes the analysis of this breaking
little more involved than in the case of the supergravity induced breaking.
One needs to include two loop corrections to the evolution of the $B_\m$
parameter and also needs to use fully one loop corrected effective
potential. Details of this analysis are presented in \cite{bkw,r1,borz}. We
follow the treatment given in \cite{borz}. 
The smallness of the $B_\m$ at the weak scale results in this scheme, in
relatively large value of $\tan\beta$ and its sign fixes the sign of $\mu$
to be positive. 
The full 1-loop corrected potential was employed in the
analysis of \cite{borz} but it was found that working with RG improved tree
level
potential also gives similar results provided one evolves soft parameters
of the supersymmetric partners up to a scale $Q_0^2\equiv (m_{\tilde{Q}}^2(X) 
m_{\tilde{U}}^2(X))^{1 \over 2}$.
We prefer to follow this approach and use the RG improved tree level
potential of eq.(\ref{vfull}) in order to determine the low energy 
parameters at the minimum.
We have however included  two loop corrections to the RG equations \cite{sp2l}
for $B_\m$ and $\Delta B_{L,H}$ in determining their values at the weak
scale. Use of RG improved tree level potential  allows
us to analytically understand the structure of neutrino masses
and mixing in a transparent way.

\vskip 0.2cm
\noindent
{\it Neutrino Mass structure}:
\vskip 0.2cm
\noindent
The three mass parameters $m_{0,1b,1\tau}$, 
eqs.(\ref{m0},~\ref{mbt}) 
with appropriate definitions control the
neutrino masses. $m_0$ is determined by solving RG equations (\ref{rgq})
along with similar ones for parameters occurring in them. We have
numerically solved them imposing eq.(\ref{squ}) as boundary conditions 
at $X$.
We evolved these equations self consistently  up to the scale
$Q_0$ defined above. The $m_0$ determined in this manner depends upon
$\m$ as well as $\tan\b$, both of which are fixed in terms of $\Lambda$.

\begin{figure}[ht]
\begin{tabular}{cc}
\epsfxsize 6 cm
\epsfysize 8 cm
\epsfbox[17 145 591 715]{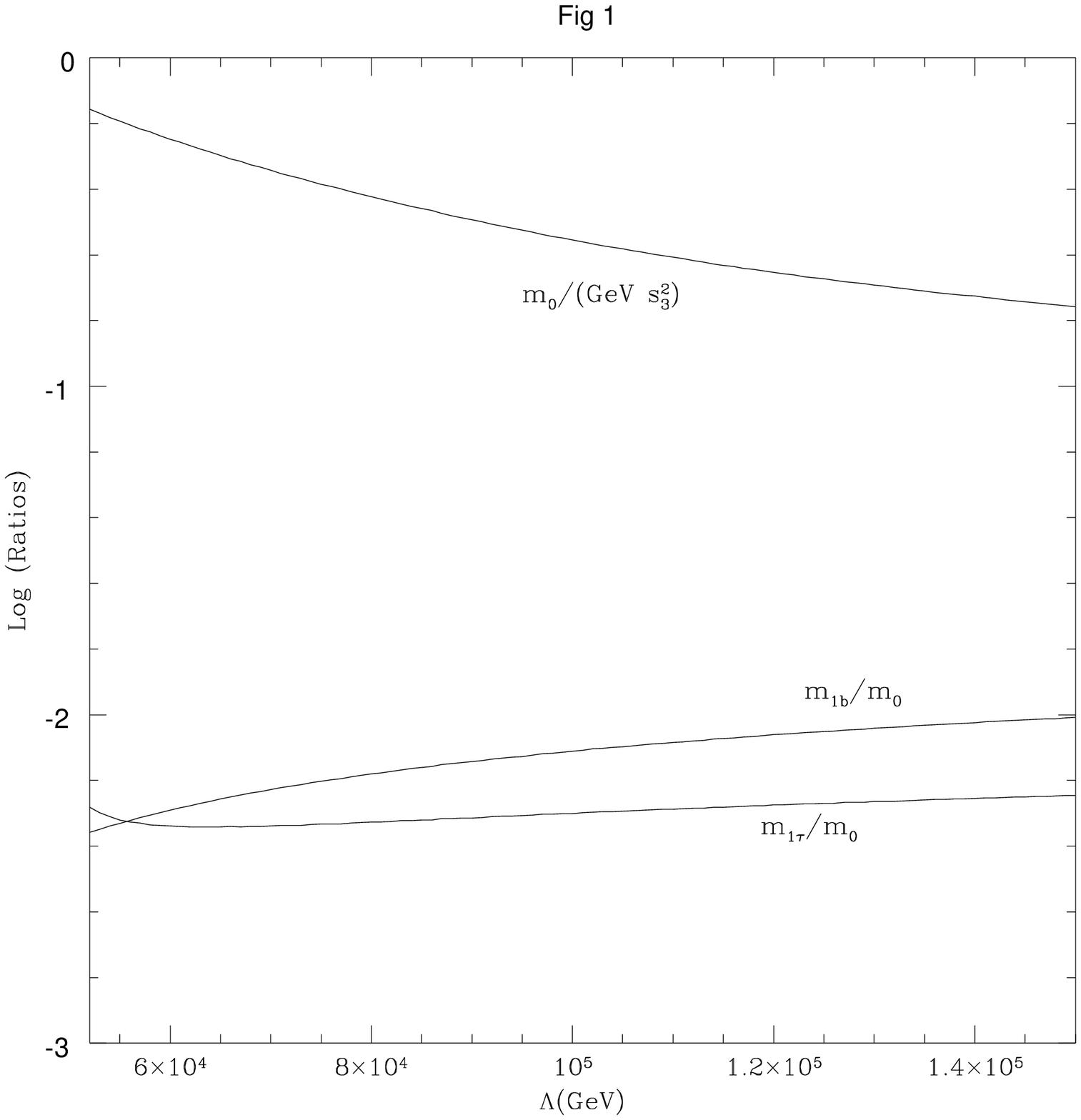}&
\epsfxsize 6 cm
\epsfysize 8 cm
\epsfbox[17 145 591 715]{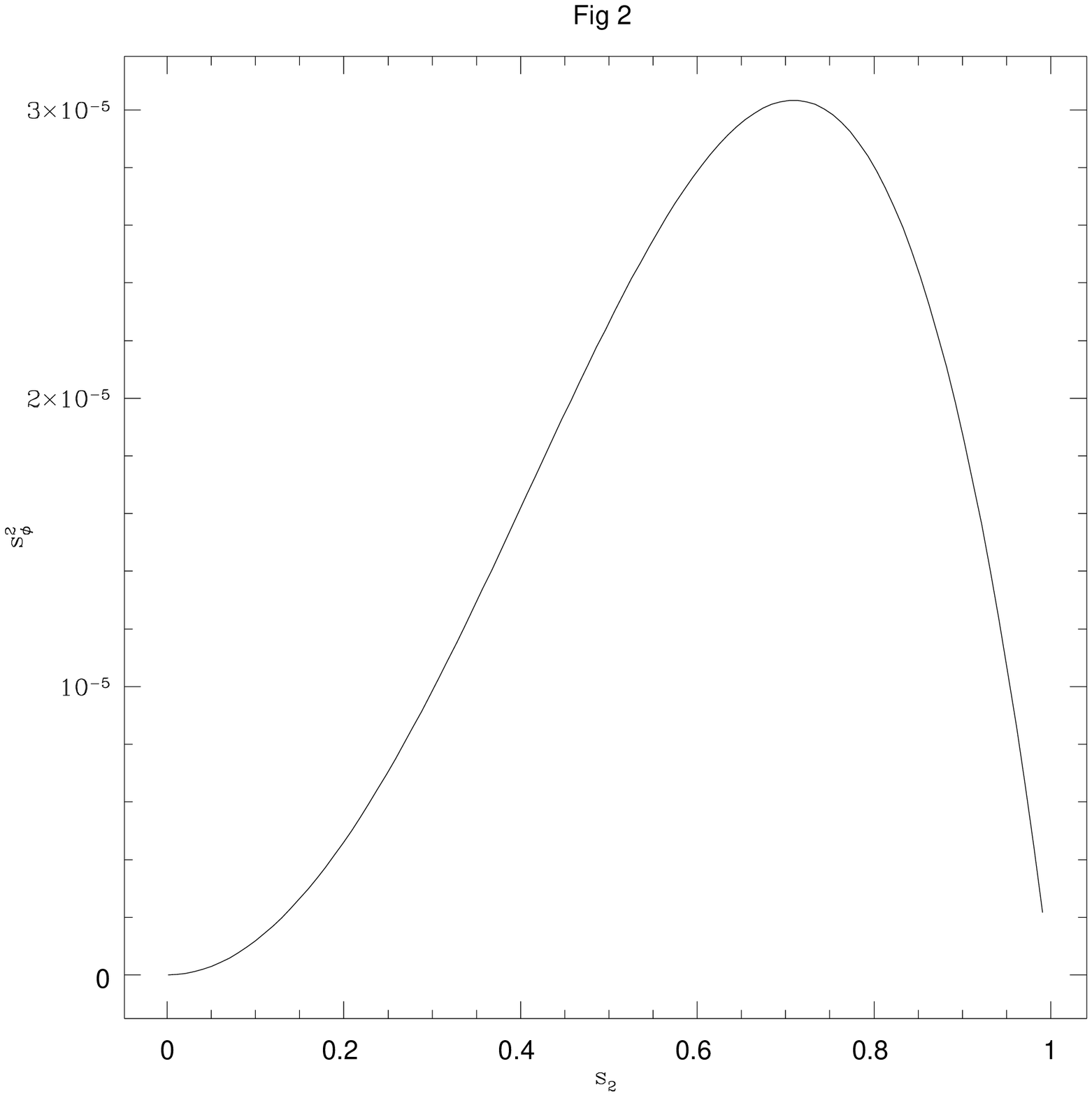}
\end{tabular}
\end{figure}

\noindent
{\bf Fig. 1}. The variations of ${m_0 \over (\GeV s_3^2)}$,
${m_{1b} \over m_0}$,${m_{1 \tau} \over m_0}$ are shown here with
respect to $\Lambda$. $m_0$ mildly depends upon $s_2$ and the displayed
curve is for $s_2 = 0.8$.\\
\noindent
{\bf Fig. 2}.The function $s_{\phi}^2$ is plotted here with respect
to $s_2$.

The loop contributions are fixed in terms of the squark and slepton masses
and mixings defined in eq. (\ref{mix}). These are determined from the
standard
$2\times 2$ matrices involving left and right squarks and slepton mixing.
The elements in these matrices are also completely fixed in terms of
$\Lambda$. All the three parameters $m_0,m_{1b},m_{1\tau}$ depend upon
an overall scale $s_3$ of the $R$ breaking. For small $s_3^2$ they
roughly scale as $s_3^2$.  The ratios $m_0/s_3^2,~m_{1b}/s_3^2,~
m_{1\tau}/s_3^2$ are thus determined by $\Lambda$ alone 
\footnote{$m_0$ also depends on $s_2$ very mildly through the
{\it vev} $\left< \tilde{\nu}_2 \right>$ which is greatly suppressed compared
to $\left< \tilde{\nu}_3 \right>$.}.
 We have displayed in Fig. 1 variations of
${m_0\over \GeV s_3^2}, {m_{1b}\over m_0}$ and ${m_{1\tau}\over m_0}$ 
with $\Lambda$. One  clearly sees  hierarchy in the loop and 
sneutrino {\it vev} induced
contributions. This hierarchy gets reflected in the neutrino masses and 
one obtains hierarchical neutrino masses independent of the
overall strength of the $R$ violating parameter $s_3$.
The mass ratio and hence the  hierarchy
among neutrino masses are seen to be less sensitive  to $\Lambda$.
The $m_0$ roughly scales linearly with $\Lambda$. But since the over all
scale of $m_0$ is set by $s_3^2$ which is also unknown, a change in
$\Lambda$ is equivalent to changing $s_3$. Thus, we may use one specific
value of $\Lambda$ and neutrino mass spectrum is then completely fixed by
three angles $s_{1,2,3}$ or equivalently by the three $R$ violating
parameters $\epsilon_i$.

\vskip 0.3cm
\noindent
The $\Delta m_{H,L}, \Delta B_{H,L}$ entering $m_0$ are determined
from the RG equations (\ref{rgq}) and are fixed in terms of $\Lambda$.
For example,
\bea \label{deltas2}
\mu\sim 397.0\GeV~ ,& ~~\tan\b\sim 46.39~ , \nonumber \\
\Delta m_H\sim 192661.23 \GeV^2~, &~\Delta m_L\sim -2392.35 \GeV^2~,\nonumber \\
\Delta B_H\sim -14.07~ \GeV~ ,&~~\Delta B_L\sim 0.12~ \GeV,
\eea
when $\Lambda=100$ \TeV. The suppression in $\Delta m_L,\Delta B_L$ is
due to color factors and larger squark masses compared to the slepton
masses in the model. It follows that the ratio $\tan \phi$ of the
sneutrino {\it vev}, eq. (\ref{phi}) gets considerably suppressed even when the
angle $s_2$ is large. We show in Fig 2. the value of $s_{\phi}^2$ as
function of $s_2$ for $\Lambda=100~ \TeV$. Note that this ratio 
is independent of the values of the other $R$ violating parameters
when $s_3$ is small.
The small value of $s_\phi$ leads to very simple expression for neutrino
masses. The neutrino mass matrix in eq.(\ref{masses}) is almost diagonal and one
finds:
\bea \label {teta}
m_{\nu_3}&\sim& m_0+m_{1b}~, \nonumber \\
{m_{\nu_2}\over m_{\nu_3}}&\sim& c_2^2 s_2^2 {m_{1\tau}\over
(m_0+m_{1b})}~\sim 
c_2^2 s_2^2~ (7.1~\times~10^{-3} - 5.6~\times~10^{-3} )~,
\nonumber \\
\theta_{23}^\n&\sim& {m_{1\tau}c_2 s_2^3\over (m_0+m_{1b})}~\sim 
\tan\theta_2 {m_{\nu_2}\over m_{\nu_3}}.
\eea
The masses are fixed in terms of $m_{0,1b,1\tau}$ which are determined 
in terms of $\Lambda$ and $s_3$. The mass ratio is fixed in terms of
$s_2$. The range indicated on the RHS in above equation corresponds to
variation in $\Lambda$ from (51 \TeV - 150 \TeV) and $\theta_{23}^\n$
represents the angle diagonalising the matrix in eq.(\ref{masses}).

\section{Neutrino masses: Phenomenology}

As discussed in the last section, the model considered here implies
hierarchical masses and large mixing without any fine tuning of the
parameters. We now try to see if the predicted spectrum can be used to
simultaneously reconcile both the solar and the atmospheric neutrino
anomalies. The model is quite constrained. Three neutrino masses and
three mixing angles get completely determined in the model in terms of
four parameters namely, $\Lambda$ and three $R$ violating angles
$s_{1,2,3}$. In particular, the angle $s_1$ characterizing the electron
number violation does not enter the muon and tau neutrino masses, see
eq.(\ref{masses}). The mixing between neutrinos is largely fixed by the matrix
$O_L$ with a small correction coming from the angle $\theta_{23}^\nu$ in
eq.(\ref {teta}). Thus one has approximately,
\bea \label{pattern}
\nu_e&\approx& N_1( c_1\nu_1 - s_1 c_2\nu_2 + s_1 s_2c_3 \nu_3) \nonumber\\ 
\nu_{\mu}&\approx& N_1(- s_1 N_2\nu_1+ c_1c_2 N_2^{-1}\nu_2
+c_1 s_2 c_3 N_2^{-1}\nu_3)\nonumber \\
\nu_{\tau}&\approx& N_1^{-1}N_2^{-1} ( -s_2 c_3 \nu_2s_1
 +c_2\nu_3) 
\eea
Note that $s_1$ ($s_2$) determines $\nu_e-\nu_{\mu}\;\; 
(\nu_{\m}-\nu_{\tau})$
mixing. We must thus require $s_2$ to be large in order to account for
the atmospheric muon neutrino deficit. The $s_1$ should be small for the
small angle MSW solution and large for the vacuum oscillation solution to
the solar neutrino problem. As we now demonstrate these constraints are
too tight and one does not obtain parameter space in case of the MMM,
allowing simultaneous solution for both these problems.

\vskip 0.5 cm

\begin{figure}[ht]
\begin{tabular}{cc}
\epsfxsize 6 cm
\epsfysize 8 cm
\epsfbox[17 145 591 715]{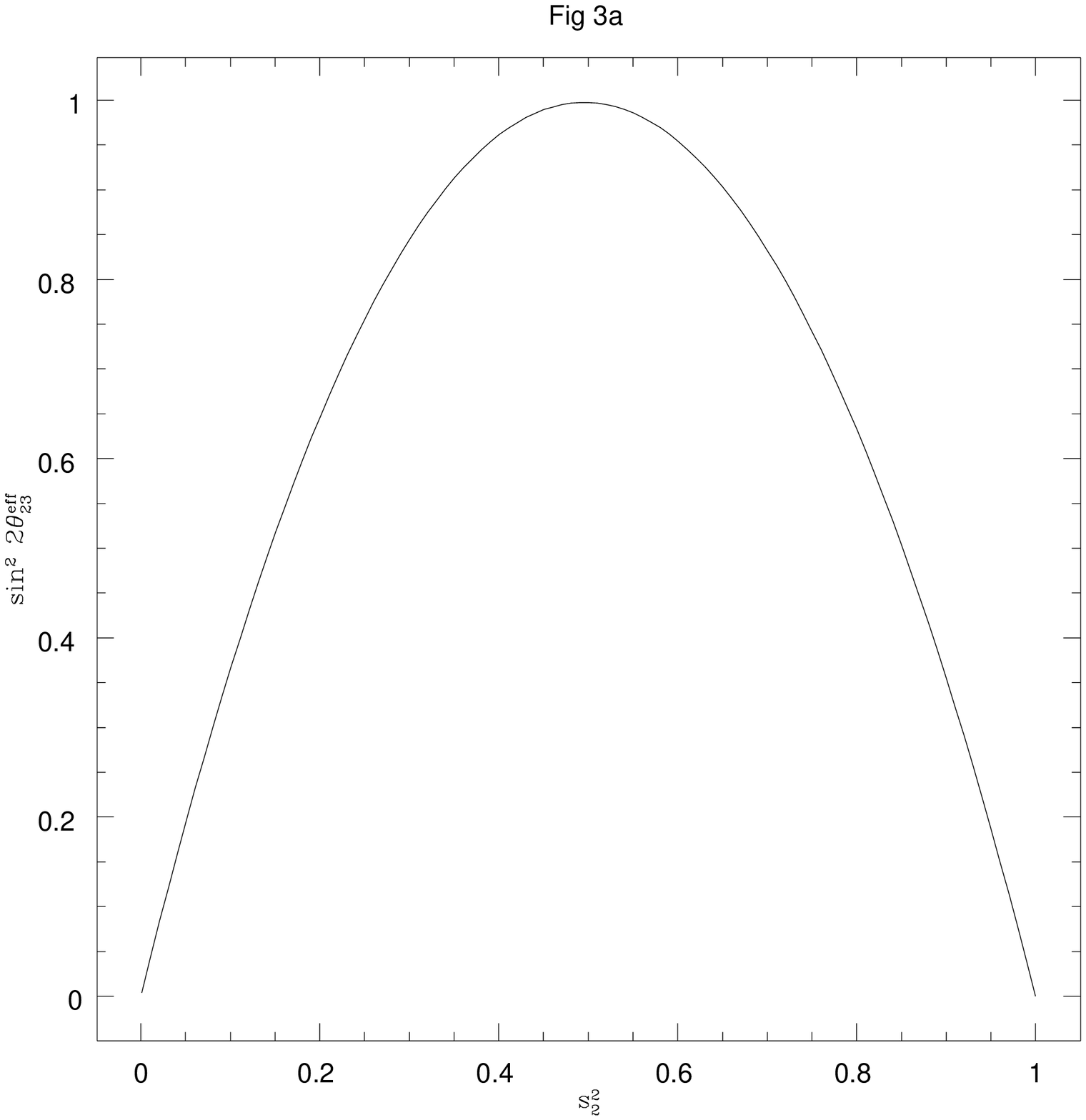}&
\epsfxsize 6 cm
\epsfysize 8 cm
\epsfbox[107 160 539 639]{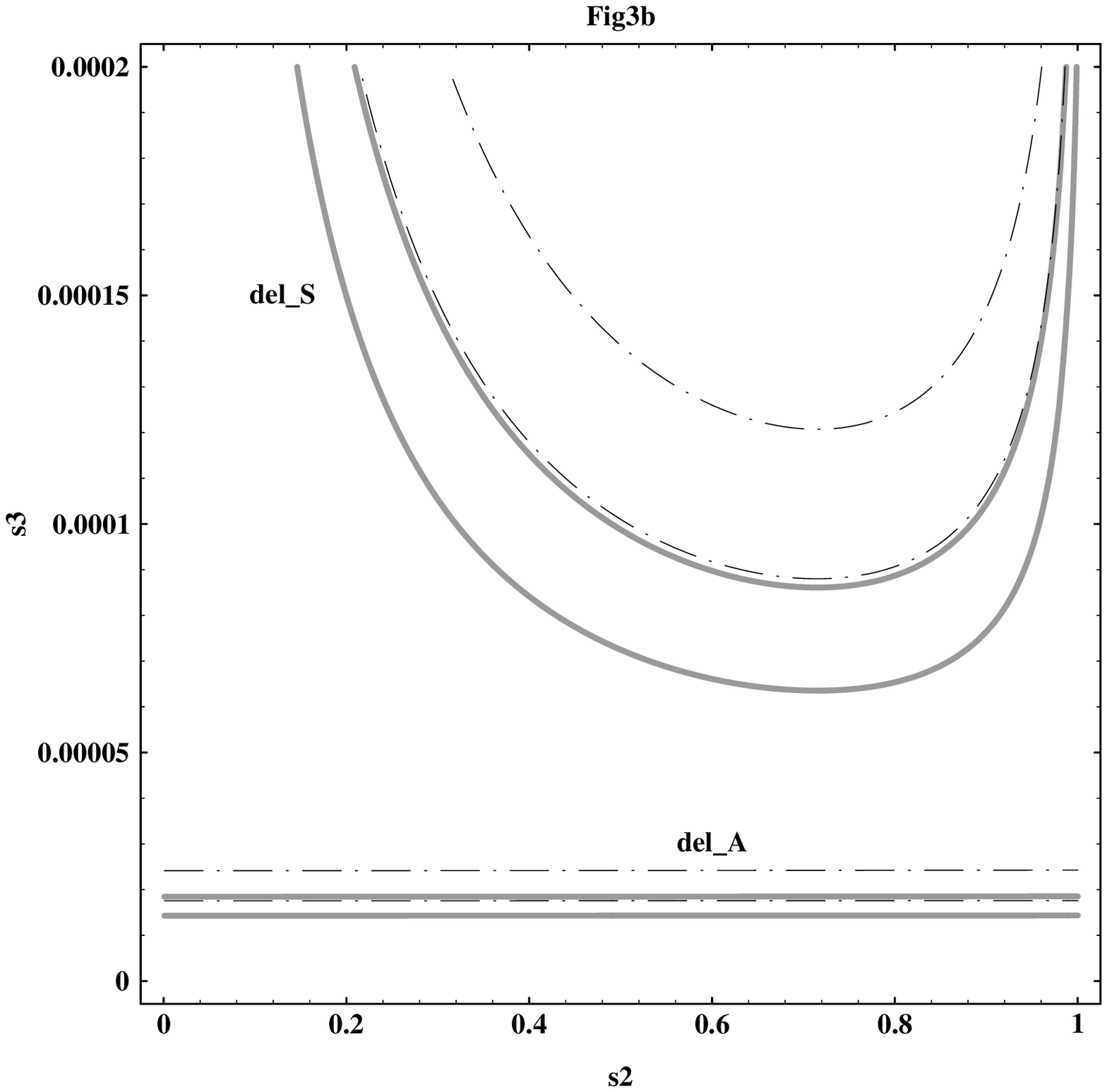}
\end{tabular}
\end{figure}

{\bf Fig. 3a}. The effective $\nu_\mu $-$\nu_\tau$  mixing angle
is plotted here with respect to $s_2^2$ for $\Lambda = 100 \TeV$
in the case of minimal messenger model.
{\bf Fig. 3b}. Contours of $\Delta m^2$ are plotted in {\bf MMM} case, 
for $\Lambda$ = 70 \TeV~(continuous lines) and $\Lambda$ = 150 \TeV ~(dash-dot).
For $\Delta_{A}$, the upper (lower) lines correspond to $3~\times~10^{-3}
\eV^2$ ($ 0.3~\times~10^{-3} \eV^2 $). For $\Delta_{S}$, the upper(lower) lines
correspond to $12~\times~10^{-6} \eV^2$ ($3~\times~10^{-6} \eV^2$).
  
\vskip 0.3 cm 

\subsection{MSW and atmospheric neutrino problem in MMM}

The  angle $s_1$ can be appropriately chosen to fix the required
mixing for the small angle MSW conversion. The angle $s_3$ which
determines the overall scale of neutrino masses is also required to be
small. In such a case,
the survival probability for the atmospheric $\nu_\m$ assumes two
generation form and one can take the restrictions on relevant parameters
from the standard analysis as reported in chapter 2.
We have determined the effective $\nu_{\mu}-\nu_{\tau}$ mixing and 
neutrino masses following from eq.(\ref {fmix}) by the procedure outlined 
in the last section. We show this mixing in Fig.(3a).
 In Fig.(3b), we show the masses 
for two values of $\Lambda = 70 \TeV, 150 \TeV$.
As seen from Fig.(3a), the $s_2 = 0.3 - 0.75 $ leads to the required $\sin^2
2\theta_{\mu\tau} = 0.8 - 1$. Fig.(3b) displays the contours 
corresponding to $\Delta_S\sim( 3. - 12.)~~10^{-6} \eV^2 $ 
and $\Delta_A= (0.3 - 3.)~~10^{-3} \eV^2$
in the $s_2-s_3$ plane. It is seen that  hierarchy among neutrino masses 
obtained in the required region is stronger than needed for a
simultaneous
solution of the solar and atmospheric neutrino problems and there
is no overlapping region in the $s_2-s_3$ plane for a combined solution.
It is of course 
possible to solve each of this problem separately and get the required
amount of mixing as well. 

\begin{figure}[ht]
\begin{tabular}{cc}
\epsfxsize 6 cm
\epsfysize 8 cm
\epsfbox[72 169 539 636]{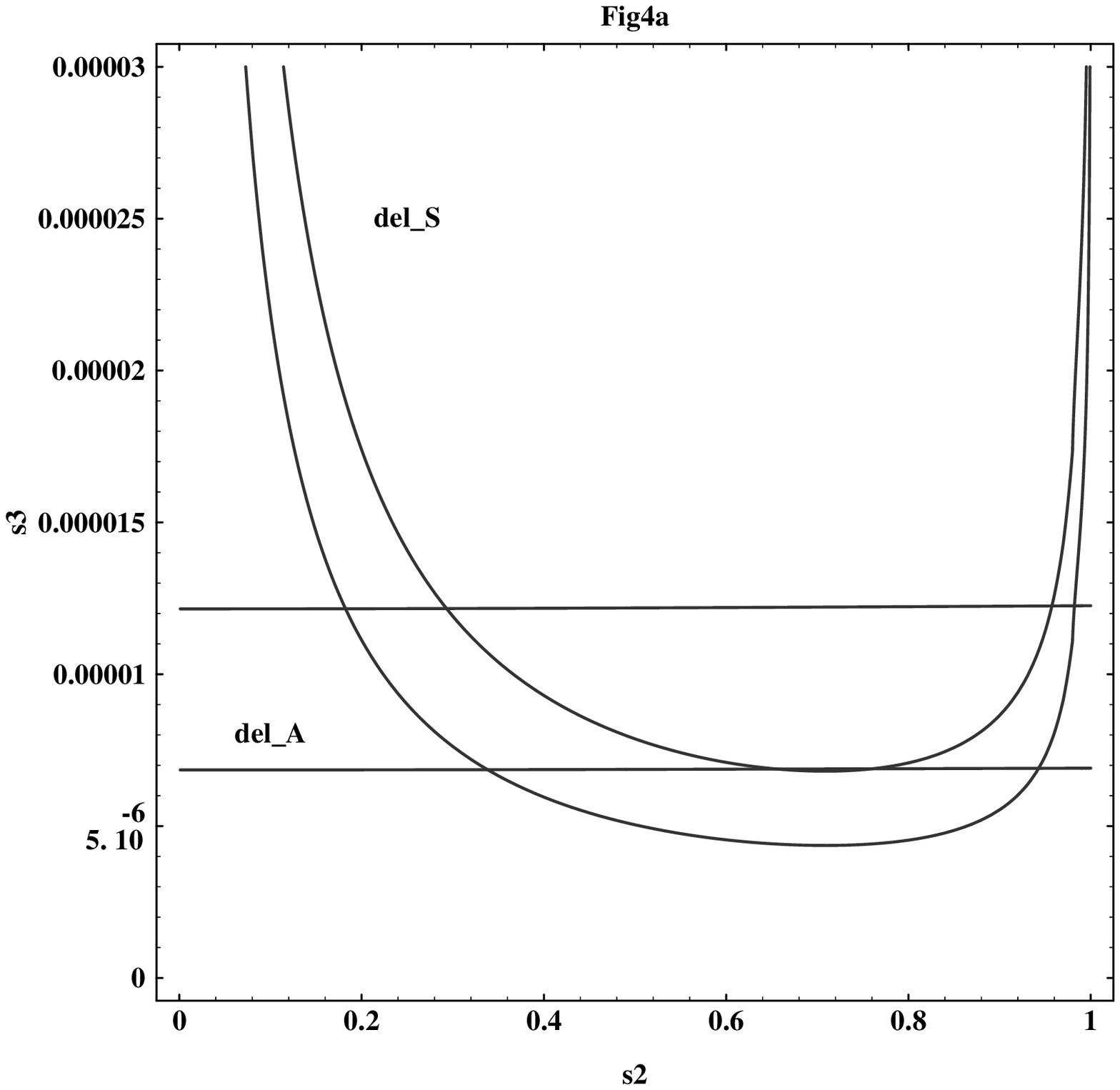}&
\epsfxsize 6 cm
\epsfysize 8 cm
\epsfbox[72 137 539 648]{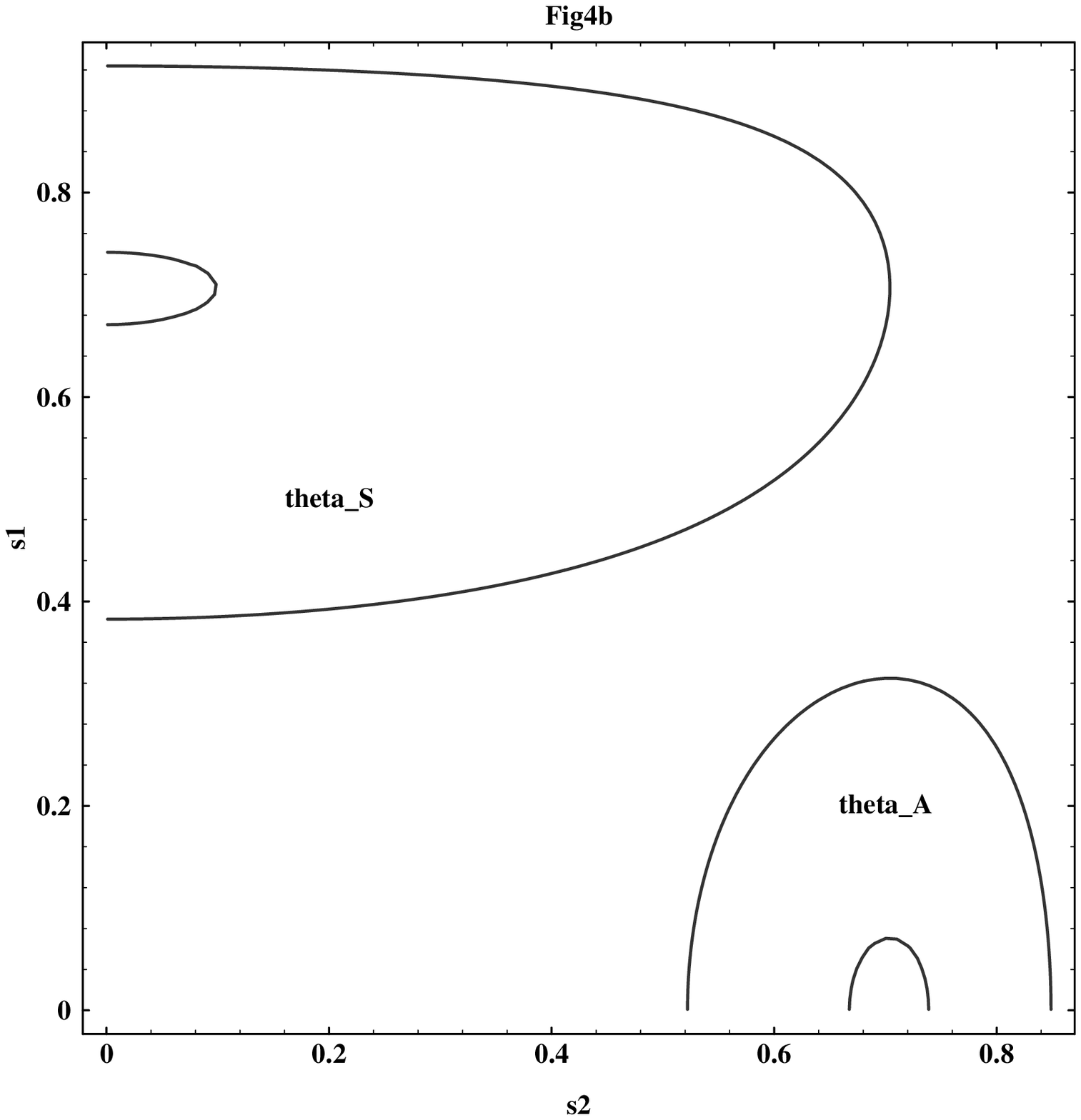}
\end{tabular}
\end{figure}
\noindent
{\bf Fig. 4a}. Contours of $\Delta m^2$ in MMM are plotted 
 for $\Lambda$ = 100 \TeV. For $\Delta_{A}$, the upper (lower) line corresponds
 to $3~\times~10^{-3} \eV^2$ ($ 0.3~\times~10^{-3} \eV^2 $). For $\Delta_{S}$, 
the upper (lower) line corresponds to $3~\times~10^{-10} \eV^2$
 ($0.5~\times~10^{-10} \eV^2$).
{\bf Fig. 4b}. The effective  $Sin ^2  2 \theta_{S}$ and 
$Sin ^2 2 \theta_{A}$ are plotted in the case of the minimal messenger model.
 The inner lines represent contours for 0.9 in both the cases whereas, 
the outer lines correspond to contours for 0.5 (0.8) for $Sin^2 2 \theta_{S}$ 
($Sin^2 2 \theta_{A})$.

\vskip 0.3 cm

\subsection{Vacuum oscillations and atmospheric neutrino problem in MMM}

Unlike in the case of the MSW interpretation, the model can nicely account for
the  hierarchies required for solving the solar and atmospheric neutrino
problems through vacuum oscillations. This is displayed in Fig.(4a) where
we show contours corresponding to $\Delta_S = (0.5 - 3)~~ 10^{-10} \eV^2$ and
$\Delta_A = (0.3 - 3)~~ 10^{-3} \eV^2$ in the
$s_2-s_3$ plane. Unlike in case of the MSW conversion, here there is a
large overlap region in $s_2-s_3$ plane which leads to the
 required values for $\Delta_{S,A}$. 
Despite this one unfortunately cannot explain both the problems
simultaneously in a phenomenologically consistent way. This is due to the
very restricted mixing structure displayed in eqs.(\ref{fmix}). 
As discussed in chapter 2, the vacuum
oscillation probability in the present case is given by,
\be
P_e=1- 4 U_{e1}^2 U_{e2}^2 \sin^2 \left({\Delta_S t \over 4 E}\right)
-2 U_{e3}^2(1-U_{e3}^2)~, \ee
where the last term comes from the averaged oscillations corresponding to
the atmospheric neutrino scale. Likewise, the muon neutrino survival
probability which determines the atmospheric neutrino flux is given by,
\be
P_\m =1- 4 U_{\m 2}^2 U_{\m 3}^2 \sin^2\left({\Delta_A t \over 4 E}\right) .
\ee
The amplitude of oscillations is controlled by two effective
angles:
\be \sin^2 2\theta_S=4 U_{e1}^2 U_{e2}^2~,\;\;\;\;
     \sin^2 2\theta_A=4 U_{\m2}^2 U_{\m 3}^2 .\ee
The matrix U appearing above is given by eq.(\ref{fmix}).
Restrictions on these angles  required for a combined solution of the
solar and atmospheric anomaly are worked out in \cite{osland} for
different values of $U_{e3}$. Independent of the 
chosen values for $U_{e 3}$ one requires,
\be \label {rest}
\sin^2 2\theta_S = 0.5 - 1~,\;\;\;\; \sin^2 2\theta_A = 0.8 - 1~.\ee
It is possible to choose these angles independently and satisfy above
equations in a generic three generation case. In our case, the 
mixings are also determined in terms of $s_{1,2}$ through eq.(\ref{ol}). We have
plotted the contours corresponding to restrictions in eq.(\ref{rest}) 
in Fig. 4b.
{\it It is seen that there is no region in $s_1-s_2$ plane for which the solar 
and vacuum mixing angles can be simultaneously large ruling out the
  possibility of reconciling atmospheric anomaly with vacuum 
solution in the case of the MMM }.

\section{Non-minimal model and neutrino anomalies}

We had restricted our analysis so far to the MMM which is characterized
by eq.(\ref{squ}) and the vanishing of the $B$ and $A$ parameters at $\Lambda$.
Apart from  predictivity, there are no strong theoretical
arguments in favour of this minimal choice. Such a choice would lead to
positive $B_\m$ at the weak scale which in turn assures $\m$ to be positive
as noted above.
One could consider variations of the MMM which in general result in 
introduction of additional low energy parameters. A class of non-minimal
models could contain more complicated messenger sector which would influence
boundary conditions in eq.~(\ref{squ}). Alternatively, one may keep the 
same messenger sector but introduce some direct coupling between messenger
and matter fields. This could result in non-zero $B_\m$ values at $X$. 
In fact, $B_\m$
gets generated \cite{bdine,pomerol} in models which  try to understand origin
of $\mu$ term in gauge mediated scenario \cite{murayama}. $B_\m$ may be 
generated in the absence of messenger-matter coupling if MSSM itself is
extended.

\begin{figure}[h]
\centerline{
\epsfxsize 8 cm
\epsfysize 8 cm
\epsfbox[73 153 540 642]{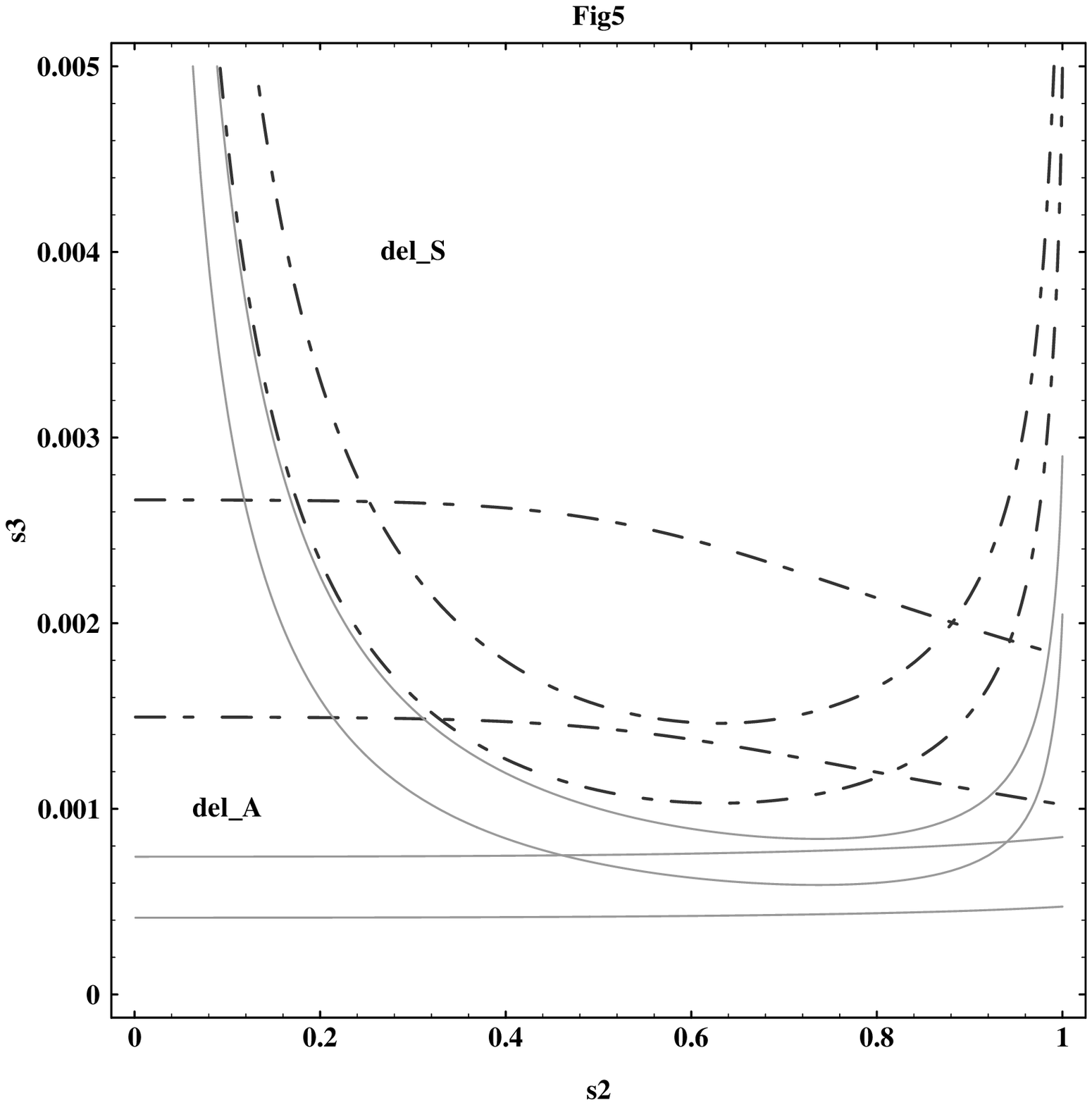}}
\end{figure}
\noindent
{\bf Fig. 5}. Contours of $\Delta m^2$ are plotted in {\bf Non- MMM} case
with $\mu < 0$, for tan $\beta$ = 50 ~(continuous lines) and
 tan $\beta$ = 40 ~ (dash-dot) with $\Lambda = 100 \TeV$. 
 For $\Delta_{A}$, the upper (lower) lines correspond to $3~\times~10^{-3}
 \eV^2$ ($ 0.3~\times~10^{-3} \eV^2 $). For $\Delta_{S}$, the upper 
(lower) lines correspond to $12~\times~10^{-6} \eV^2$ ($3~\times~10^{-6}
 \eV^2$).

We shall not consider any specific model here, but would adopt a
purely phenomenological attitude to point out possible ways which can
allow simultaneous solutions  of the solar and atmospheric neutrino
anomalies. It turns out that prediction of $m_0$ is quite sensitive
to the sign of $\mu$ term which is fixed to be positive in MMM. This 
follows from eq.(\ref{omegas}) which shows that two contributions
 to $\left< \tilde{\nu} \right>$ add or cancel depending on the sign of
$\mu$. We may thus consider a slightly less 
restrictive form of the MMM in which we regard the value and sign of $B_\m$  
as independent parameters to be determined phenomenologically. Typically,
the $B_\m$ value remains positive due to the strong Yukawa sector coming from
the higher loop terms in the RGE for $B_\m$. One should thus generate $B_\m$
 sufficiently
negative at the $\Lambda$ scale, so that, it remains negative even at the
weak scale. The typical value of $B_\m$ at the $\Lambda$ scale should be around,

\be
B_\m (X) \approx - 12~ \tilde{\alpha}_3~ (Q_0)~ M_3 (Q_0)~ Y_t~ t
\approx - 7~ 10^{-4}~ M_3 (Q_0)~ t
\ee

where, the magnitude of RHS denotes the dominant positive 
contribution to $B_\m$ with $Y_t$ denoting the top Yukawa coupling. A 
negative $B_\m$ of this order 
at the $X$ scale can now generate a negative $\m$ 
at the weak scale.  We still assume that the mechanism
responsible for generation of $B$ parameters does not distinguish between 
leptonic and the Higgs doublet $H_1$ and hence 
$B_\m$ and $B_i$ coincide at the scale $X$. Due to this, sneutrino
{\it vevs} are still characterized by the differences in eq.(\ref{deltas1}) 
and hence are suppressed. The boundary conditions on soft
masses are still assumed to be given by eq.(\ref{squ}).
This particular scenario is now
characterized by  parameters $\Lambda$ and $B_\m$. As follows from the
minimum equation, (\ref{mini}), one may regard the value of $\tan\beta$
and sign of $\mu$ as independent parameters instead of $B_\m$.
The magnitude of $\m$ is determined in terms of these parameters 
by eq.(\ref{mini}).  It is now possible to simultaneously account for the
 atmospheric neutrino deficit and have the MSW conversion for the
 solar neutrinos.  This is depicted in Fig.~(5) in which
we show contours (in $s_2-s_3$ plane) corresponding to
$\Delta_S= 3 - 12 ~~ 10^{-6} \eV^2,\Delta_A= 0.3 - 3 ~~10^{-3} \eV^2$ for
negative $\mu$ and two representative
values of $\tan\beta=40,50$. The magnitude of $\mu$ gets fixed 
by eq.~(\ref{mini}) to 379.28 ~\GeV~and 382.03 ~\GeV~in the respective cases.
The corresponding $B_\m$ value at the $X$ scale 
is approximately given as $B_\m (X) \approx -13.10 ~\GeV~ 
(\tan \b = 40); ~~ -7.06 ~\GeV~ (\tan \b = 50).$ It is seen that 
now there is a considerable overlap where two mass scales arise 
simultaneously. As mentioned before, these masses are independent 
of the value of $s_1$ which can be chosen in the required range 
namely, $$s_1 = 0.0225 - 0.071 $$ to allow MSW conversion.
 The angle $\theta_A$
relevant for the atmospheric anomaly coincides roughly with $s_2$ and
as follows from Fig.(5), one can simultaneously account for mixing as well
as masses needed to solve the atmospheric and the solar neutrino
problems. 

\section{Conclusions}

The supersymmetric standard model  contains natural source of lepton
number violation and hence of neutrino masses. The resulting neutrino
  mass pattern is quite constrained if source of lepton number
 violation is provided by soft bilinear operators and 
if the SUSY breaking is introduced through gauge mediated interactions. 
This scenario has the virtue that
one can obtain hierarchical masses and large mixing in the neutrino sector.
The hierarchy in masses results from hierarchy in the two different sources
of neutrino masses while large mixing can be linked to ratio of
$R$ violating parameters $\epsilon_i$. 
Overall scale of neutrino mass is  set by $s_3$ and by
the Yukawa couplings of the $b$ and $\tau$ . Neutrino masses
are thus naturally suppressed and hierarchical. One however needs to
assume relatively suppressed $R$ violation, i.e. $s_3\sim 10^{-3}$
in order to obtain the mass scale relevant for the atmospheric neutrino
anomaly. This requires that $\e_2\sim \e_3\sim 1\GeV$ when $\m\sim 1
\TeV$.

In case of the minimal gauge mediated model, the three neutrino
masses and three mixing angles are 
controlled by five  parameters $\Lambda$, $x$ and $s_{1,2,3}$. This proves to
be quite constraining and does not allow one to obtain simultaneous
solution of the solar and atmospheric neutrino anomalies. In the actual
analysis, we chose $x= {1 \over 2}$. A smaller value of $x$ would not change
this conclusion in view of the mild dependence of the boundary conditions,
eq.(\ref{squ}) on $x$.  However, 
a non-minimal version which allows negative $\mu$ parameter is capable of
accommodating the MSW effect and atmospheric neutrino anomaly. The number
of parameters needed are still less than in the models based 
on the minimal supergravity scenario. 

\subsection{RG equations}

In this subsection we present the two-loop parts of the RGE for
the parameters $B_\m,\Delta B_H,\Delta B_L$ used in the 
calculations in the MMM model.

\bea
\label{twoloopb}
{d B_\m^{2-loop} \over dt}&=& 30 \tilde{\a}_2(t) M_2(t) + {9 \over 5}
\tilde{\a}_2(t) \tilde{\a}_1(t) \le M_1(t) + M_2(t) \ri
+ {207 \over 25} \tilde{\a}_1^2(t) M_1(t) \nonumber \\
&+& 16 \tilde{\a}_3(t) M_3(t) Y_t(t) + 
{12 \over 15} \tilde{\a}_1(t) M_1(t) Y_t(t) + 
16 \tilde{\a}_3(t) M_3(t) Y_b(t) \nonumber \\
&-& {12 \over 30} \tilde{\a}_1(t) M_1(t)
Y_b(t) + {12 \over 10} \tilde{\a}_1(t) M_1(t) Y_\tau(t) - 
16 \tilde{\a}_3(t) A_t(t) Y_t(t) \nonumber \\
&-& {12 \over 15} \tilde{\a}_1(t) A_t(t) Y_t(t) - 
16 \tilde{\a}_3(t) A_b(t) Y_b(t) + {12 \over 30} \tilde{\a}_1(t) Y_b(t) A_b(t)
\nonumber\\ 
&-& {12 \over 10} \tilde{\a}_1(t) A_\tau (t) Y_\tau(t) + 
6 A_\tau(t) Y_\tau^2(t) + 6 \le A_t(t) 
+ A_b(t) \ri Y_t(t) Y_b(t) \nonumber \\
&+& 18 A_t(t) Y_t(t)^2 + 18 Y_b(t)^2 A_b(t) 
\eea 

\bea
\label{twoloopbl}
{d \Delta B_L^{2-loop} \over dt}&=&6 Y_\tau^2(t) A_\tau(t) + 3 Y_\tau(t)Y_b (t)
\le A_b(t) + A_\tau(t) \ri  \nonumber \\
&-& {6 \over 5} \tilde{\a}_1(t) A_\tau(t) Y_\tau(t) 
+ {6 \over 5} \tilde{\a}_1(t) M_1(t) Y_\tau(t)
\eea

\bea
\label{twoloopbh}
{d \Delta B_H^{2-loop} \over dt}&=& 3 Y_\tau(t) Y_b(t) \le A_\tau(t) +
A_b(t) \ri-18 Y_b^2(t) A_b(t) \nonumber \\
&-& 3 Y_b(t) Y_t(t)\le A_t(t)+ A_b(t)\ri
- 6 Y_b(t) A_b(t) \le {8 \over 3 } \tilde{\a}_3(t) - {1 \over 15} 
\tilde{\a}_1(t) \ri \nonumber \\
&-& 6 Y_b(t) \le {8 \over 3} \tilde{\a}_3(t) M_3(t) -
{1 \over 15} \tilde{\a}_1(t) M_1(t) \ri
\eea
\section{Appendix}
We present here the derivation of the tree level mass matrix, 
eq.(\ref{mtree}). In the presence of R-violating couplings neutrinos 
mix with neutralinos. In the bilinear R-parity violating scenarios,
this mixing takes place with $\e_i$ couplings and the sneutrino
{\it vevs}. In the Weyl basis, the Lagrangian describing the 
neutrino-neutralino mass matrix is given by,
\be
\lm_{mass} = - {1 \over 2} \Si_0^T {\cal M}_0 \Si_0 + H. c
\ee
where in the two component notation, $\Si_0$ is a column vector of
neutrinos and neutralinos,
\be
\Si^T_0 = \le \n_e, \n_\m, \n_\tau, -i \l_1, -i \l_3, \si_{H_1}^0, 
\si_{H_2}^0
\ri 
\ee
The mass matrix has the following  general structure which is of
see-saw type:
\be
{\cal M}_0 = \le \ba{cc} 0 & m \\ m^T & M_4 \ea \ri 
\ee
Here the sub-matrix $m$ is of dimension $ 3 \times 4$ and has the
following structure:
\be
m = \le \ba{cccc} -{g_1 \over \sqrt{2}} \omega_1 & 
{g_2 \over \sqrt{2}} \omega_1 & 0& \e_1 \\
 -{g_1 \over \sqrt{2}} \omega_2 & {g_2 \over \sqrt{2}} \omega_2 & 0& \e_2 \\
 -{g_1 \over \sqrt{2}} \omega_3 & {g_2 \over \sqrt{2}} \omega_3 & 0& \e_3 \ea \ri
\ee
with the $\omega_i$ representing the sneutrino {\it vevs}.
$M_4$ is the standard $4 \times 4$ neutralino mass matrix of the MSSM
which has the following form:
\be
M_4 = \le \ba{cccc} M_1 &0& -{g_1 \over \sqrt{2}} v_1 & {g_1 \over \sqrt{2}} v_2 \\
0&M_2& -{g_2 \over \sqrt{2}} v_1 & {g_2 \over \sqrt{2}} v_2 \\
 -{g_1 \over \sqrt{2}} v_1 & {g_1 \over \sqrt{2}} v_1&0& - \m \\
 -{g_2 \over \sqrt{2}} v_2 & {g_2 \over \sqrt{2}} v_2& -\m & 0 \ea \ri 
\ee 
The effective $3 \times 3$ neutrino mass matrix is obtained by block diagonalising
the above matrix. It has the form :
\bea
m_{eff} &=& - m M_4^{-1} m^T \nonumber \\
&=& {\mu (M_1 g^2+ M_2 g'^2) \over D}  \le \ba{ccc} 
\Lambda_1^2& \Lambda_1 \Lambda_2 & \Lambda_1 \Lambda_3 \\
\Lambda_1 \Lambda_2 & \Lambda_2^2 & \Lambda_2 \Lambda_3 \\
\Lambda_1 \Lambda_3 & \Lambda_2 \Lambda_3 & \Lambda_3^2 \\
\ea \ri
\eea
The vector $\vec{\Lambda}$ is defined as,
\be
\vec{\Lambda} = \m \vec{\omega} - v_1 \vec{\e}.
\ee
$D$ is given by,
\be
D = 2(-\m~ M_1~ M_2  +2~ M_W^2 c_\b  s_\b (M_1 + M_2~ tan^2 \theta_W)).
\ee
Here, the matrix $m_{eff}$ is written in the basis where $\e_i$ are not rotated
away from the superpotential. The rotated form of this matrix is given in the
text and takes the form $M_0$, eq.(\ref{mtree}), in the charged lepton 
mass basis. From $m_{eff}$ we can easily see that it has only one eigenvalue,
even in the presence of the first generation sneutrino {\it vev}.  This is
a generic result of all the R-parity violating models \cite{asmarek,marekpil}.

%% file: chap5.tex
\chapter{Trilinear R violation and Neutrino Masses}

\section{Introduction}
In the previous chapter we have been concerned with neutrino masses
in the presence of only bilinear R-parity violating couplings. As we
have seen bilinear R violation provides a very constrained framework
to accommodate simultaneous solutions for both solar and atmospheric
neutrino problems. In this chapter we pursue an alternative framework
where we consider trilinear R parity violation in the superpotential. 
We will study the structure of neutrino mass matrix in this framework
and the existence of simultaneous solutions for the solar and 
atmospheric neutrino problems within these models. 

One can immediately see that these models definitely allow more freedom
compared to the bilinear case as the $\l'$ and $\l$ are large in number.
The $\l'$ couplings are 27 in number whereas the $\l$ couplings are
9 in number. Thus due to the large freedom they offer we would
expect that simultaneous solutions for solar and atmospheric neutrino 
problems could be accommodated in these models for some definite 
choice of the parameters.  But, in addition to bringing in larger freedom
these models also bring in large amount of arbitrariness with them 
which would make calculations cumbersome. To overcome this arbitrariness
and make these models predictive,
the following choices are generally made in literature:\\
1). Assume $\l'$, $\l$ couplings to be hierarchical. This choice is
generally made by motivating that $\l'$ couplings if present in the
superpotential are likely to have the same pattern as the standard
Yukawa couplings which give masses to the fermions \cite{chun1}.\\
2). Assume only a sub-set of the $\l',\l$  couplings is non-zero. 
Fix the values of these couplings numerically,  requiring that
the neutrino mass matrix determined by this sub-set of couplings
would give rise to the required pattern. This analysis can then
be extended to all possible sub-sets \cite{arclosada}.

\noindent
In this thesis we do not make any of the above choices. We instead 
make the assumption that all the trilinear couplings are present 
in the superpotential are free parameters, but they are typically of 
the same magnitude. This is a valid assumption as the R-parity violating
couplings are required to be suppressed compared to the regular 
Yukawa couplings to give correct order of neutrino masses 
to understand neutrino anomalies. This assumption has been earlier 
followed by Drees et.al \cite{mdrees} in the same context. In
their work, neutrinos attain masses at the 1-loop level due to 
the presence of trilinear interactions. But, as we have seen in 
chapter 3, in any realistic model of supersymmetry breaking, trilinear
interactions would also modify the weak scale soft potential through
RG scaling. This would give additional `tree level' mass to the 
neutrinos. In this chapter \cite{mine}, we reanalyse the neutrino mass spectrum
in these models taking in to consideration the `tree level' contribution
to the neutrino masses also. This would significantly modify the neutrino
mass spectrum as we will see below.

\section{Sneutrino vevs and Neutrino Masses }

For definiteness,  we would consider only trilinear $\l'$ couplings 
to be present in the superpotential. Such an assumption is made
for simplicity and we would comment on the inclusion of $\l$ 
couplings later on. The superpotential in this case is given as,

\be \label{wl}
W =  h^u_{ij} Q_i U_j^c H_2  + h^d_{ii} Q_i d_i^c H_1 + h^e_{ii} L_i e_i^c H_1
+ \m H_1 H_2  + \l'_{ijk} L_i Q_j D_k^c,
\ee

\noindent
where we have used the standard notation specified in chapter 1.  As we
have mentioned earlier, the presence of non-zero $\l'$ couplings would
induce two separate contributions to the neutrino mass. One contribution 
arises due to a soft term linear in the sneutrino {\it vev}. 
This soft term gets generated through the loops but leads to a sneutrino
{\it vev} in the `tree level' of the renormalisation group improved
low energy effective potential. This contribution indirectly generates
neutrino masses through their mixing with the gauginos. A majorana 
mass term for the light neutrinos is also generated directly by the
loop diagrams involving squarks and sleptons \cite{hs,tata}. We shall
refer to these two contributions as RG induced tree level (or simply
tree level) and loop level masses respectively. 

\vskip 0.3 cm
\noindent
{\it The framework }
\vskip 0.3 cm
In the present chapter, we choose to  work in the mSUGRA inspired
MSSM. In this case the soft terms are added at the high scale
$\sim M_{GUT}$. In the limit all $\l'$ couplings are chosen to be
similar in magnitude, the analysis of the neutrino mass matrix would
become simpler. It is now possible to study the neutrino mass matrix
analytically. The two contributions to the neutrino mass matrix 
are given as below. 

\subsection{RG induced Tree Level Mass:}

In the mSUGRA inspired MSSM framework, the structure of the superpotential 
dictates the structure of the soft SUSY breaking terms. Thus, with 
only trilinear L-violating interactions, the
soft terms do not contain bilinear terms at a high scale. But as
we have seen in chapter 3, they are nevertheless generated at the
 weak scale  and should be retained in the scalar potential at this scale. 
The relevant part of the soft scalar potential is now given as:
\bea \label{softpot}
V_{soft}&=& m_{L_i}^2 \mid \tilde{\n}_i \mid ^2 + m_{H_1}^2 
\mid H_1^0 \mid^2 + m_{H_2}^2 \mid H_2^0 \mid^2  + \left[ m_{\n_i H_1}^2 
\tilde{\n}^{\star}_i H^0_1 \right. \nonumber \\
& & \left. - \m\;B_\m H_1^0 H_2^0 - B_{\e_i} \tilde{\n}_{i} H_2^0  + h.c
\right]+ .... \;\;\; ,
\eea

where, we have retained only neutral fields and used standard notation
with $B_{\e_i}$ and $m_{\n_i H_1}^2$  representing the bilinear
lepton number violating soft terms.  The weak scale value of the
lepton number violating soft parameters is determined by the RGE given
in chapter 3. We reproduce them here:

\bea \label{rge}
{d B_{\e_i} \over dt} & = & B_{\e_i} \left (- {1 \over 2} Y^{\tau} - {3 \over 2}
Y^t + {3 \over 2} \tilde{\a_2} + {3 \over 10} \tilde{\a_1} \right ) -
{3 \over 16 \pi^2}~~ \m~~ h^D_{k}~~ \l'_{ikk}
 \left( {1 \over 2} B_\m + A^{\l'}_{ikk} \right ), \nonumber\\
{d m^2_{\n_i H_1} \over dt} & = & m^2_{\n_i H_1} \left( - 2 Y^{\tau}
- {3 \over 2} Y^b \right) - \left(3 \over 32 \pi^2 \right)~~ h^D_{k} ~~
\l'_{ikk} \left( m_{H_1}^2 + m_{L_i}^2 \right.\\ \nonumber
&~~& \left. +~~  2 ~~{m^2}_{kk}^{Q} + 2 ~~A^{\l'}_{ikk}
A^D_{kk} + 2~~ {m^2}^{D^c}_{kk} \right),
\eea
and the standard MSSM RGE for the parameters on the RHS 
\footnote {The standard RGE for the soft parameters appearing on the
RHS of the above equations do get modified in the presence of the
$\l'$ couplings. But as mentioned in chapter 3, 
in the limit of very small $\l'$ couplings, as
will be required by our model, the presence of these couplings would not
modify the running of the soft parameters appreciably.}.
In the above, we have confined ourselves with the same notation as
in chapter 3. Since we allow only  trilinear interactions in 
$W_{\not L}$, $m_{\n_i H_1}^2 = B_{\e_i} = 0 $ at high scale.
In the presence of non-zero $\l'_{ijk}$ and with the above boundary
conditions, the parameters $m_{\n_i H_1}^2$ and $B_{\e_i}$ have 
low energy solutions of the form given in chapter 3. Using the 
solutions for $\l'_{ijj}$ and $h^D_{jj}$,~ it is convenient to 
parameterise these solutions as \footnote{The assumption that all $\l'$ 
couplings are similar in magnitude is made at the weak scale. Such
an assumption at the high scale would in general lead to hierarchical
$\l'$ at the weak scale.},
\bea \label{sol}
B_{\e_i}&=& \l'_{ipp} h^D_{p} \k_{ip}, \nonumber \\
m_{\n_i H_1}^2&=&\l'_{ipp} h^D_{p} \k'_{ip}\;\;\;.
\eea
Here $p$ is summed over generations. The parameters $\k$ and $\k'$
represent the running of  the parameters present in the RGE's
 from the high scale to the weak scale.

The above soft potential would now give rise to sneutrino {\it vevs},
which are described by the stationary value conditions for the above
soft potential. These are given as,
\be
\label{omega}
< \tilde{\n_i} >~~   = {B_{\e_i} v_2 - m^2_{\n_i H_1} v_1 \over m_{L_i}^2 
+ {1 \over 2} m_Z^2~ cos 2 \beta }\;\;\;,
\ee
where $v_1$ and $v_2$ stand for the {\it vevs} of the Higgs fields $H_1^0$ and
$H_2^0$ respectively.  
The sneutrino {\it vevs} so generated will now mix the neutrinos with the 
neutralinos thus giving rise to a tree level neutrino mass matrix 
as has been described in the previous chapter. This matrix has the 
structure:
\bea
\label{treecon}
{\cal M}^0_{ij}&=&{\m (cg^2 + g'^2) ~ < \tilde{\n_i} >~ < \tilde{\n_j} > 
\over 2 ( -c \m M_2 + 2 M_w^2 c_\b s_\b (c + tan \theta_w^2 ))} \;\; ,
\eea

where  $c = { 5 g'^2 \over 3 g^2 } $ and $g (g') $ denotes $ SU(2)(U(1)) $
gauge coupling.

\subsection{Loop Level Mass }
As we have seen earlier, the trilinear couplings in the 
superpotential would also give rise to
a loop induced neutrino mass with the down squark and anti-squark pairs
being exchanged in the loops along with their ordinary partners 
\cite{hs,babu}.
This mass can be written as,
\be
\label{mloop}
{\cal M}^{l}_{ij} = {3 \over 16 \pi^2 } \l'_{ipk} \l'_{jkp} ~ v_1~ h^D_{k} ~sin 
\phi_p\; cos \phi_p~ ln {M_{2p}^2 \over M_{1p}^2}\;\;\;. 
\ee
In the above, $sin\phi_p\; cos\phi_p$ determines the mixing of the
 squark-antisquark pairs and $M_{1p}^2$ and $M_{2p}^2$ represent
the eigenvalues of the standard 2$\times$ 2 mass matrix of the 
down squark system \cite{nilles}. The indices $p$ and $k$ are
 summed over. 
The mixing $sin\phi_p\; cos\phi_p$ is proportional to $h^D_p$ and thus 
one can write the loop mass as,
\be 
\label{mloop1}
{\cal M}^l_{ij} = \l'_{ipk}~ \l'_{jkp}~ h^D_{k}~ h^D_{p}~ m_{loop}\;\; .
\ee
\be
\label{expmloop}
\mbox{Explicitly,}~~m_{loop} \equiv {3 ~v_1 \over 16 \pi^2}~~ {sin \phi_{p}~ cos \phi_{p} 
\over h^D_p} ~~ln {M_{2p}^2 \over M_{1p}^2} \sim {3 ~v_1^2 \over 16 \pi^2}~~
 {1 \over M_{SUSY}},
\ee
\noindent
with $M_{SUSY} \sim 1$ TeV referring to the typical scale of 
SUSY breaking. Note that
$m_{loop}$ defined above is independent of the R violating couplings and is 
solely determined by the  parameters of the minimal supersymmetric
standard model (MSSM).

\section{Neutrino Masses and Mixing}
We now make a simplifying approximation which allows us to discuss
neutrino masses and mixing analytically. It is seen
from the RG eqs.(\ref{rge}) that the parameters $\k_{ik}$,
$\k'_{ik}$, defined in
eq. (\ref{sol}) are independent of generation structure in the limit in
which generation dependence of the scalar masses $m_{L_i}^2$, $m_{Q_i}^2$
and soft parameters $A_{ikk}^{\l'}$ and $A_{ikk}^{D}$
is neglected. Since we are assuming the universal boundary conditions, this
is true in the leading order in which  the $Q^2$ dependence of the
parameters multiplying $\lambda'_{ikk} h_k^D$ in eq.(\ref{rge}) is
neglected. $Q^2$ dependence in these parameters generated through the
gauge couplings will also be flavour blind though Yukawa couplings
will lead to some  generation
dependent corrections. But their impact on the
the conclusions based on the analytic approximation below is not
expected to be significant \footnote{A rough estimate gives 
$\sim 20\%$ variation in the parameter space due to this approximation.}.
The neglect of the generation 
dependence of $\kappa_{ik},\kappa'_{ik}$ allows us to rewrite 
eq.(\ref{treecon}) as,
\be
\label{mnotp}
{\cal M}^0_{ij} \equiv m_0 a_i a_j \;\; ,
\ee
where~ $a_i \equiv \l'_{ikk}~ h^D_{k}$ and
$m_0$ is now completely determined by the standard MSSM parameters and the
dependence of the R-violating parameters gets factored out
 as in eq.(\ref{mloop1}). $m_0$ can be determined by solving the relevant RGE. 
 Roughly, $m_0$ is given by,
\be 
\label{expmnot}
m_0 \sim \left( {3 \over 4 \pi^2} \right)^2 {v^2 \over M_{SUSY}} 
\left( ln {M_{GUT}^2 \over M_Z^2} \right)^2\;\; , \ee
where $M_{GUT}=3\cdot 10^{16} \GeV$.

\noindent
Let us rewrite the loop induced mass matrix as,
\bea
{\cal M}^l_{ij}&=&m_{loop}~ \l'_{ilk}~ \l'_{ikl}~ h_{k}^D~ 
h_{l}^D \nonumber \\
&=& m_{loop}~ a_i~ a_j + m_{loop}~ h_{2}^D~ h_{3}^D ~ A_{ij} + 
O( h_{1}^D, h_{2}^{D^2} )\;\; , 
\eea
where  
$ A_{ij} = \l'_{i23} \l'_{j32} + \l'_{i32} \l'_{j23} - \l'_{i22} \l'_{j33} - 
 \l'_{i33} \l'_{j22}.\;\;$ 
Neglecting O($h_{1}^D$,$h_{2}^{D\;2}$) corrections to the loop 
induced mass matrix, the total mass matrix is given by,
\bea 
\label{mtotal2}
{\cal M}_{ij}^\n&\approx&(m_0 + m_{loop} )~ a_i a_j + m_{loop}~ h_{2}^D~
h_{3}^D~
 A_{ij} \nonumber \\
&\approx& {\cal M}'_{ij} + m_{loop}~ h_{2}^D~ h_{3}^D~ A_{ij} \;\; .
\eea
The matrix ${\cal M}'$ can be easily diagonalised using
 a unitary transformation,
\be
U^T {\cal M'} U = diag~ (0,0,m_3)\;\;,
\ee
\be
\label{mthree}
\mbox{where}~~m_3 \sim (m_0 + m_{loop})~~(~\l^{'2}_{333} + \l^{'2}_{233} +
 \l^{'2}_{133} )~~ {h_{3}^D}^2 , 
\ee
\be
\mbox{and}~~U = \left(
\ba{ccc}
c_2&s_2 c_3&s_2 s_3\\
-s_2&c_2 c_3&c_2 s_3\\
0&-s_3&c_3\\ \ea \right)\;\;,
\ee
with
$s_2={a_1 \over \sqrt{a_1^2 + a_2^2}}~\;\mbox{and}\; s_3={(a_1^2 + a_2^2)
^{1 \over 2} \over \sqrt{a_1^2 + a_2^2 + a_3^2}}\;\; $. 

\noindent
The total mass matrix is now given by,
\be
U^T {\cal M}^{\n} U \approx m_3 ~\left( ~diag~(0,0,1) +
\e A' \right) 
\ee
where
\be
\ba{cc}
A' = U^T A~ U \;\mbox{and}&\;\;
\e~ A'_{ij} \approx {m_{loop} \over m_3 }~h_2^D~ h_3^D~ A'_{ij}
\approx {m_{loop} \over m_0 }~{h_2^D \over h_3^D}\;\;. 
\ea
\ee
The last equality follows under the assumption that $\l'_{ijk}$ are
similar in magnitude and $m_{loop} \ll m_0$.

\noindent
If $U'$ is a rotation in the 12 plane with an angle $\theta_1$ 
defined by $\tan 2 \th_1 = { 2 A'_{12} \over A'_{22} - A'_{11}}\;\;$,
then,
\bea
U'^T U^T {\cal M}^{\n} U U'&= &m_3 \left(  \ba{ccc}
\e \d_1&0&c_1 \e A'_{13} - s_1 \e A'_{23}\\
0& \e \d_2&s_1 \e A'_{13} - c_1 \e A'_{23}\\
c_1 \e A'_{13} - s_1 \e A'_{23}&s_1 \e A'_{13} - c_1 \e A'_{23}&1+\e A'_{33}
 \ea
\right)\;\;' 
\nonumber \\
&\approx& m_3~diag( \e \d_1, \e \d_2, 1)\;\;.
\eea
where $\d_1$ and $\d_2$ are parameters generically of O($\l^{'2}$) if all
$\l'_{ijk}$ are assumed to be similar in magnitude. 
The off-diagonal elements will generate additional mixing in the model. But,
as $\e A' \ll 1$, we can neglect these off-diagonal elements. The eigenvalues
in this approximation are given as,
\be \label{evalues}
m_{\n_1}\sim \e~ m_3~\d_1 \;\;\; ; \;\;\; 
m_{\n_2}\sim \e~ m_3~\d_2 \;\;\; ; \;\;\; 
m_{\n_3}\sim m_3\;\; ,
\ee
As a consequence, 
\be \label{ratio}
{m_{\nu_2}\over m_{\nu_3}}\sim {m_s\over m_b}{m_{loop}\over
m_0}\left({\delta_2\over \Sigma_i \lambda_{i33}^{'2 }}\right)
\ee
The last factor in the above is of O(1) and the remaining part is controlled
completely by the standard parameters of the MSSM.
Eq.(\ref{ratio}) may be regarded as a generic prediction of the model.
It is seen from eqs. (\ref{expmloop},\ref{expmnot}) that typically,
\be
{m_{loop} \over m_0} \sim {\pi^2 \over 3 \left( ln~ 
{M_{GUT}^2 \over M_Z^2} \right)^2 } \sim  10^{-3}
\ee
Thus the neutrino mass ratio in eq.(\ref{ratio}) is suppressed considerably
compared to the hierarchy obtained when sneutrino {\it vev} contribution is 
completely neglected. As we have mentioned earlier,  
the authors of \cite{mdrees} have not considered the tree level 
contribution to the neutrinos and thus have achieved a hierarchy 
of neutrino masses proportional to the Yukawa couplings:
\be
{m_{\n_2} \over m_{\n_3}} = {m_s \over m_b},
\ee
where $m_s(m_b)$ represents the strange(bottom) quark mass. But, as we
have seen in the above, the RG induced tree level mass alters the hierarchy
drastically as the tree level mass is almost thousand times larger than
the loop mass for most of the parameter space. 
The exact value of the additional suppression factor due to the
tree level mass is dependent on MSSM parameters and we will calculate 
it in the next section.

The mixing among neutrinos is governed by $K=U~U'$.
The angles appearing in $K$ are determined by 
the ratios of the trilinear couplings and 
hence can be naturally large. Thus, as in supersymmetric model with
purely bilinear $R$ violation which we have studied in the
previous chapter, one gets hierarchical masses
and large mixing without fine tuning in any parameters.\\

\section{Neutrino Masses : Phenomenology:}

We now discuss the phenomenological implications of neutrino masses,
eq.(\ref{evalues}) and mixing. Due to hierarchy in 
masses, one could simultaneously solve the solar and atmospheric $\n$
problems provided, $m_{\n_1} \sim m_{\n_2} \sim 10^{-5}$ eV and 
$m_{\n_3} \sim 10^{-1} - 10^{-2}$ eV. 

In order to determine these masses exactly, we have numerically integrated
the RGE eq.(\ref{rge}) along with similar equations
 for the parameters appearing in them.
We have imposed the standard universal boundary condition and required 
radiative breaking of the $SU(2) \times U(1)$ symmetry. Solution of the RGE
determines both $m_{loop}$ (eq.(\ref{mloop1})) and
 $m_0$ (eq.(\ref{mnotp})). We display these in Fig.1 as a function of
$\m$ for positive (negative) $\m$ and
$\tan \b = 2.1 $,  $M_2 = $400 $\GeV$ ( $M_2 = $200 $\GeV$ ).
 The ratio ${m_{loop} \over m_0}$ is
quite sensitive to the sign of $\m$. For $\m >0$, this ratio is found to
be rather small, typically  
$\sim 10^{-2} - 10^{-3}$, while it can be much larger for
 $\m <0$. There exists a region with negative  $\m$ in which ${m_{loop} \over
m_0}\geq 1$. In this region, two contributions to the sneutrino {\it vev} in eq.
(\ref{omega}) cancel and $m_0$ gets suppressed. Barring this region, the
${m_{loop} \over m_0}$ is seen to be around $\sim 10^{-1}-10^{-2}$
 for negative $\mu$ leading to
\be
{m_{\n_2} \over m_{\n_3}} \sim {m_s\over m_b} {m_{loop}\over m_0}\sim 
2~(10^{-3}-10^{-4})\ee
For $m_{\n_3} \sim 10^{-1} - 10^{-2} \eV$, one thus
obtains $m_{\n_2} \sim m_{\n_1} \sim 2~(10^{-4} - 10^{-6}) \eV$ which is
in the right range required  to solve the solar neutrino problem through
vacuum oscillations.
The typical value of $m_0 \sim \GeV$ found in Fig.1 implies through
eq.(\ref{mthree}), $\l' \sim 10^{-4}$.
Thus, one needs to choose all $\l'_{ijk}$ of this order. Once this is done,
one automatically obtains solar neutrino scale for some range in the MSSM
parameters.

While, hierarchy needed for the vacuum solution follows more naturally,
one could also obtain scales relevant to the MSW conversion. This happens
for very specific region of parameters with negative $\mu$ in which two
contributions to sneutrino {\it vev}, eq. (\ref{omega}), cancel. As already
mentioned, ${m_{loop} \over m_0}$ can be in large this region. 
One  then recovers the result of \cite{mdrees}, namely,
 eq.(\ref{ratio}) which allows
MSW solution for the solar neutrino problem.

\begin{figure}[h]
\begin{tabular}{cc}
\epsfxsize 6 cm
\epsfysize 8 cm
\epsfbox[74 228 536 566]{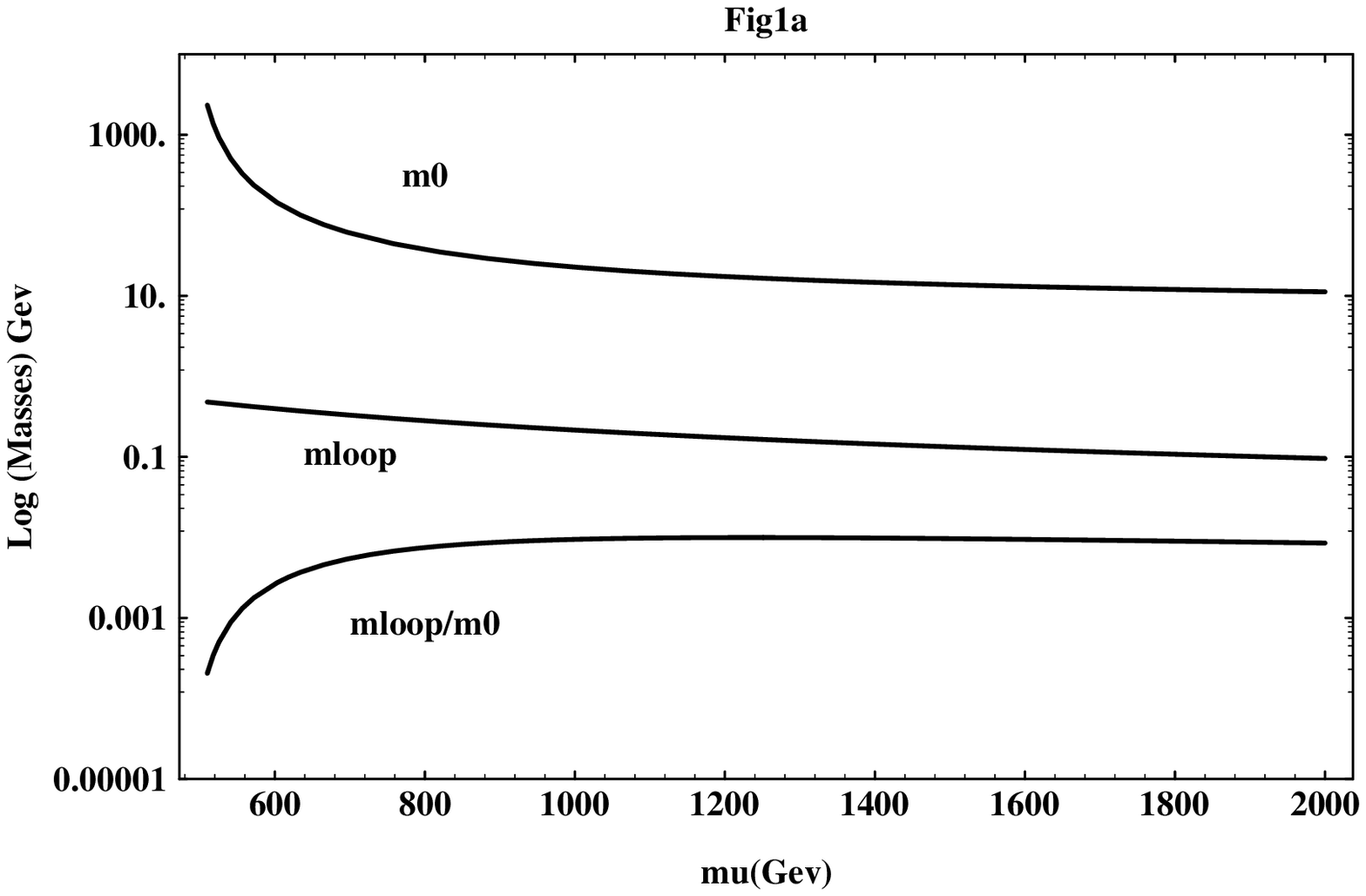}&
\epsfxsize 6 cm
\epsfysize 8 cm
\epsfbox[73 227 536 574]{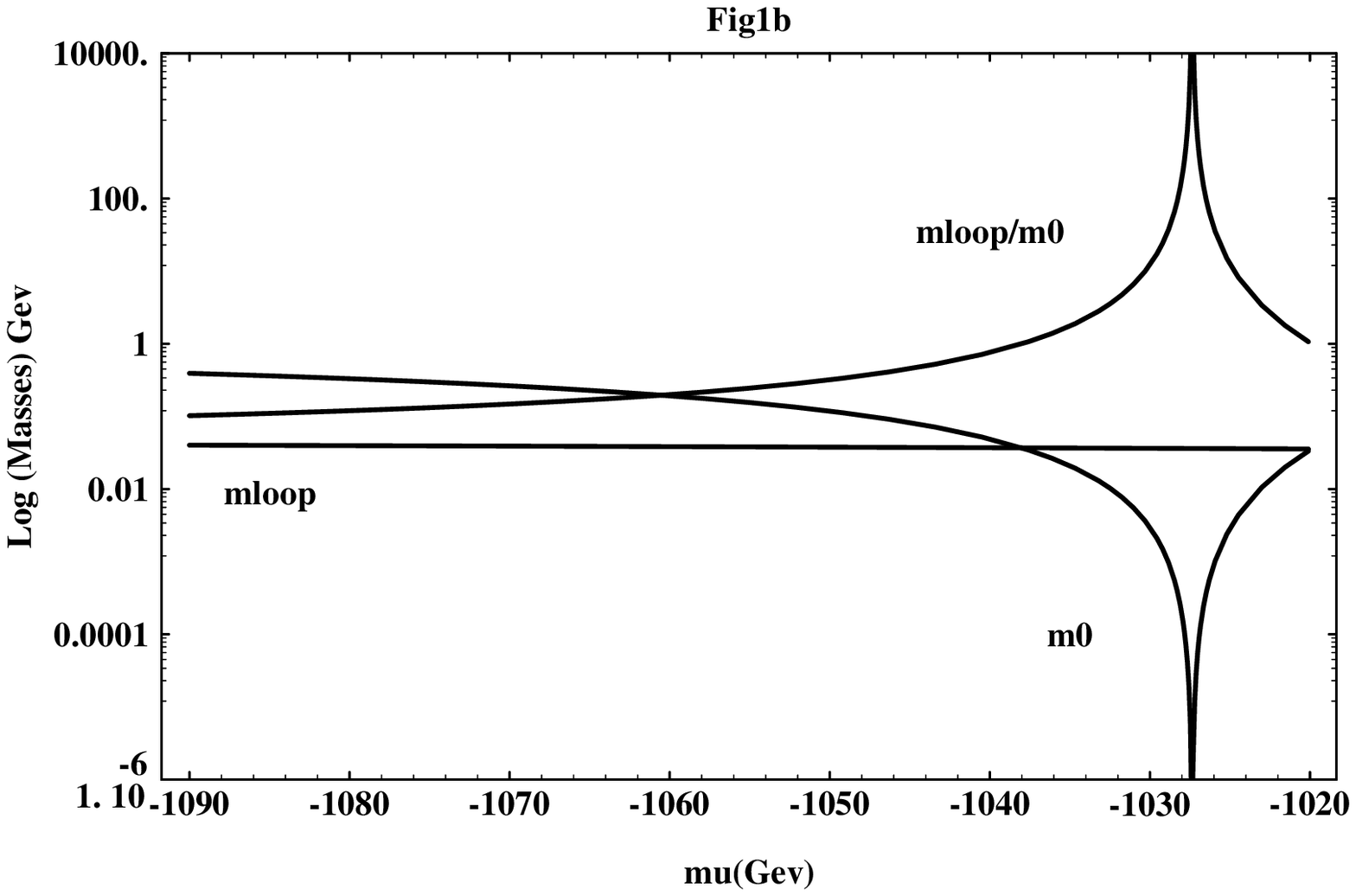}
\end{tabular}
\end{figure}

{\bf Fig. 1a}.The absolute values of tree level contribution,
$m_0$, the loop level contribution, $m_{loop}$  and  their ratio ${m_{loop}
\over m_0}$ are plotted with respect to $\m$ (positive) for
$M_2 = 200\;GeV$, $A$=0 and tan$\beta$ = 2.1 . The $m_0$ and
$m_{loop}$ are defined in the text.\\
{\bf Figure 1b}.Same as in Fig. (1a) but  for $M_2 = 400 \;GeV $
and $\m$ (negative).

\vskip 0.3 cm 

We showed in Figures {\bf 1a, 1b}  neutrino mass ratio for specific values of
 $M_2$ and tan$\b$.  Qualitatively similar results follow for other
 values of these parameters.  We have displayed in Table.1 values for 
the MSSM parameters 
and what they imply for ${m_{\n_2} \over m_{\n_3}}$. We have shown
illustrative values of the parameters which lead to the vacuum as well as
MSW solution. The latter arise only for limited parameter range  
corresponding to
cancellations in eq.(\ref{omega}). The former is a more generic
possibility
which arise for larger region with both positive and negative values of
$\mu$.
The MSW solution in the present context will have to be restricted
to the large angle solution if one does not want to impose any discrete
symmetries or fine tune $\l'$s. It should be noted that this solution is
preferred by the present data from super-Kamiokande experiments as we 
have seen chapter 2.

\textwidth 16.0truecm
\begin{center}
\begin{tabular}{|c|c|c|c|c|c|}
\hline
~~~~ m~~~~  &$~~~~ M_2~~~~  $&~~~~ $\m$~~~~  &~~~~ $m_0$~~~~ 
& ~~~~$m_{loop}~~~~$ 
&~~~~$ratio$~~~~ \\
 ~~~~GeV~~~~ & ~~~~GeV~~~~ & ~~~~GeV~~~~ & ~~~~GeV~~~~ 
& ~~~~GeV~~~~ & ~~~~${m_{\n_2}
\over
m_{\n_3}}$~~~~ \\
\hline
 1312 &200&1225&-16.94&-0.1689&$1.8~10^{-4}$\\
 3898 &300&-3400&-3.238&0.0394&$2.3~10^{-4}$\\
 3921 &350&-3450&-2.655&0.0376&$2.6~10^{-4}$\\
\hline
 157.2 &300&-777.5&-0.0470&0.0487&$-1.9~~10^{-2}$\\
 225.7 &400&-1038&-0.0368&0.0369&$-1.8~ 10^{-2}$\\
\hline
\end{tabular}
\end{center}
\vskip 0.5truecm
{\bf Table1:} {\sl The values of $m_0$,  
$m_{loop}$ and ratio of the eigenvalues, ${m_{\n_2} \over m_{\n_3}}$
  for various values of the standard MSSM parameters m, $M_2$ 
and $\m$ for tan $\beta = 2.1$ , $A$=0. }

\vskip 0.4cm

The constraints on mixing matrix K, implied by the
experimental results are also easy to satisfy keeping all the $\l'_{ijk}$
similar in magnitude. As we have seen in chapter 2, the solar and 
atmospheric neutrino problems can be approximately treated in terms 
of mixing among two generations
(12) and (23) respectively as long as $K_{e3}$ is small (as implied by
CHOOZ) and $m_{\n_3}$ and $m_{\n_2}$ are hierarchical. The constraint 
on relevant mixing following from the experiments can then be 
translated to constraints on elements of K and are given as follows:
\bea
0.6 ~\leq~ 4~ K_{\m 3}^2~ (1 - K_{\m 3}^2) & = & s_3^4~ \sin^2 2 \theta_2 
~+ ~ c_2^2~ \sin^2 2 \theta_3~ \leq ~1. \nonumber \\
K_{e3} &~\leq~& 0.18 \\ \nonumber
0.8 ~\leq~ 4~ K_{e1}^2~ K_{e2}^2 &=& 4 (c_1 c_2 - s_1 s_2 c_3)^2~~
 (s_1 c_2 + s_2 c_1 c_3 )^2 ~\leq~ 1.
\eea
It is possible to satisfy all these constraints by choosing for example,
$$c_3 = s_3 = s_1 = c_1 = {1 \over \sqrt 2 }~;~~ s_2 = 0.254 $$
The relative smallness of $s_2$ required here does not imply significant
fine tuning and can be easily obtained, e.g. by choosing,
$${\l'_{133} \over \l'_{233}} \sim {1 \over 4}.$$ 
It is to be emphasized that much smaller value of $K_{e3}$ than at the present
level will force fine tuning and cannot be accommodated in the model naturally.

We have so far concentrated on the $\l'_{ijk}$ couplings alone. The analogous
discussion can be carried out for $\l_{ijk}$ couplings appearing in the
eq.(\ref{wl}). Here also, the tree level contribution to neutrino masses will
dominate over the loop contribution although the structure of mixing matrix
will differ slightly due to the anti-symmetry of the couplings $\l_{ijk}$ 
in indices $i$ and $j$.\\ 

\section{Conclusions}

We have discussed in detail the structure of
 neutrino masses and 
mixing in MSSM in the presence of trilinear R-violating couplings, 
specifically
$\l'_{ijk}$. Noteworthy feature of the present analysis is that it is
possible to obtain the required neutrino mass pattern under fairly general
assumption of  all the $\l'_{ijk}$ being of equal magnitudes.
It is quite interesting that hierarchy
among neutrino masses is controlled by few parameters in MSSM and is
largely independent of the trilinear $R$ violating couplings under
the assumption that all the trilinear couplings are equal in magnitude.
Thus one could understand the required neutrino mass ratio without being 
specific about the exact values of large number of the trilinear
couplings. This `model-independence' is an attractive feature of the
scenario discussed here.

The key difference of the present work compared to many of the other
works is  proper inclusion of the sneutrino {\it vev}
contribution. While we had to resort to specific
case of the minimal supergravity model for calculational purpose, the
sneutrino {\it vev}
contribution would arise in any other scheme with $\l'_{ikk} \neq 0$ at 
a high scale such as $M_{GUT}$. Such contribution thus cannot be
neglected a priori. On the contrary, the inclusion of this
contribution makes the model more
interesting and fairly predictive in spite of the presence of large number
of unknown couplings. The model prefers simultaneous solutions for
solar and atmospheric neutrino problems with vacuum oscillations 
for the solar neutrino problem. However, Large Angle MSW also can be
accommodated in regions of the parameter space where sneutrino {\it vev}
is suppressed.

%% file: chap6.tex
\chapter{Neutrino Mass constraints on R violation and HERA anomaly}

\section{Introduction}
In the last two chapters we were interested in studying the
structure of neutrino mass matrix in the presence of explicit 
lepton number violation and the possibilities of having simultaneous
solutions to solar and atmospheric neutrino anomalies. We have
motivated such a study by claiming that since lepton number violation
naturally occurs in a supersymmetric extension of Standard Model
\footnote{When we say natural here, we mean it is not disallowed by
gauge symmetries as in SM.}, it would be the right framework to generate
neutrino masses, as required for solutions of neutrino anomalies. But
lepton number violation has not been observed in nature ( except
for a possible neutrino mass ), whereas the
presence of lepton number violating couplings can lead to various
interesting processes like $K^+ \rightarrow \pi^+ \n \bar{\n}$, the
violation of weak universality, neutrinoless double beta decay etc. 
Most of these processes have not been observed yet in nature and there
exist stringent experimental limits on these processes \cite{driener}.
These experimental limits can be converted into limits on lepton number
violating couplings themselves. The limits on various couplings so obtained
have been reviewed in \cite{driener,GB}. 

The limits discussed above also have an interesting feature associated
with them, which is the so called `single coupling scheme'. In obtaining
the limit on a particular coupling, one usually assumes that the particular
coupling is the most dominant one compared to the other lepton number 
violating couplings \cite{dimohall}. This is to facilitate a meaningful
analysis. But the single coupling scheme has its own drawbacks. The single
coupling scheme makes the analysis basis dependent \cite{kaustabh}. This 
basis dependence comes due to the CKM matrix in the quark sector. In such a
case, one coupling in one basis may correspond to several couplings in
another basis. 

We have also seen that the presence of trilinear lepton number violating
couplings in the superpotential would also lead to majorana masses for
the neutrinos. The neutrino masses are themselves constrained
by kinematic limits on them. These limits can be used
to constrain the lepton number violating couplings present in the 
superpotential. Thus in the present chapter, we do a sort of 
`reverse analysis' to what has been done so far in this thesis. We choose
$\l'_{1ij}$ couplings for our analysis and study the limits on these
couplings from neutrino mass constraints \footnote{This is because
the electron neutrino mass is most stringently constrained compared
to other neutrino masses.} . 

As has been mentioned in the earlier chapter, neutrinos attain mass
both at the tree level as well as at the 1-loop level in models with
trilinear R violation. The present analysis in the literature
 \cite{rohini,masero} of the 
constraints arriving from neutrino masses has neglected the important
and dominant tree level  contribution to the neutrino masses which would 
occur when any of the lepton number violating term is present in the
superpotential. But as we have seen in chapter 5, in any realistic model 
where supersymmetry is broken at a high scale, RG evolution strongly 
modifies the neutrino mass by giving rise to a tree level mass 
to the neutrino, which we call `RG induced tree level mass'.  
Incorporation of this additional contribution can change 
 the already obtained \cite{rohini,masero}
limits significantly. The aim of this chapter is to systematically
derive these constraints and discuss its implications. The restriction
to only $\l'_{1ij}$ couplings is also interesting because these 
couplings have been 
relevant in attempts to understand the HERA anomalous results, which
we will discuss later in the chapter. 

\section{ Basis choice and definition of $\lambda'_{ijk}$ :}
Since we consider only trilinear R-violation, the lepton number 
violating part of the  superpotential  is given as follows:

\be
\label{wr}
W_{\not{L}} = \lambda'_{ijk}~ L'_iQ'_{j} D'^{c}_{k}
\ee
The above is written in the current basis of the quarks. But,
as has been mentioned above, in order to meaningfully constrain 
the trilinear coupling, it is usually assumed that
only a single coupling is non-zero at a time. While the physics implied
by these couplings is basis independent, the said assumption makes the 
constraints on ${\lambda}'_{ijk}$ basis dependent since a non-zero 
$\lambda'$ in one basis correspond to several non-zero $\lambda'$'s
in the other. The trilinear couplings in eq.~(\ref{wr}) can be rewritten
 \cite{kaustabh} in the quark mass basis as follows:
\begin{equation}
\label{basis1}
W_{\not{R}} = \lambda'_{ijk} (-\nu_id_{l}  K_{lj}+e_i u_j )d_k^c,
\end{equation}
where the above terms  are now in the quark  mass basis and $K$ denotes the
Kobayashi-Maskawa matrix.  Even in the mass basis one could choose 
a different definition for the trilinear couplings:
\begin{equation}
\label{lbar}
\lambda^{\nu}_{ijk}\equiv K_{jl}\lambda'_{ilk} 
\end{equation}
and rewrite (\ref{basis1}) as,
\begin{equation}
\label{basis2}
W_{\not{R}} =  \lambda^{\nu}_{ijk}(-\nu_id_{j}+e_i K^{\dagger}_{lj}u_l )d_k^c
\end{equation}
With the first choice, a single non-zero ${\lambda}'_{ijk}$ can lead to
 tree level flavour violations in the neutral
sector \cite{kaustabh} while this is not so if only
 one $\lambda^{\nu}_{ijk}\; (j\neq k)$
is non-zero. Clearly, there are other equivalent definitions 
of trilinear couplings which are between the two extreme cases
given in eqs.(\ref{basis1}) and (\ref{basis2}). 
The basis dependence can be clearly seen here as 
 the first coupling $\l'_{121}$ is constrained 
severely by the neutrino mass but the coupling $\l^{\n}_{121}$ is not. 

The generation of RG induced tree level mass in the presence of
trilinear lepton number violating couplings has already been 
discussed in chapter 5. However, for the sake of completeness we
repeat it here again.  The crucial terms to 
our analysis are the soft supersymmetry breaking terms in the scalar potential.
 The lepton number violating parts of these terms can be written as:
 
\begin{equation}
\label{vsoft}
V_{soft} = A^{\nu}_{ij} \lambda^{\nu}_{ij}~ D_{i} D_{j}^{c} \tilde{\nu} -
B_{\nu}~ \tilde{\nu} H_{2}^0 + m^{2}_{\nu H_{1}^0}~ \tilde{\nu}^{\star}  H_{1}^0
+ c.c + \cdots ~,
\end{equation}
where we have explicitly written only those terms which are responsible
for generating the electron neutrino ($\nu_{1} \equiv \nu$) mass and 
$\lambda^{\nu}_{ij}\equiv \lambda^{\nu}_{1ij}$. Note that the presence of only
 trilinear terms in the superpotential generate only the first 
terms in eq.(\ref{vsoft}) at the high scale
in conventional supergravity based scenario. The remaining terms are
 however generated at the weak scale and should  therefore be retained. 
In this basis, the general RGE presented in Chapter 3 become,
\begin{eqnarray}
\label{rg}
\frac{dB_{\nu}}{dt}& =& -~ \frac{3}{2} B_{\nu}~\left(Y_{t}^{U} - \tilde{\alpha}
_{2} - \frac{1}{5} \tilde{\alpha}_{1}\right) - \frac{3 \mu}{16 \pi^{2}} 
\lambda^{\nu}_{kk} h^{D}_{k} \left(A^{\nu}_{kk} + \frac{1}{2} B_{\mu} \right)~, \\
\frac{d m_{\nu H_{1}}^{2}}{dt}& = &-~ \frac{1}{2} m_{\nu H_{1}}^{2} 
\left(3 Y_{l}^{D} + Y_{l}^{E} \right)  
-\frac{3}{32 \pi^{2}} \lambda^{\nu}_{kk}
h^{D}_{k} \left( m_{H_{1}}^{2} + m_{\tilde{\nu}}^{2} \right. \nonumber \\
& & \left. +2~ A^{\nu}_{kk} A_{k}^{D} + 
2~  m_{k}^{Q{2}} +~ 2 ~m_{k}^{D^{c}2 } \right)~\\
\frac{d \lambda^{\nu}_{kk}}{dt}& = &  \lambda^{\nu}_{kk}\left(- 3 Y^D_k
 - {1 \over 2} Y^U_k  + {7 \over 30} \tilde \alpha_1 + {3 \over 2}
 \tilde \alpha_2 + {8 \over 3} \tilde \alpha_3 \right), \\
\frac{d A^{\nu}_{kk}}{dt}& =& - {3 \over 2} A^{\nu}_{kk} Y_k^D -
{9 \over 2} A_k^D Y_k^D - {1 \over 2 } A_k^U Y_k^U \nonumber \\
&-& \frac{16}{3} \tilde \alpha_{3} M_{3}  - 3 \tilde \alpha_{2}
M_{2} - \frac{7}{15}\tilde  \alpha_{1} M_{1}~. 
\end{eqnarray}

The rest of the parameters carry the same meaning as in chapter 3.
We have kept only leading order terms in $\lambda^{\nu}_{kk}$ 
in writing eqs.(\ref{rg}). The other parameters appearing 
in eqs.(\ref{rg}) satisfy the standard RG equations to 
this order in $\lambda^{\nu}_{ij}$.  The second terms in 
eqs. (\ref{rg}) generate non-zero  
$B_{\nu}$ and $m_{\nu H_{1}}^{2}$ at $Q_{0}$.
These terms in eq.(\ref{vsoft}) generate a non-zero {\it vev}  
$\langle\tilde{\nu}\rangle$ which can be determined by minimizing
the full scalar potential. These are given by,

\begin{equation}
\label{vev}
\langle \tilde{\nu} \rangle \sim {\displaystyle{{B_\nu v_{2} - m_{\nu H_{1}}^2 v_{1}}} \over 
{m_{\tilde{\nu}}^{2} + \frac{1}{2} M_{z}^{2} \cos 2 \beta}}~,
\end{equation}
$m_{\tilde{\nu}}^{2}$ is the soft SUSY breaking sneutrino mass.
This in turn leads to the following neutrino mass:
\be \label{treem0}
(m_{\nu_{e}})_{tree} = {\mu (c g^2+g'^2)(<\tilde{\nu}>^2)
 \over 2(-c\m M_2 +2 M_W^2 c_\b  s_\b~(c+ tan^2\theta_W))}.
\ee   
where $c\simeq 5 {g'}^{2}/3 g^2$, $g^2$ and ${g'}^{2}$
are gauge couplings and $\tan\beta = \langle H_{2}^{0} \rangle/
\langle H_1^0 \rangle  = v_2/v_1$.
It follows from the eqs. (\ref{rg}, \ref{vev})
 that the $(m_{\nu_{e}})_{tree}$ 
involves the combination $(\lambda^{\nu}_{kk} h_{k}^{D})^{2}$.
The trilinear interactions in eq.(\ref{basis1}) lead also to the following
$m_{\nu_{e}}$ at the one-loop level:
\begin{equation}
(m_{\nu_{e}})_{loop} \simeq -~ \frac{(\lambda^{\nu}_{kk})^{2}}{16 \pi^{2}}
m_{k}^{D} \sin\phi_{k} \cos\phi_{k}~ ln \frac{M_{2 k}^{2}}{M_{1 k}^{2}}~,
\end{equation}
where we have implicitly assumed that only one $\lambda'_{1jk}$ is
non-zero at a time. $\phi_{k}$ and $M_{2,1 ~ k}^{2}$ respectively denote 
the mixing among squarks $\tilde{d}_{k},\tilde{d}^{c}_{k}$ and their masses. 
The mixing $\phi_{k}$  is proportional to $m_{k}^{D}$. As a result 
just like the tree level mass, the $(m_{\nu_{e}})_{loop}$ also scales as 
$(\lambda^{\nu}_{kk} h_{k}^{D})^{2}$. $m_{\nu_e}$ therefore provides a
bound on $\lambda^{\nu}_{kk}$ which can be converted to a bound on
$\lambda'_{1lk}$ from ~eq.(\ref{lbar}). The resulting bounds become 
stronger with increase in $\tan\beta$
due to the fact that $B_{\nu}, m_{\nu H_{1}}^{2} $ 
involve $h_{k}^{D} = m_k^D/v \cos\beta$ .  
For the same reason, bounds  display strong hierarchy, typically,
\begin{equation}
\frac{(\lambda^{\nu}_{kk})_{max}}{(\lambda^{\nu}_{ll})_{max}}\sim
\frac{m_{l}^{D}} {m_{k}^{D}}~.
\end{equation}
It also follows from eq.(\ref{lbar}) that,
\begin{equation}
\label{kratio}
\frac{(\lambda'_{1kk})_{max}}{(\lambda'_{1lk})_{max}}\sim\frac{K_{lk}}
{K_{kk}}~.
\end{equation}
\noindent
It is seen from last two equations that the $\lambda'_{133}, 
(\lambda'_{131})$ is constrained most (least) by $m_{\nu_{e}}$.  

\begin{figure}[ht]
\begin{tabular}{cc}
\epsfxsize 6 cm
\epsfysize 8 cm
\epsfbox[17 145 591 715]{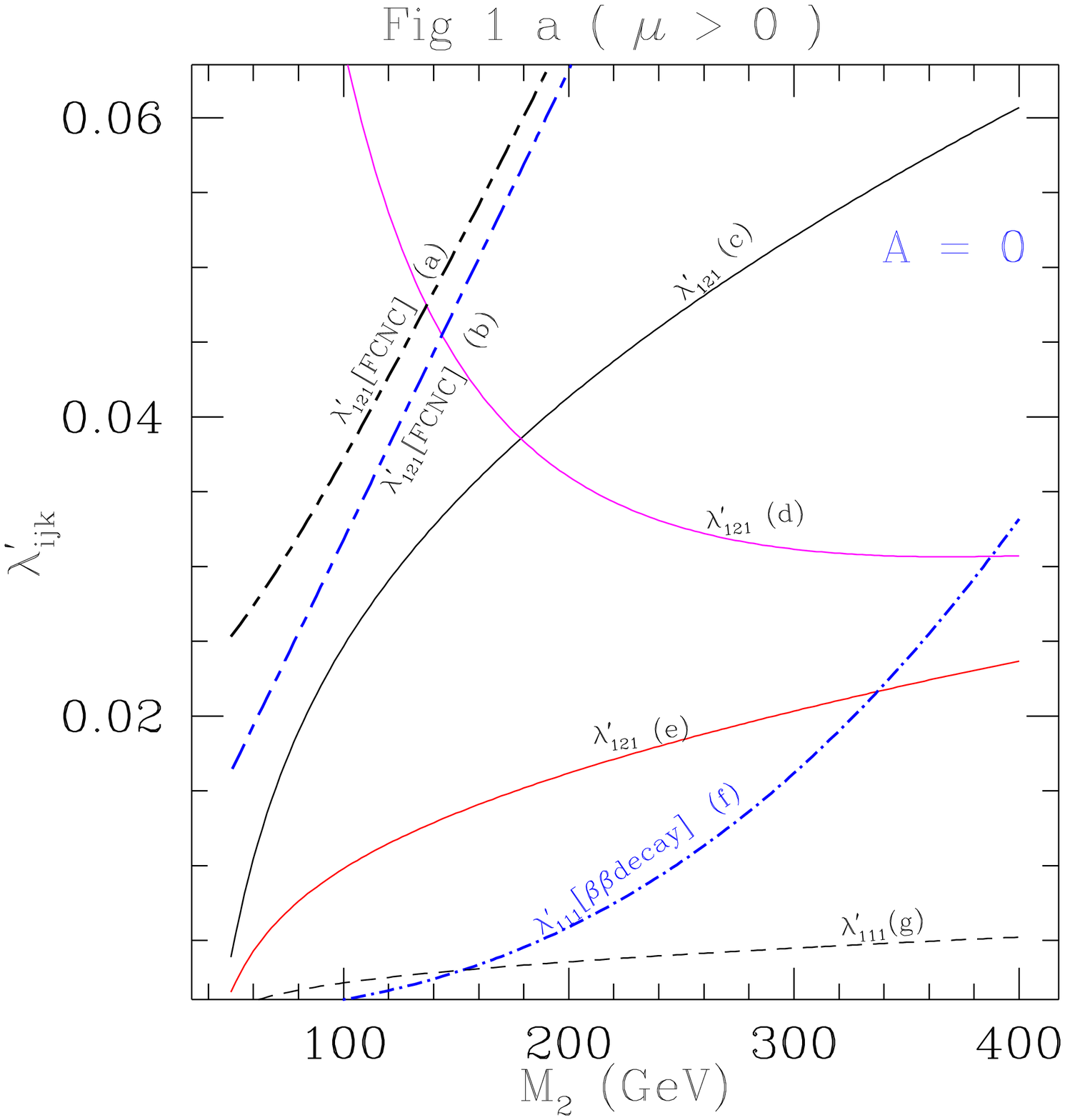}&
\epsfxsize 6 cm
\epsfysize 8 cm
\epsfbox[17 145 591 715]{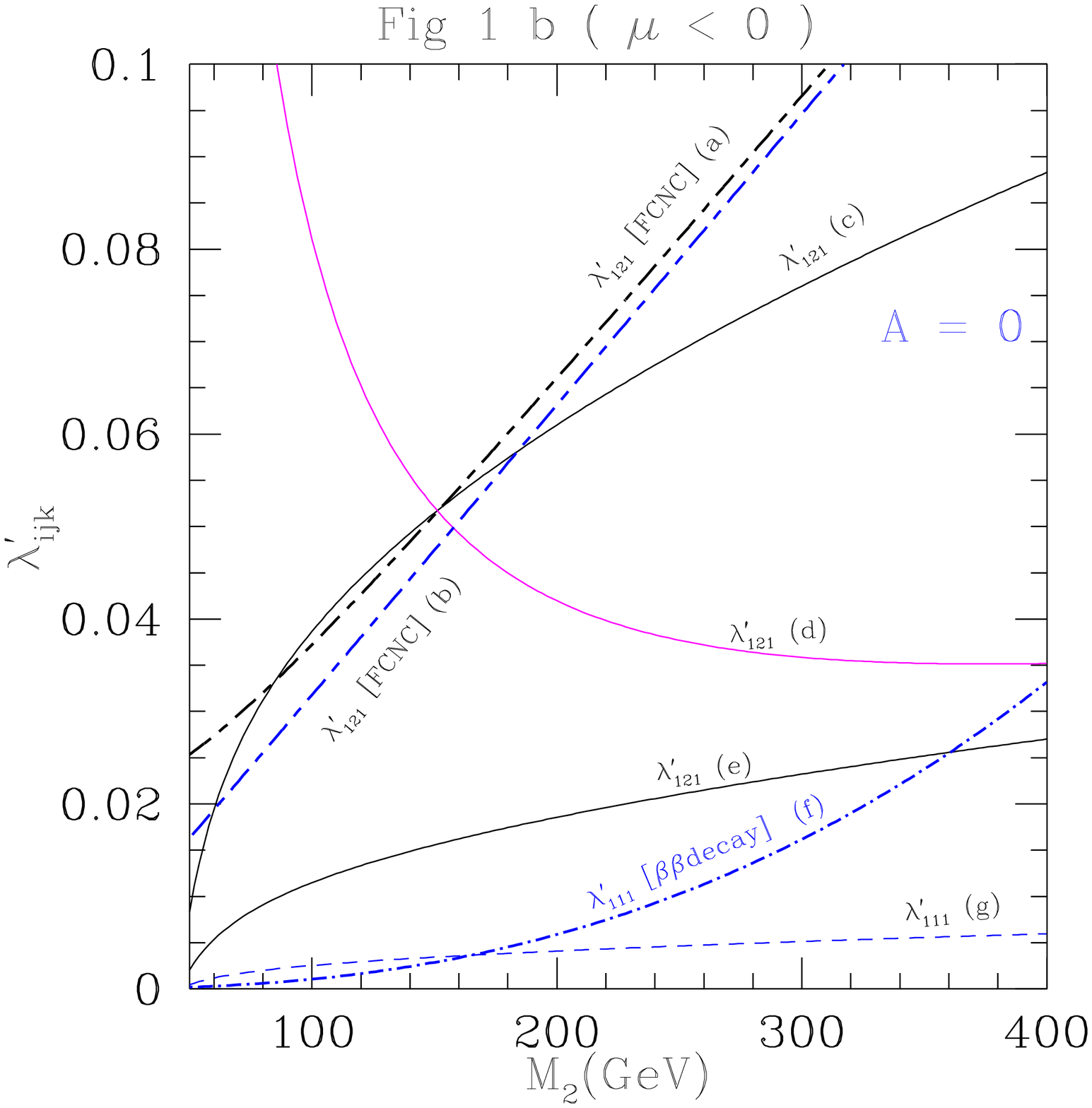}
\end{tabular}
\end{figure}

{\bf Figure 1a}. FCNC constraints  on 
$\lambda^{\prime}_{121}$ for a)~$m = 200$ GeV and 
b)~ $m = 50$ GeV for $tan\beta = 40$.
Neutrino mass constraints on $\lambda^{\prime}_{121}$ for 
c)~$m = 50$ GeV, $tan\beta = 15$; d)~$m = 200$ GeV, $tan\beta = 40$ and	 
e)~$m = 50$GeV, $tan\beta = 40$.	 
f)~ $\lambda^{\prime}_{111}$ from neutrino less $\beta\beta$ decay 
g)~$\lambda^{\prime}_{111}$ from neutrino mass constraints for $m = 50$ GeV 
and $\tan\beta = 40$.
{\bf Figure 1b}. Same as of Figure {\bf 1a)} but with $\mu <0$ 

\vskip 0.2cm

\section{Bounds on trilinear couplings :}

We now present bounds on the couplings $\lambda'_{1jk}$ obtained from
the experimental limit on the neutrino mass
$m_{\nu_e}$ \cite{hera1}. In the present framework, 
the neutrino obtains a majorana
mass and hence stringent constraint on  $m_{\nu_e}$ would follow
from the non-observation of neutrinoless $\b\b$ decay. We shall use
somewhat conservative value of $m_{\nu_e}\leq 2.0~$eV in numerically
discussing the bounds on different couplings which lead to 
$m_{\nu_e}$. In order to obtain bound on the relevant coupling, we have  
numerically solved eqs.(\ref{rg}) along with
the other standard equations for the parameters appearing on the RHS of
eqs. (\ref{rg}) as discussed in Chapter 3.  

We work with R violating version of the MSSM with a standard set of soft 
supersymmetry breaking terms specified at a high scale near 
$ M_{GUT} = 3 \times 10^{16}~ \GeV$. We confine ourselves to
the scenario with radiative breaking of the $SU(2)\times U(1)$ symmetry.  
All  parameters in eq.(\ref{vev}) are specified at a low scale $Q_{0}$ .  
We have chosen $Q_{0}$ to be $M_{Z}$ \footnote{
Note that change in $Q_0$ can alter some of the bounds
obtained by minimization of the tree level potential and more detailed
treatment should include 1-loop corrected potential, see G. Gamberini, G.
Ridolfi and F. Zwirner, \np{B331}{90}{331}}. The parameters  
$B_{\nu},m_{\nu H_{1}}^{2}$ are assumed to be zero at $M_{GUT}$ . 
Their values at $Q_{0}$ depend upon the standard parameters of MSSM  
which we take to be gaugino (gravitino) mass, $M_2\;(m)$, $\tan\beta$ and  
universal trilinear strength $A$. $B_{\nu}$ and $m_{\nu H_{1}}^{2}$ determine 
$(m_{\nu})_{tree}$ in terms of these parameters and  
$\lambda^{\nu}_{kk} (tz)$, where, $tz = 2~\ln({M_{GUT} \over M_Z})$.  
The 1-loop contribution also gets fixed in terms of these parameters  
 since $\phi_{k}$ and $M_{2,1~k}^{2}$ appearing in eq.(12) are  
determined using the standard $2 \times 2$  mixing matrix for   
$\tilde{d}_{k},\tilde{d}^{c}_{k}$  system.  

\begin{center}
\begin{tabular} {|l|l|l|l|}
\hline
& &$~~m=50~GeV~~$ &$~~m=50~GeV~~$ \\
$~~\lambda'_{ijk}~~$& $~~$Previous limits$~~$ &$~~M_2=150~GeV~~$ &$~~M_2=175~GeV~~ $ \\
\hline \hline
\quad 111 &\quad 0.021\cite{db}    &\quad 0.0017           &\quad 0.0018    \\
\quad 121 &\quad 0.06\cite{kaustabh} &\quad 0.0077           &\quad 0.0085     \\
\quad 131 &\quad 1.3\cite{barger} &\quad 0.48           &\quad 0.54     \\
\quad 112 &\quad 0.15\cite{barger} &\quad 4 $~10^{-4}$ &\quad 4.5 $~10^{-4}$ \\
\quad 122 &\quad 0.06\cite{kaustabh} &\quad 8.9 $~10^{-5}$ &\quad 9.9 $~10^{-5}$\\
\quad 132 &\quad 1.05\cite{sridhar1}&\quad 2.1 $~10^{-3}$ &\quad 2.4 $~10^{-3}$ \\
\quad 113 &\quad 0.15\cite{barger} &\quad 3.1 $~10^{-4}$ &\quad 3.4 $~10^{-4}$\\
\quad 123 &\quad 0.06\cite{kaustabh} &\quad 7 $~10^{-5}$&\quad 7.7 $~10^{-5}$ \\
\quad 133 &\quad 0.0029\cite{rohini} &\quad 2.8 $~10^{-6}$ &\quad 3.1 $~10^{-6}$ \\
\hline
\end{tabular} \\
\end{center}
{\bf Table 1.} Limits on single $\lambda'_{ijk}$ following from 
the electron neutrino mass for $A=0$, $\tan\beta=30$ and $\mu>0$.
 The limits become stronger for  larger $\tan\beta$ and weaker for
 negative $\mu$. The existing  limits mentioned in the table are for 
the relevant squark mass $\sim 500  \GeV$ and gluino mass $M_3=500~\GeV$ 
in case of $\lambda'_{111}$.

\vskip 0.2cm

The bounds on different couplings are displayed in Figs. 1 and 2. 
Apart from being dependent on SUSY parameters, these bounds are quite
sensitive to the chosen sign of the $\mu$ parameter. This is due to the
fact that for one (namely negative) sign of $\mu$, two terms appearing 
in the sneutrino {\it vev}, eq.(\ref{vev}) cancel 
while they add for positive sign.
 The suppression in sneutrino {\it vev} 
occurring in the first case weakens the bound on $\lambda'$. 
 Fig. 1 displays bounds on couplings
 $\lambda'_{111}$ and $\lambda'_{121}$ obtained by demanding $m_{\nu_e}\equiv
(m_{\nu_e})_{tree}+(m_{\nu_e})_{loop}\leq 2.0 \eV$. Curves for three
representative values of $\tan\beta$ and universal gravitino mass
$m$ are shown. For comparison, we also display in the same figure
the existing bounds on these couplings. The most
stringent bound on $\lambda'_{111}$ is derived from neutrinoless
double beta decay \cite{db}
and on  $\lambda'_{121}$ from the process $K^+\rightarrow \pi^+\nu \bar{\nu}$  
\cite{kaustabh} .  These are shown as function of $M_2$ in the same figure 
using MSSM expressions for the relevant squark masses.
 It is seen that the bounds derived here are comparable or better 
(in case of  larger values of $M_2$ ) than the already existing ones.

\begin{figure}[h]
\centerline{
\epsfxsize 8 cm
\epsfysize 8 cm
\epsfbox[18 146 591 715]{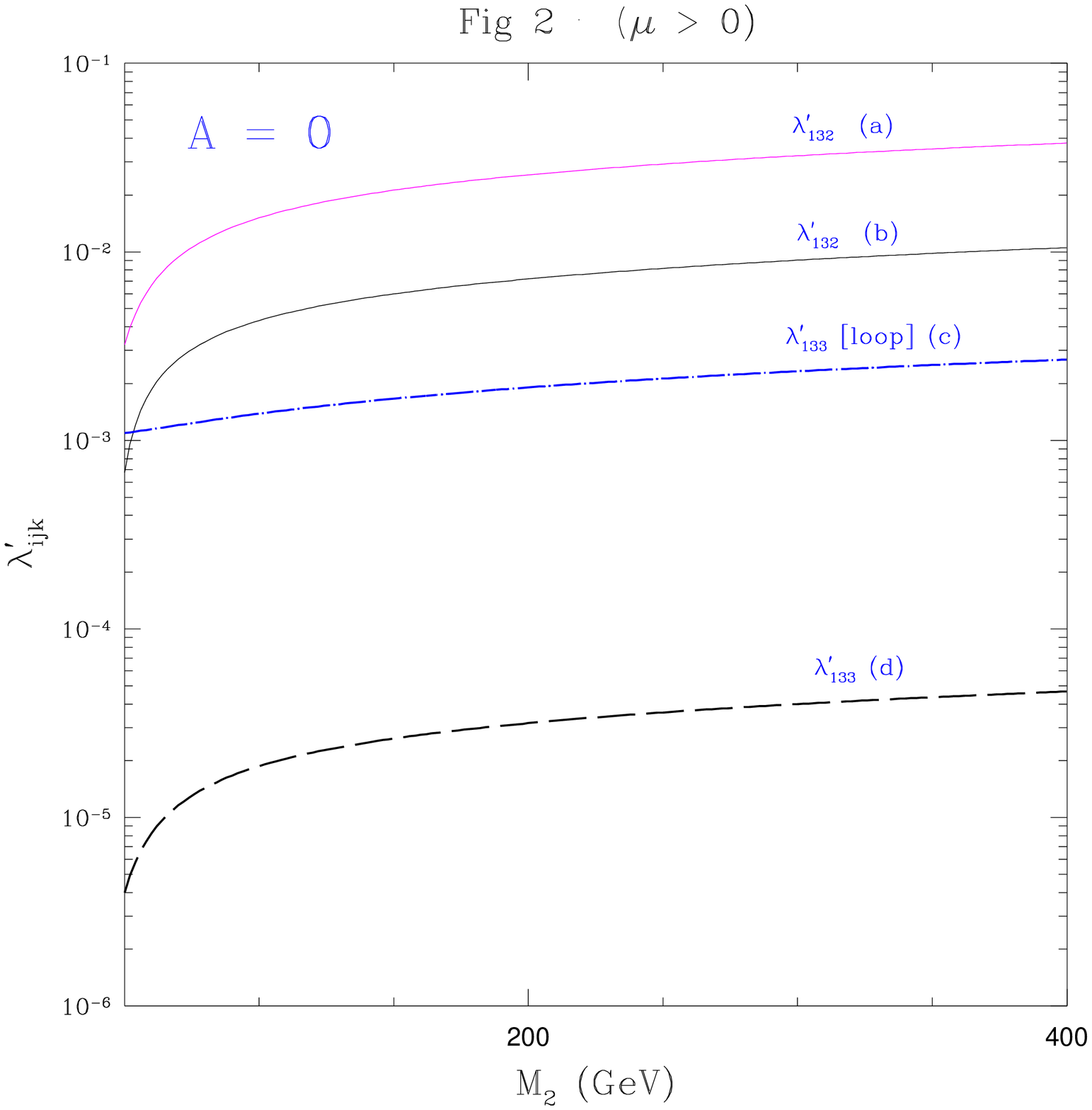}}
\end{figure}
\noindent
{\bf Figure 2}.  Neutrino mass constraints on $\lambda^{\prime}_{132}$ for 
a)~$tan\beta = 5$ and b)~$tan\beta = 25$; on $\lambda^{\prime}_{133}$ 
for $tan\beta = 5$ c)~considering only loop contributions and d)~loop 
as well as sneutrino {\it vev} contributions are shown for $m=50~GeV$. 
\vskip 0.3cm

Fig 2 shows the bounds on $\lambda'_{133}$ and $\lambda'_{132}$ for
 $\mu > 0$. The $\lambda'_{133}$ is constrained most and 
is required to be as low as $ 3\times 10^{-6}$ even for $\tan\beta=5$. For
comparison we also show the limit on $\lambda'_{133}$ which would follow
if sneutrino {\it vev} is completely neglected as in previous works
\cite{rohini,masero}. It is clear
that inclusion of this {\it vev} drastically alters the bound. We also display
limit on $\lambda'_{132}$ which was poorly constrained in the previous
analysis \cite{sridhar1}. 
\noindent
The bounds on other couplings $\lambda'_{1jk}$ not explicitly
 displayed in above figures can
be read off from eq.(\ref{kratio}). These constraints are listed in a table for
two representative values $m = 50 \GeV, M_2 = 150 \GeV$ and 
$m= 50\GeV, M_2 = 175 \GeV$ and $\tan\beta=30$.
These respectively correspond in MSSM to $m_{\tilde{d}_R}\sim 474, 553
\GeV$. It is seen that the neutrino mass considerably improves
on the existing constraints. This table is given to illustrate the
importance of the bounds following from the neutrino mass. It should however be
kept in mind that these bounds are  sensitive to the choice
of the MSSM parameters a fact which becomes evident through figures (1-2).

Thus, we have investigated the restrictions on the $R$ parity violating
trilinear couplings, specifically the $\lambda'_{1jk}$ implied by neutrino
mass. These restrictions have been discussed earlier \cite{rohini,masero}
 but earlier discussions either neglected very
 important and dominant contribution due
to the induced sneutrino {\it vev} or did not incorporate the effect of
the quark mixing. We have attempted  here a complete  analysis 
which incorporates these and also pays careful attention to the dependence
of the bounds on the choice of basis. Unlike the earlier works, we have
also  numerically discussed the bounds in terms of 
parameters of the MSSM. 

Many of the discussions and in particular
the generation of sneutrino {\it vev} in the presence of a single trilinear
coupling would be true in a more general set up, {\em e.g.} in
case when soft terms arise from the gauge mediated supersymmetry breaking.
The main point following from our analysis is that the electron 
neutrino mass provides much stronger constraints on $R$ violation
and associated phenomenology than has been hitherto realized.

\section{HERA anomalies}

The HERA experiments at DESY,  collide electrons (or positrons) with
protons. The electrons typically carry an energy of $\sim 30$ GeV,
whereas the protons carry an energy of around $\sim 820 (920)$ GeV. 
In addition to measuring the structure functions of the protons through
the DIS (Deep Inelastic Process), these experiments are also ideal 
to verify R-violating interactions. The experiment consists of two 
general purpose detectors, H1 and ZEUS. In 1997, anomalous events have been 
reported by the H1 and the ZEUS detectors \cite{hera} at HERA in the 
deep inelastic $e^+ p$ scattering. They claimed to have found a resonance
for a leptoquark of mass $\sim 200$ GeV. This has generated considerable 
excitement within  the community \cite{s2} as, if substantiated these 
results could form strong evidence for physics beyond the SM and may be
for supersymmetry. The work presented here is based on ref.\cite{he2} 
which was motivated from the results discussed above. However, this 
evidence for the leptoquark did not stand the test of time and the 
subsequent data from HERA \cite{cash} were consistent with the 
Standard Model expectations. We describe here the work in 
ref.\cite{he2} for the sake of completeness, though it is now only
of academic interest. To this extent, we consider here data reported 
by HERA in 1997.

The  available data when taken seriously allowed for two 
possible interpretations: ({\em i})
 The presence of some lepton number violating contact interaction 
\cite{cont} or ({\em ii}) production of a resonance in the 
$e^+q$ channel. MSSM with R-violation \cite{s2,s1,dp1}
provides a natural theoretical framework to incorporate the second possibility
although an alternative in terms of a scalar leptoquark \cite{lepto} is
open.

The supersymmetric interpretation of the HERA events assumes 
that the excess events seen at
HERA are due to resonant production and  subsequent decay 
of the squark to $e^+ q$. Three
possibilities have been considered in this context\cite{s2,s1,dp1}:
$e^+_Rd_R\rightarrow \tilde{c}_L,e^+_Rd_R\rightarrow
\tilde{t}_L,e^+_Rs_R\rightarrow \tilde{t}_L$. 
In  analyzing these scenarios \cite{s2,s1,dp1} it has been implicitly
 assumed that the squark masses are free parameters of the
model. While this would be true in the most general situation, 
specific model dependence can alter some of the conclusions. Our
aim is to show that the very minimal model dependent assumption on the
charm squark mass necessarily requires large $\lambda'_{121}$ to
understand HERA events and this large coupling by itself is ruled
out from other constraints like neutrino mass bounds which we have
obtained in the previous section.

The specific assumption that we make and which leads to the above
conclusion is that the charm 
squark mass squared is positive at the unification scale. 
This assumption is  true in the radiative electro-weak breaking
 scenario with 
universal boundary conditions at the GUT scale, but it can also be true in
a much more
general context.  We shall  first assume that the gaugino masses
are unified at $M_{GUT}$ but demonstrate later that 
the removal of this assumption does not
significantly change the basic conclusion.
The argument leading to the above conclusion is largely insensitive 
to the details of the radiative $SU(2)\times U(1)$ breaking in  the MSSM and
runs as follows.

Consider the following $R$ violating couplings:
\begin{equation}
\label{wr1}
W_{R} = \lambda'_{ijk} (-\nu_id_{l}  K_{lj}+e_i u_j )d_k^c
\end{equation}
The above terms are defined in the quark mass basis and $K$ denotes the
 Kobayashi-Maskawa 
matrix. The charm squark interpretation of the HERA anomaly requires 
$\lambda'_{121}$ to be non-zero. 
The HERA data can be explained provided
\begin{equation}
\lambda'_{121}\sim \frac{0.025-0.034}{B^{1/2} }
\ee
The number in the numerator of eq.(2) is indicative  of the required range and
depends upon the weightage given to the different experiments as well as on 
the next to leading order QCD
corrections \cite{kz}.  In the following, we shall take \footnote{ 
The lower limit is obtained when H1 and ZEUS data from 1997
run are also included while the upper limit corresponds to inclusion of
H1 data alone. In both cases, 30\% increase in the relevant cross 
section due to next to leading order corrections \cite{kz} 
is assumed.}
the value 0.025 for the numerator in the RHS of eq.(2). 
$B$ refers to the branching ratio for the squark decay to $q e^+$. 
This decay would take place through the coupling in eq.(1) itself. 
$B$ is also influenced by the $R$ conserving decays of the charm squark to
an $s$ ($c$)  and a chargino  (neutralino) . The $\lambda'_{121}$ and 
the  parameters $\mu,M_2,\tan\beta$ determine $B$ in the MSSM.
HERA data can be reconciled if for a region in these parameters
({\em i}) eq.(2) is satisfied, ({\em ii}) $\lambda'_{121}$ is
consistent with other constraints \cite{kaustabh,db,apv} 
due to $R$ breaking and ({\em iii})
charm squark has a mass around 180- 220 GeV. 

In supergravity based models, the charm squark mass at the weak scale is 
governed by the gauge couplings and 
the gaugino masses. Its value at $Q_0=200 \GeV$ is given in the limit of
neglecting Kobayashi-Maskawa  and $\tilde{c}_L-\tilde{c}_R$ mixing by
 \footnote{ We also neglect the effect of additional trilinear $R$ violating
couplings on the running of the charm squark mass. Their inclusion does
not significantly alter the charm squark mass even when they 
are large, see e.g.  K. Cheung, D. Dicus and B. Dutta \cite{dic}.}
\be
m_{\tilde{c}_L}^2 (Q_0)\approx m_{\tilde{c}_L}^2 (M_{GUT})+
 8.83  M_2^2+ 1/2~ M_Z^2 ~\cos 2\beta~ (1-4/3 \sin^2\theta_W)
\ee
where we have assumed that the gauge couplings and the gaugino masses are 
unified at the GUT scale, $M_{GUT}=3\times 10^{16}\GeV$ and chosen
 $\alpha_s(M_Z)=0.12$. 
The $M_2$ in eq.(3) is the value of the wino
mass at the weak scale
identified here with $M_Z$. The last term in
the above equation is a (-ve ) contribution from the D-term.
It follows that  the charm squark mass provides  strong upper bound on
$M_2$ as long as $m_{\tilde{c}_L}^2 (M_{GUT})>0$:
\be
M_2\leq 74.04 \GeV \left({m_{\tilde{c_L}}\over 220 \GeV}\right )
\left ( 1- 0.06 \cos 2\beta \left({220 \GeV \over m_{\tilde{c_L}}}\right
)^2
\right)^{1/2}
\ee

The branching ratio $B$ is determined in the MSSM by the following widths \cite{kk}:
\bea
\Gamma (\tilde{c}_L\rightarrow e^+ c)&=& 
{\lambda'^2_{121}\over 16 \pi}
m_{\tilde{c}_L}  \\ 
\Gamma (\tilde{c}_L\rightarrow \chi^0_i c)&=& \frac{\alpha}{2~m_{\tilde{c}_L}^3}
\lambda^{\frac{1}{2}} (m_{\tilde{c}_L}^2, m_{c}^2, m_{\chi^0_i}^2)\nonumber \\
&& \left[~( \mid F_L \mid^2 + \mid F_R \mid^2 ) (m_{\tilde{c}_L}^2 - m_{c}^2 - 
m_{\chi^0_i}^2 ) - 4 m_c m_{\chi ^0_i} Re(F_R 
F_L^\star)~\right] \\ \nonumber
\Gamma (\tilde{c}_L\rightarrow \chi^+_i s)&=& \frac{\alpha}{4 sin^2 \theta_W
m_{\tilde{c}_L}^3} \lambda^{\frac{1}{2}}(m_{\tilde{c}_L}^2, m_{s}^2, 
m_{\chi^+_i}^2) \nonumber \\
&& \left[~( \mid G_L \mid^2 + \mid G_R \mid^2 ) 
(m_{\tilde{c}_L}^2 - m_{s}^2 - m_{\chi^+_i}^ 2 ) -
 4 m_s m_{\chi ^+_i} Re(G_R G_L^\star)~\right] \\ \nonumber
{\mbox where},
F_L&=& \frac{m_c N'^\star_{i4}}{2~ m_W~ sin \theta_W ~sin \beta}, \\ \nonumber
F_R&=& e_c N'_{i1} + \frac{\frac{1}{2} - e_c sin^2 \theta_W}{ cos
\theta_W~ sin \theta_W } N'_{i2}, \\ \nonumber
G_L&=& - \frac{m_s U^\star_{k2} }{\sqrt{2}~ m_W~ cos \beta},\\ \nonumber
G_R&=& V_{k1}.
\eea
 We have adopted the same notation as in \cite{kk}.
From the expression for B in terms of the above decay widths, and the HERA
constraint, eq.(2), one can solve for the allowed $\lambda'_{121}$.
The contours in the $\mu-M_2$
plane for different values of $\lambda'_{121}$ are displayed in fig 1a
($\tan\beta$=1) and fig 1b ($\tan\beta=40$). The horizontal lines in these
figures show the upper bound on $M_2$, eq.(4). We also display, the curves
corresponding to two representative values of the chargino masses namely
45 and 85 GeV. The later is the present experimental bound obtained
assuming $R$ conservation. This need not hold in the presence of $R$
violation. It is seen from fig.1b that for chargino mass around 85 GeV,
the bound on $M_2$ by itself rules out charm squark interpretation
for large tan $\beta$ independent of the value of $\lambda'_{121}$ 
\footnote{Specifically, the upper bound on $M_2$ can be reconciled
with the chargino mass of 85 GeV or more only if $\tan\beta\leq 2.5$.}.
But irrespective of the value of $\tan\beta$ and the chargino mass one
needs very large $\lambda'_{121} \geq 0.13 $ in order to satisfy the
bound on $M_2$ coming
from the charm squark mass. This strong bound on $\lambda'_{121}$
 arises because of
the following reason. For $M_2\leq 74 \GeV$, at least one of the charginos
is sufficiently light and contributes dominantly to the $\tilde{c}_L$
decay. This reduces $B$ \footnote{ Reduction in the branching ratio for 
charm squark decay in case of the minimal model was also noticed 
in \cite{dp1}.} and results in large value for $\lambda'_{121}$
due to eq.(2). In contrast, the chargino decay is suppressed kinematically
for $\tan \beta\sim 1$ if $M_2> 200\GeV$. This results in smaller 
allowed value as seen from the figure. But these are in conflict 
with the charm squark mass.

\vskip 0.3 cm
Let us now see if one could make large $\lambda'_{121}$ consistent with
other constraints. The strong constraints come from  atomic parity violation
 \cite{apv}, the decay $K^+\rightarrow
\pi \nu \bar{\nu}$ \cite{kaustabh} and the electron neutrino mass  which
we have derived above. 
The data from Cs on the relevant weak charge have been argued 
\cite{s2,dp2} to
imply
\be
\lambda'_{121}\leq 0.074
\ee
at 2$\sigma$ level in conflict with the large value required here.
In principle, the extra contribution due to
charm squark to atomic parity violation can be canceled by a similar
contribution from the scalar bottom or strange squark but the existing
constraints on the relevant couplings make this cancelation 
 difficult \cite{dp2}.
Thus, one cannot easily avoid the atomic parity violation constraint strictly
in the MSSM but this can be done by postulating new physics, e.g. the
presence of an extra $Z$ \cite{dp2}.

\begin{figure}[ht]
\begin{tabular}{cc}
\epsfxsize 6 cm
\epsfysize 8 cm
\epsfbox[73 141 539 656]{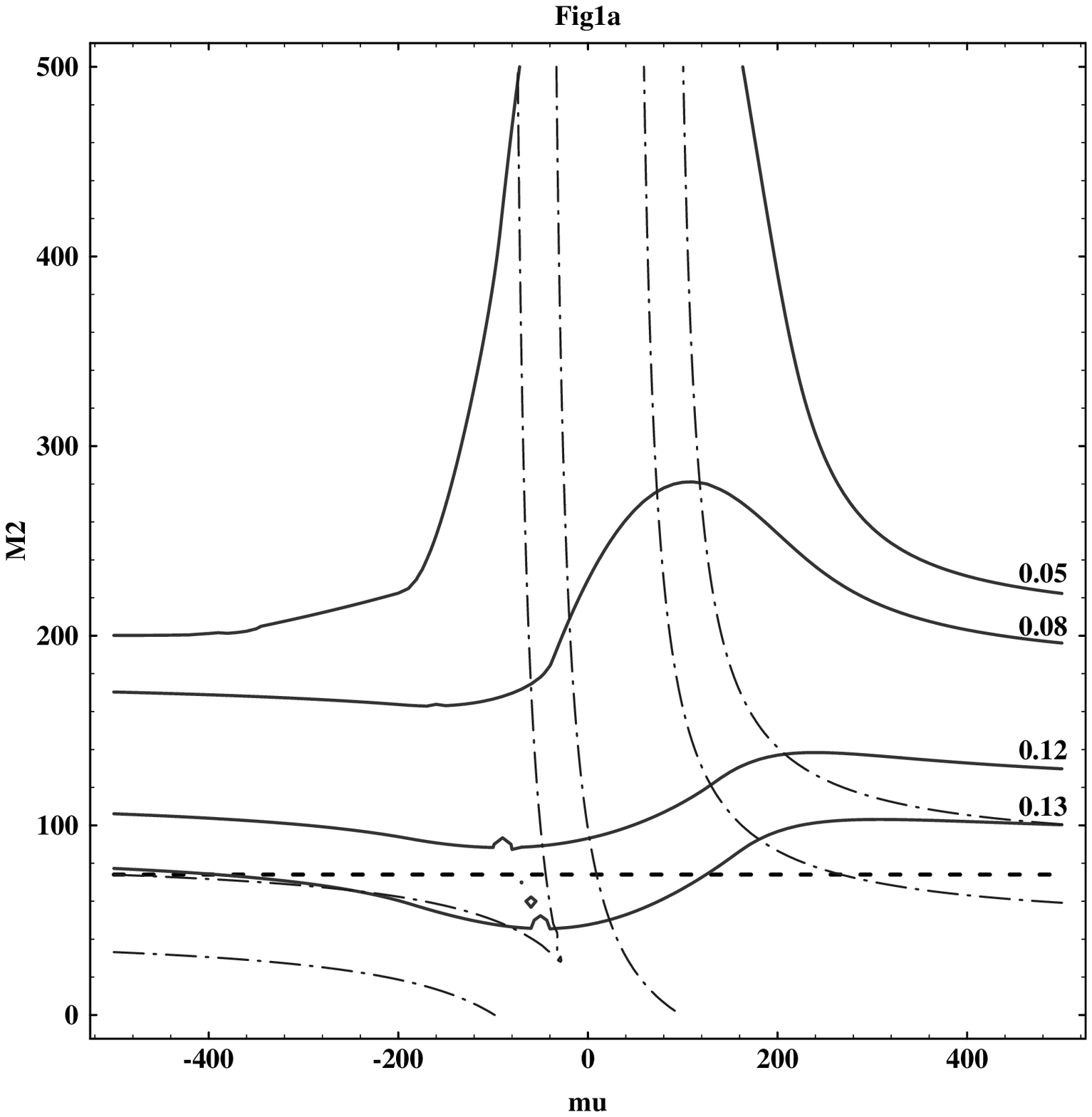}&
\epsfxsize 6 cm
\epsfysize 8 cm
\epsfbox[73 141 539 651]{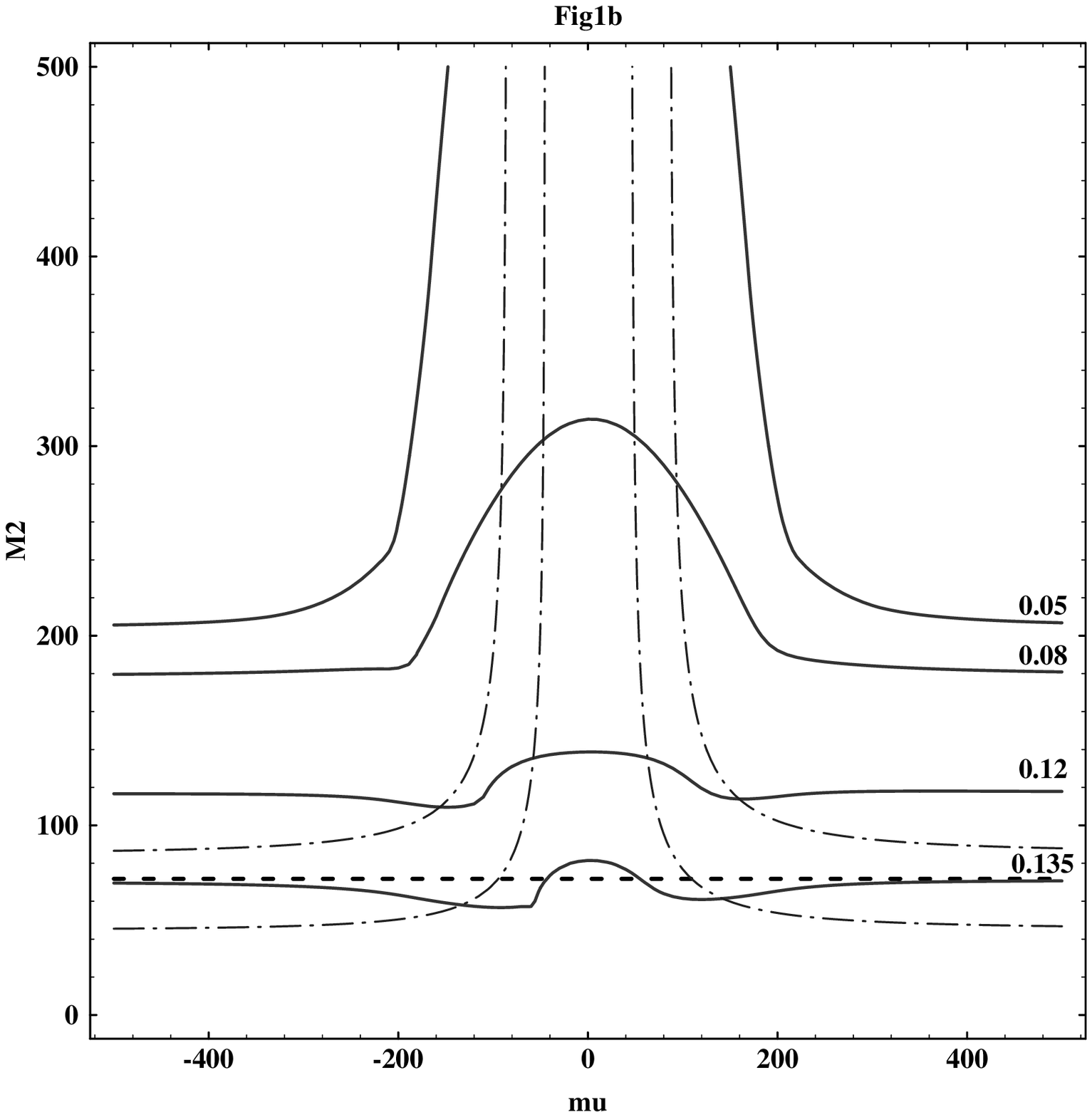}
\end{tabular}
\end{figure}
\noindent
{\bf Fig 1 a:} The contours (continuous lines ) of 
constant $\lambda'_{121}$  
obtained by imposing HERA constraint, eq.(2). The contours are for
values  0.05, 0.08, 0.12, and 0.13. 
The horizontal dashed line represents the bound on $M_2 $ 
coming from requiring $m_{\tilde{c}_L}=220 \GeV$. 
The vertical dash-dot lines represent the bounds on 
the chargino mass, the upper one for a mass of 85 \GeV and the
lower one for a mass of 45 \GeV.  All the
above are computed for tan$\beta$ = 1.
{\bf Fig 1 b:} Same as fig 1a but  for tan$\beta$ = 40 and
$\lambda'_{121}$=0.05, 0.08, 0.12, and 0.135.

The other significant constraint comes from the decay $K^+\rightarrow \pi
\nu \bar{\nu}$ which implies \cite{kaustabh}
\be
\lambda'_{121}\leq 0.02 \left ({m_{\tilde{d}_R} \over 200 \GeV }\right ).
\ee
The electron neutrino mass also gives similar constraint in the same
parameter range. The question of choice of the basis becomes
 relevant in
the discussion of these constraints. This is particularly so when one assumes 
only one
$\lambda'_{ijk}$ to be non-zero. For $\lambda'_{121}$ defined in the mass
basis as in eq.(1) the above constraint is unavoidable if rest of the
couplings are zero. This basis choice is
natural from the point of view of interpreting HERA results but is not
unique. One may redefine the couplings as 
$$ \bar{\lambda'}_{ijk}\equiv K_{jl}\lambda'_{ilk} $$
and rewrite eq.(1) as follows:
\begin{equation}
\label{wr2}
W_{R} =  \bar{\lambda'}_{ijk}(-\nu_id_{j}+e_i K^\dagger_{lj}u_l )d_k^c
\end{equation}
HERA result would now require $\bar{\lambda'}_{121}$ to be large. If this is
the only non-zero $\bar{\lambda'}_{ijk}$ then there will not be any
constraint on $\bar{\lambda'}_{121}$ from the neutrino mass or from 
the $K^+\rightarrow \pi
\nu \bar{\nu}$ decay \cite{kaustabh}. But eq.(10) will now 
generate a contribution to 
the neutrinoless double beta decay which is also severely constrained.
Specifically, one has \cite{db}
\be
 K^{\dagger}_{12} \bar{\lambda'}_{121} \leq 2.2 \times 10^{-3}
\left({m_{\tilde{u}_L}\over 200
 \GeV}\right)^2 \left({ m_{\tilde{g}}\over 200
 \GeV}\right)^{1/2} \ee
This clearly does not allow $\bar{\lambda'}_{121}$ of O(0.1). Thus, 
notwithstanding basis dependence one has problem in accommodating large
value for the relevant coupling. An alternative is to allow more than one 
non-zero
$\bar{\lambda'}_{ijk}$. It is seen from eq.(10) that 
$\bar{\lambda'}_{1j1}$ (j=1,2,3) 
contribute to the neutrinoless $\b\b$ decay and simultaneous presence of
these may lead to cancellations. Eq.(11) now gets replaced by
\be ( \bar{\lambda'}_{111}+ K^{\dagger}_{12} \bar{\lambda'}_{121}+
K^{\dagger}_{13}\bar{\lambda'}_{131} ) \leq 2.2 \times 10^{-3} 
\left({m_{\tilde{u}_L}\over 200
 \GeV}\right)^2 \left({m_{\tilde{g}}\over 200
 \GeV}\right)^{1/2} \ee
With $\bar{\lambda'}_{121} \sim 0.13$, cancellation between the  last two terms
is unlikely as it requires $\bar{\lambda'}_{131} \sim 2$. The first two 
terms can cancel but the $\bar{\lambda'}_{111} $ is independently
constrained from the neutrino mass. Its presence generates a large
contribution to the electron neutrino mass induced through 
neutrino-gaugino mixing which can be given as,
\be
m_{\nu}\sim {g^2\over m_{SUSY}} <\tilde{\nu}>^2
\ee
The value of the induced sneutrino {\it vev} is sensitive to the MSSM
 parameters as we have seen above, 
but can be approximately written as, 
\be
<\tilde{\nu}> \sim {9~\bar{\lambda}_{111}\over 16~
\pi^2}~m_d~~ln \left( \frac{M_{GUT}^2}{M_Z^2} \right)
\ee
Requiring $m_{\nu}\leq 2 eV$ leads for $m_{SUSY}\sim 100 \GeV$ to 
$$ \bar{\lambda}_{111}\leq .04$$
It is seen that cancellations between the first two terms in
eq.(12) are feasible
and can allow $\bar{\lambda}_{121}\sim 0.13$ if this fine tuning is accepted.
It must be added that the bound in the previous equation is quite
sensitive to the MSSM parameters and for a large range in these
parameters, the actual bound can be stronger than the
generic bound displayed above.

While  wino and zino control the
decay of the charm squark, its mass is mainly controlled by the large
radiative corrections driven by the gluino mass. The unification of the
gaugino mass parameters relates the two and leads to the above
difficulty. Thus  giving up this unification may open up a
possibility of reconciling 
HERA events . Let us treat   
the gaugino masses $M_{1,2,3}$ at $M_Z$ 
 as independent parameters. Then integration of the RG equation for the
 charm-squark from $M_{GUT}$ to $Q_0=200 \GeV$ leads to  
\bea
m_{\tilde{c}_L}^2 (Q_0)&\approx& m_{\tilde{c}_L}^2 (M_{GUT})+
 0.77  M_3^2 +0.70 M_2^2 + 0.024 M_1^2 \nonumber \\
&+&1/2~ M_Z^2~ \cos 2\beta~ (1-4/3 \sin^2\theta_W)
\eea
If gaugino masses were to be unified at $M_{GUT}$ then $M_3\sim 3.25 M_2$
and $M_1\sim 0.5 M_2$. Even in the absence of such unification, the
physical gluino mass $m_{\tilde g}\sim (1+4.2 {\alpha_s\over \pi}
)M_3$ must be greater than
the charm squark mass if large $\lambda'_{121}$ is to be avoided.
This follows since in the converse case, the charm squark would
predominantly decay to a gluino and a quark. This decay being governed
by strong coupling, would dominate the other decays and would reduce
$B$. The later is given in case of $m_{\tilde{c}_L}\gg
m_{\tilde{g}}$  by
$$
B\sim {3~\lambda'^2_{121}\over 32~ \pi~\alpha_s} \sim 2.5 \times
\;10^{-3}
\left({\lambda'_{121}\over 0.1}\right)^2$$
Such a tiny value of $B$ would need unacceptably large $\lambda'_{121}$.
It therefore follows that  one must suppress the squark decay to gluino
 kinematically by
requiring $m_{\tilde {g}}\geq m_{\tilde {c}_L}$. Given this bound on
 $M_3$ it follows from
eq.(15) that 
\be
M_2\leq 170 \GeV
\ee
if $m_{\tilde{c}_L}\sim 220 \GeV$. This bound on $M_2$ is  weaker than
 the one in the case of the 
gaugino mass unification, eq.(4). But it nevertheless cannot suppress
the 
decay of squarks to chargino kinematically. It follows
\footnote{Note that fig.1 is based on the assumption of $M_1=0.5 M_2$ but
does not use any relation between $M_2$ and $M_3$.}
 from Fig. 1  that one now  approximately needs 
$ \lambda'_{121}\geq 0.08$. This value is close to the 
2$\sigma$ limit coming from the atomic parity violation but 
one would still need some cancellations to satisfy other
constraints as discussed above.
Thus giving up
unification helps only partially.

An alternative possibility is to allow for a -ve (mass)$^2$ for the charm
squark at the unification scale. In view of the large positive
contribution induced by the gluino mass such negative (mass)$^2$ need not lead
to colour breaking and may be consistent phenomenologically. In fact  a
 -ve (mass)$^2$ for top squark has been considered
in the literature in a different context.
 The universality is a simplifying feature of 
MSSM but it does not follow from any general principle. It
 does not hold in a large class of string based models
which may allow negative (mass)$^2$ for some sfermions as well
\cite{string}. Such masses  can also arise  when  SUSY is
broken by an anomalous $U(1)$ \cite{u1} with some of the sparticles
 having -ve charge under this $U(1)$. 

The large radiative corrections induced through the running
in squark masses from a high $\sim M_{GUT}$ to the weak
scale has played an important role in this analysis. In
contrast to the supergravity based models, this
running is over a much smaller range in models with gauge
mediated supersymmetric breaking. But in these models, the
initial value of the charm squark (mass)$^2$ is positive and
large with the result that these models are
incompatible with the charm squark interpretation of HERA 
anomaly even  without the radiative corrections \footnote{
See, K. Cheung, D. Dicus and B. Dutta \cite{dic}.}.

The interpretation of HERA events in terms of stop may not suffer from the
above mentioned difficulty encountered for the charm interpretation for two
reasons. Firstly, the stop mass is reduced compared to the charm squark
mass due to the  possible large
$\tilde{t}_L-\tilde{t}_R$ mixing as well due to the large top coupling.
Secondly, this mass also
involves one more parameter (the trilinear couplings $A$) compared to the
charm squark mass. Thus while this is a less constrained possibility,
imposition of the requirement that $m_{\tilde{t}}\sim 200 \GeV$ would
certainly lead to more a constrained parameter space than considered 
in model independent studies \cite{s1}.

In summary, we have shown that the charm squark interpretation of HERA events
is possible only for large $\lambda'_{121}\sim O( 0.1)$ in a large class of
supersymmetric standard models characterized by a positive charm squark 
(mass)$^2$ at the GUT scale. The simplest and the most popular
minimal supergravity model with universal boundary condition falls in
this class. The required large value of $R$ violating parameter
is difficult to admit without postulating new physics and /or fine
tuned cancellations due to constraints coming from the R-violating 
processes like neutrino mass, $K^+ \rightarrow \pi^+ \n \bar{\n}$ etc. 
However, the later data from HERA experiments did not substantiate
their claim for a resonance. Treating the deviations at 
$200$ GeV as statistical fluctuations, the HERA collaborations have
now put constraints on the mass and generic couplings of leptoquarks.
These bounds are now compatible with the results obtained from 
Tevatron \cite{cash}.

%% file: chap7.tex
\chapter{Conclusions}

The highly successful Standard Model has to be incorporated
in larger theories to have a deeper understanding of Nature.
Supersymmetric Standard Models have been built to solve
the hierarchy problem which is encountered whenever Standard
Model is being incorporated into a larger theory. Experimentally,
one of the first signals for physics beyond Standard Model is
provided by the atmospheric neutrino experiments which indicate
the existence of atleast one non-zero neutrino mass. An interpretation
of both solar and atmospheric neutrino anomalies interms of neutrino
oscillations would lead to a specific pattern of the neutrino mass
matrix. In this thesis, we studied the possibilities of generating
such a pattern within the context of Supersymmetric Standard Models. 

Neutrino masses can be generated in the Supersymmetric Standard Model
as it naturally allows for lepton number violation. Models which violate
lepton number are classified into bilinear
and trilinear lepton number violating models based on the dimensionality
of the couplings in the superpotential. Firstly, we studied in detail
models which violate lepton number only through bilinear 
(dimensionful) couplings. In these models, only three R-parity 
violating couplings determine the scale of the neutrino masses. The
masses are naturally small due to Yukawa suppression if universal 
boundary conditions are assumed at the high scale. 
After presenting a general analytical  structure of the neutrino 
mass matrix in these models, we have studied them in detail 
in the framework the Minimal Messenger Model (MMM) of gauge mediated 
supersymmetry breaking. The bilinear R-violating model within
MMM, is very economical with  essentially only four parameters, the three
R-parity violating couplings and the supersymmetry breaking scale $\Lambda$,
determining the three masses and three mixing angles of the neutrinos. 
We have shown that these models are too restrictive to provide 
simultaneous solutions for solar and atmospheric neutrino problems. 
For simultaneous solutions with MSW conversion, we have numerically
shown that these models do not provide the required mass hierarchy
in the neutrino mass eigenvalues. On the other hand, for simultaneous
solutions with vacuum oscillations, we have shown that these models 
cannot accommodate the two large mixings required. 
Thus,  extensions of the minimal 
messenger model are inevitable for existence of simultaneous solutions
for solar and atmospheric neutrino anomalies within these models. A
simple extension which would allow for a negative $\m$ parameter would
be able to provide the required  framework for simultaneous solutions. 

The other class of models we have considered are models with 
trilinear lepton number violation. These models, unlike the bilinear
case, offer much larger freedom due to the larger number of parameters
present. One of the major results of this thesis is the effect
of Renormalisation Group Evolution on the neutrino mass spectrum 
in the presence of only trilinear (dimensionless) lepton number
violating couplings in the model. RG evolution generates non-zero 
soft bilinear lepton number violating couplings at the weak scale
though these couplings are absent at the high scale. These would
generate vacuum expectation values ({\it vevs}) to the sneutrinos leading
to non-zero mixing between neutralinos and neutrinos. A `tree level'
neutrino mass is thus generated, which we call the `RG induced tree
level' mass. Analyses of the neutrino mass matrix in these models
presented in the literature have been neglecting  this important
contribution which can drastically alter the neutrino mass spectrum
in these models. We have analysed the neutrino mass spectrum taking
into consideration RG induced tree level mass contribution in addition
to the standard 1-loop induced masses within the framework of mSUGRA 
inspired MSSM. Our results show that the model favours vacuum solution
to the solar neutrino problem while simultaneously allowing solutions
for the atmospheric neutrino problem. However, there exist regions
in the parameter space where the hierarchy in the neutrino mass 
eigenvalues is reduced due to a suppression of the tree level mass.
In these regions, Large Angle MSW solution is preferred which is
favoured by the latest data from super-Kamiokande. 

In addition to studying the neutrino mass spectrum in the R-parity 
violating scenarios, one can instead consider bounds on the 
lepton number violating couplings due to direct experimental 
limits on the neutrino masses.  Since neutrinos attain majorana
masses in this case, the most stringent limit on them comes from
the non-observation of neutrinoless double beta decay. Using this
experimental limit,  bounds on the R-parity violating couplings can be derived. 
The existing limits do not consider the RG effects on the neutrino
mass. But, as we mentioned above, the RG induced tree level mass 
would alter the neutrino mass spectrum drastically. We have included
this effect and re-derived bounds on the lepton number violating
couplings. In addition to this, we have also considered the effect 
of CKM matrix and discussed the  basis dependence of these bounds.
We have subsequently used these bounds in our discussion of the
charm squark interpretation of the now defunct HERA anomalies. However,
our analysis is of general nature and can be redone if a relevant
case arises in the future colliders.

\vskip 0.4cm 
\noindent
{\underline {\sc General properties of neutrino masses in R-violating theories}}
\vskip 0.4cm 
\noindent
So far, in this thesis, we have concentrated on specific models of 
R-violation and supersymmetry breaking to study the structure of 
neutrino mass matrix. Though some of the properties 
of the neutrino mass matrix (essentially related to solutions of 
solar and atmospheric neutrino problems) depend on specific models 
of R-violation and type of supersymmetry breaking, most of the 
properties hold true in any of the R-violating scenarios. We 
summarise them as follows.

\vskip 0.4cm
\noindent
{\it (i)~R -violating couplings}
\vskip 0.2cm
\noindent
Majorana masses for the neutrinos are produced in these models due to the
presence of lepton number violating couplings. The neutrino masses 
are proportional to the square of these couplings. In the limit R-violating
couplings are set to zero, there are no neutrino masses in these models 
and they represent the standard MSSM with R-parity. 

\vskip 0.4cm
\noindent
{\it (ii) Hierarchical masses}
\vskip 0.2cm
\noindent
Neutrinos attain masses both at the tree level as well as at the
1-loop level. The tree level mass is generated indirectly through
the bilinear soft $L$ violating terms in the scalar potential. They 
are either present `originally' in the scalar potential or 
generated at the weak scale through RG scaling. The total neutrino
mass matrix is sum of the tree level and 1-loop contributions. At the
tree level only one combination of the neutrinos attains mass. 
The other combinations attain mass at the 1-loop level only. The tree 
level mass is much larger compared to the 1-loop contribution for 
most of the R-violating theories. The hierarchy in the neutrino mass
spectrum is typically characterised by ${m_{1-loop} \over m_{tree}}$.
Since this factor is much smaller  than one, large hierarchy in the 
neutrino mass spectrum is natural in these models. In some theories
it is possible to find regions in parameter space where the tree 
level mass is suppressed and becomes comparable to the 1-loop level
mass.

\vskip 0.4cm
\noindent
{\it (iii) Large Mixing}
\vskip 0.2cm
\noindent
In most of these models, by  suitable analytical approximations one can
decouple the neutrino mass matrix into R-violation dependent part
and R-violation independent part. The R-violation dependent part 
determines the mixing in most of the cases. The mixing angles are
dependent on ratios of the R-violating parameters. When R-violating
parameters are equal in magnitude large mixing is natural in the
model \footnote{ This may not hold true in some models with only $\l$
couplings.}. But this does not guarantee us that two large mixing angles
can be simultaneously accommodated in these models. 

\vskip 0.4cm
\noindent
{\it (iv) Role of Yukawas}\\
\vskip 0.2cm
\noindent
The R-violating parameters which determine the neutrino masses are
always accompanied by down quark or charged lepton Yukawa couplings.
Thus in the limit of vanishing charged lepton and down quark Yukawa
couplings, neutrino masses also vanish in these theories (In models
with only dimensionful R violation this is true only in the case
of universal boundary conditions at the high scale). This dependence
on the Yukawa couplings can be understood in terms of a $U(1)$ 
symmetry. In the absence of down quark, charged lepton Yukawas and
the $\m$ term, the superpotential is invariant under the $U(1)$ with 
charges for the superfields:
\be
L~=~l\;\;\;;H_2~=~-l \;\;\;;  E^c~=~-2~l; \;\; D^c~=~-l ;\;\; U^c~=~l 
\ee
and the rest of the superfields carry zero charge. $l$ is an integer.
The neutrino remains massless in this case. The Yukawas and the $\m$
term break this symmetry and thus accompany the R-violating terms
in the neutrino mass formulae. 

\vskip 0.4cm
\noindent
{\it (v) Small R violation}\\
\vskip 0.2cm
\noindent

Neutrinos are the only standard model fermions which attain mass through
the soft sector in these theories. Whereas the loop induced masses 
are suppressed by the Yukawa couplings and are small, the tree level  
neutrino mass can be in general very large $\sim O(M_{susy})$
in these models. These large neutrino masses have to be suppressed to
be phenomenologically meaningful. 
These masses can be suppressed either in a natural way or by a 
choice of suppression of the parameters:\\
\noindent
(a) By a suitable choice of the boundary conditions for bilinear
lepton number violating soft terms at the high scale
where supersymmetry is broken, one can bring down the neutrino masses
to O(100 MeV) in these models \footnote{This may not generally hold 
true for some models with only trilinear $\l$ couplings.}. 
For example, by choosing these terms
to be zero at the high scale (as naturally happens in models with only
dimensionless lepton number violation) or to be of the same magnitude
as the other bilinear terms like $B_\m$ at the high scale (as happens
in models with universal boundary conditions) leads to bilinear
lepton number violating soft terms which are only Yukawa suppressed 
at the weak scale due to RG scaling. This kind of suppression can 
generally lead to O(100 MeV) neutrino masses. 

\noindent
(b) One can always `choose' the two contributions to the sneutrino
{\it vev} to cancel each other. This naturally happens for some parameter
space in mSUGRA inspired MSSM. But, in model independent analyses, which
we discuss below the two parameters which contribute to the sneutrino
{\it vev} or the sneutrino {\it vev} in general itself are free parameters
of the model. In these cases, there is no natural choice of the sneutrino
{\it vev} and it has to be chosen small in order to have a small neutrino
mass. 

Both the above methods would not be able to suppress the neutrino mass
sufficiently to give the right scale of $\sim$ O(eV). Thus one has to
choose extremely small R-violation to generate O(eV) neutrino masses
in these models. This is possible as the neutrino masses are directly
proportional to the R-violating couplings in these models. For example,
to accommodate solutions of solar and atmospheric neutrino problems, 
bilinear R-parity violating parameters are required to be O($10^{-5}$)$\times
M_{susy}$. Similarly the dimensionless $\l'$ couplings are typically 
required to be of O($10^{-4}$) in these models. Some models have been 
explored in literature where a natural way of such small R-violation
can be achieved  \cite{shafi,achunsmir,nirpo,nelson}.

\vskip 0.4cm
\noindent
{\it (vi) Experimental Signatures}\\
\vskip 0.2cm 
\noindent
R-violating theories allow for completely different experimental
signatures in comparison to the standard MSSM. The additional 
L violating couplings have characteristic signatures some of which
we have already seen in chapter 6. A comprehensive study of 
experimental signatures of R-violating theories  has been recently 
presented in \cite{lhouch}. The models presented in this thesis
have restricted the R-violating parameter space as well as the
standard supersymmetry soft parameters so as to provide simultaneous
solutions to solar and atmospheric neutrino problems. Thus these
models would have different mass spectrum and decay signatures. An
analysis of this type combining neutrino masses and experimental 
signatures have been done in case of some models recently \cite{valle,
bissou}. The other most important consequence of R-violation is that
the LSP is no longer stable. One has to look for a new cold dark matter
candidate in these models. 

Thus we see that R-parity violating theories provide a natural framework
where small neutrino masses can be realised. In these models neutrino
masses are calculable in terms of few basic parameters and thus can
predict the existence of solutions for solar and atmospheric 
neutrino masses within specific models of supersymmetry breaking. 
Both bilinear and trilinear R-violating models have various 
interesting features associated with them making studies within these
theories important and essential.